\documentclass[journal]{IEEEtran}
\usepackage{cite}
\usepackage{amsmath,amssymb,amsfonts}
\usepackage{algorithm}               
\usepackage{algorithmic}             
\usepackage{graphicx}
\usepackage{textcomp}
\usepackage{xcolor}
\usepackage{setspace}
\usepackage{subfigure}
\usepackage{multirow} 
\usepackage{diagbox}
\usepackage{color}
\usepackage{stfloats}

\begin{document}
	\bstctlcite{IEEEexample:BSTcontrol}
	
	\title{Reliability-Latency-Rate Tradeoff in Low-Latency Communications with Finite-Blocklength Coding}

\author{Lintao Li,~\IEEEmembership{Graduate Student Member,~IEEE,}
Wei Chen,~\IEEEmembership{Senior Member,~IEEE,} \\
Petar Popovski,~\IEEEmembership{Fellow,~IEEE}, and Khaled B. Letaief,~\IEEEmembership{Fellow,~IEEE} \vspace{-6mm}

\thanks{Manuscript received 13 September 2023; revised 7 April 2024 and 6 August 2024; accepted 9 October 2024. This work is supported in part by the NSFC/RGC Joint Research Scheme under Grant No. 62261160390/N\_HKUST656/22, in part by the Villum Investigator Grant ``WATER” from the Velux Foundations, Denmark, in part by the National Natural Science Foundation of China under Grant No. 62471276, in part by the Research Grants Council under the Areas of Excellence Scheme under Grant AoE/E-601/22-R, and in part by the National Key Research and Development Program of China under Grant 2018YFA0701601. (\emph{Corresponding author: Wei Chen}.)}

\thanks{Lintao Li and Wei Chen are with the Department of Electronic Engineering, Tsinghua University, Beijing 100084, China. They are also with the State Key Laboratory of Space Network and Communications and the Beijing National Research Center for Information Science and Technology, Beijing 100084, China (email: llt20@mails.tsinghua.edu.cn; wchen@tsinghua.edu.cn).

	Petar Popovski is with the
Department of Electronic Systems, Aalborg University, 9220 Aalborg, Denmark
(e-mail: petarp@es.aau.dk). 

Khaled B. Letaief is with the Department of Electronic and Computer Engineering, Hong Kong University of Science and Technology, Clear Water Bay, Hong Kong (email: eekhaled@ust.hk).

© 2024 IEEE. Personal use of this material is permitted. Permission
from IEEE must be obtained for all other uses, in any current or future media, including reprinting/republishing this material for advertising or promotional purposes, creating new collective works, for resale or redistribution to servers or lists, or reuse of any copyrighted component of this work in other works.}}

\maketitle

\begin{abstract}

Low-latency communication plays an increasingly important role in delay-sensitive applications by ensuring the real-time information exchange. However, due to the constraint on the maximum instantaneous power, guaranteeing bounded latency is challenging. In this paper, we investigate the reliability-latency-rate tradeoff in low-latency communication systems with finite-blocklength coding (FBC). Specifically, we are interested in the fundamental tradeoff between error probability, delay-violation probability (DVP), and service rate. Based on the effective capacity (EC), we present the gain-conservation equations to characterize the reliability-latency-rate tradeoffs in low-latency communication systems. In particular, we investigate the low-latency transmissions over an additive white Gaussian noise (AWGN) channel and a Nakagami-$m$ fading channel. By defining the service rate gain, reliability gain, and real-time gain, we conduct an asymptotic analysis to reveal the fundamental reliability-latency-rate tradeoff of ultra-reliable and low-latency communications in the high signal-to-noise-ratio (SNR) regime. To analytically evaluate and optimize the quality-of-service-constrained throughput of low-latency communication systems adopting FBC, an EC-approximation method is conceived to derive the closed-form expression of that throughput. Our results may offer some insights into the efficient scheduling of low-latency wireless communications, in which statistical latency and reliability metrics are crucial.
\end{abstract}


\section{Introduction}

With the tremendous development of communication technologies, the flow of wireless data transmission has experienced an exponential growth in recent years, supported by a higher transmission rate and ubiquitous connectivity. Under this trend, multiple delay-sensitive applications have garnered significant attention in the fifth-generation (5G) communication networks. Furthermore, ongoing research on the six-generation (6G) communication networks indicates that there will be more stringent requirements for delay and reliability in 6G applications \cite{6g}. As one of the essential requirements for mission-critical and emerging applications in 5G and 6G, low-latency communication has been a popular topic under active consideration in the research community.

In the realm of low-latency communications, ultra-reliable and low-latency communications (URLLC) represents a typical application scenario. URLLC aims to achieve an ultra-low and bounded delay to support mission-critical applications. For the next-generation URLLC, the latency requirement will be more stringent \cite{extreme}. Additionally, providing real-time user experiences requires communication latency to be tightly controlled or deterministic \cite{6Gtime}. However, due to the fading nature of wireless channels, ensuring a bounded delay is usually challenging \cite{iccli1}, \cite{hlicc}. Furthermore, limited maximum power brings more difficulty of achieving a bounded delay \cite{icc1}. As a result, statistical quality-of-service (QoS) metrics have become the prevalent choice for assessing reliability and latency in URLLC \cite{URLLC12}. Statistical QoS is typically characterized by the queue length or delay thresholds and the associated queue length or delay violation probability \cite{urllcproc}.  As mentioned in \cite{bennis}, focusing solely on average-based metrics without considering the tail behavior of wireless systems makes it difficult to satisfy the low-latency and reliability requirements. Several studies have explored the statistical delay in closed-loop control \cite{stdelay1} and edge computing systems \cite{stdelay2}. Therefore, incorporating statistical QoS analysis into URLLC and other low-latency systems is essential for characterizing performance limits and optimizing QoS metrics accordingly.

To analyze the QoS of communication systems, the first step is to formulate a proper model, which takes QoS metrics into account. In low-latency communications, the primary concern lies in the probabilities of queue length and delay exceeding certain thresholds \cite{bennis}, both of which are critical indicators of QoS performance. To this end, the concepts of effective bandwidth (EB) and effective capacity (EC) are utilized to perform asymptotic QoS analysis via a link-layer queuing model. With the help of EB and EC, the queue-length-violation probability (QVP) and delay-violation probability (DVP) can be expressed concisely. EB refers to the minimum constant service rate required for a given arrival process to meet the QoS constraints \cite{eb}. EC is the dual concept of EB, representing the maximum constant source rate that can be supported by the given service process to meet the QoS requirement \cite{ec}. Based on EB, EC, and Shannon's formula, several studies focused on the performance analysis and resource allocation designs under statistical QoS constraints in various communication systems with different channel and arrival models, including independent Rayleigh fading channel \cite{rayleigh1}, time-correlated Rayleigh fading channel \cite{classic},  and Nakagami-$m$ channel \cite{pc1}. Specifically, the high-SNR slope of EC was analyzed in \cite{rayleigh1} under the assumption of infinite blocklength.

With a proper analysis model established, the next step is to reveal the relationship between rate and reliability in the finite blocklength (FBL) regime. A commonly quoted requirement for 5G is achieving a latency of less than 1 ms and a packet loss probability no greater than $10^{-5}$ \cite{URLLC12}. To satisfy these requirements, the assumption that the blocklength shall tend to infinity has to be revised given the limited bandwidth. In particular, Shannon's formula, which is asymptotically accurate with the infinite blocklength, is not accurate enough in the FBL regime. Thus, the precise approximation of the capacity with finite-blocklength coding (FBC) is expected to play a central role in addressing this problem. A milestone work \cite{codingrate} analyzed the maximum coding rate in the FBL regime. The normal approximation was also proposed in \cite{codingrate} to estimate the maximum coding rate for the AWGN channel in the FBL regime. This conclusion inspired further studies into the analysis of the coding rate for various communication scenarios in the FBL regime, including Gilbert-Elliott channels \cite{fbl1}, quasi-static multiple-antenna fading channels \cite{Yang2014}, and coherent block fading channels \cite{fbl4}. Additionally, Lancho, Koch, and Durisi proposed a high-SNR approximation of the maximum coding rate in the FBL regime based on unitary space-time modulation \cite{highsnr}. Under the FBL assumption, the properties of EC differ from those derived under the infinite blocklength assumption.

\begin{table}[t]
	\centering 
	\caption{Main Notation} 
	\begin{center}  
		\begin{tabular}{|c|c|} 
			\hline  
			\multicolumn{1}{|c|}{\textbf{Symbol}} & \multicolumn{1}{c|}{\textbf{Definition}} \\ \hline  
			$|h[n]|^2$ & channel power gain  \\ \hline  
			$T$ & the number of time slots in a frame\\ \hline
			$f(\cdot)$ & distribution of 	$|h[n]|^2$ \\ \hline
			$m$, $\Omega$ & parameters of Nakagami-$m$ fading \\ \hline
			$\gamma$ & transmitted SNR \\ \hline
			$\epsilon$ & error probability \\ \hline
			$N$ & blocklength \\ \hline
			$R_N^{\epsilon}[n]$ & maximum coding rate \\ \hline
			$\tilde{R}_N^{\epsilon}[n]$ & normal approximation \\ \hline
			$G(\cdot,\cdot)$ & error term of normal approximation \\ \hline
			$\theta$ & QoS exponent \\ \hline
			$\alpha_A(\theta)$ & effective bandwidth \\ \hline
			$\alpha_S(\theta)$ & effective capacity \\ \hline
			$\Lambda(\theta)$ & normalized effective capacity   \\ \hline
			$\chi$ & queue-length-violation probability \\ \hline 
			$\delta$ & delay-violation probability \\ \hline
			$\Psi(\gamma)$ & function of $\gamma$, $\lim\limits_{\gamma\to+\infty}\Psi(\gamma)=+\infty$ \\ \hline
			$\varrho$ & revised QoS exponent \\ \hline
			$s_{\infty}$ & high-SNR slope of normalize EC \\ \hline
			$\zeta$ & service-rate gain \\ \hline
			$\varpi$ & reliability gain \\ \hline
			$\tau$ & real-time gain \\ \hline
			$\Xi(\cdot)$ & power allocation scheme \\ \hline
		\end{tabular}  
	\end{center}  
\end{table}

Combining the FBC analysis with the EC-based model, many studies tried to characterize and optimize the performance of low-latency communication systems in terms of reliability, latency, and rate. There are mainly two lines of work on this topic. The first line of work utilized EC to derive the QVP and DVP of the low-latency systems and then optimized the system performance. The authors of \cite{fbleb1} validated that the DVP derived from EC serves as the upper bound of the actual DVP for certain types of arrival processes, even with small delay thresholds. This finding highlighted the potential applicability of the EC-based model in URLLC systems. Building on this insight, a joint uplink and downlink resource configuration problem was formulated in \cite{she3}, aiming to enhance the spectrum usage efficiency while maintaining stringent reliability and latency requirements. In contrast to these works, another line of research focused on maximizing EC to improve QoS-constrained throughput in the FBL regime. EC maximization has been explored in various kinds of fading channels, including Rayleigh \cite{ecresouce1,thirdorder2}, Rician \cite{zxjsac}, and Nakagami-$m$ \cite{8466036}, each considering different constraints and conditions in the FBL regime. In \cite{eur}, the author calculated the throughput for a simple automatic repeat request (ARQ) mechanism in the FBL regime, which was subsequently applied in \cite{huarq}. Despite the growing body of research using EC and FBC for QoS optimization in low-latency systems, the theoretical performance limit of EC in the FBL regime remains an open problem. Additionally, the presence of the channel dispersion term in the FBC rate complicates the derivation of an analytical expression for EC, posing additional challenges in assessing QoS-constrained throughput and further refining EC optimization.

As mentioned above, although many works have sought to optimize the reliability, latency, and service rate in low-latency systems, there remains a gap in the comprehensive analysis of the fundamental tradeoff, namely the reliability-latency-rate tradeoff, in the FBL regime. In our previous work \cite{YLTCOM}, the gain-conservation equation was derived to characterize the relationships between the reliability gain, real-time gain, and service-rate gain. However,  \cite{YLTCOM} only considered bounded random arrivals, which is equivalent to the deterministic arrival under the high-SNR assumption. Moreover, the conclusions in \cite{YLTCOM} did not account for the influence of high SNR on the FBC rate. Therefore, there is still a need for a rigorous analysis of the reliability-latency-rate tradeoff in low-latency communication systems, particularly with FBC and random arrivals.

\begin{figure*}[t]
	\centerline{\includegraphics[width=18cm]{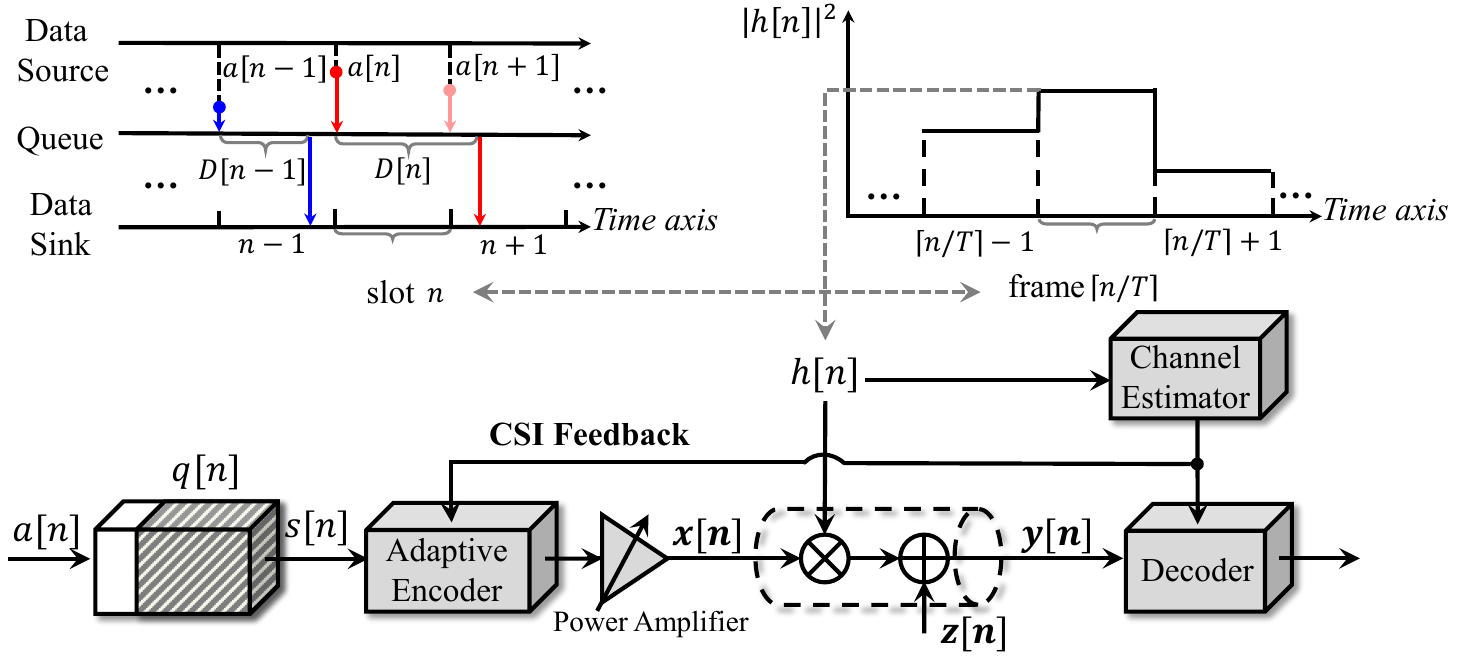}}
	\caption{System Model.}
	\label{sys}
\end{figure*}

In this paper, we characterize the fundamental tradeoff between reliability, latency, and rate in low-latency systems with FBC. Specifically, we begin by analyzing the high-SNR slope of EC in the FBL regime. We consider two classical scenarios including transmission over an AWGN channel and a Nakagami-$m$ fading channel. First, we consider the case in which the error probability is fixed to determine the service-rate gain, defined as the derivative of EC with respect to the logarithm of the SNR. Based on this result, we further explore the case in which the error probability is a function of SNR. The reliability gain and real-time gain are then defined as the derivatives of logarithms of error probability and DVP with respect to the logarithm of the SNR. Subsequently, we derive the revised gain-conservation equations for both AWGN and fading channels. Additionally, a detailed discussion of the tradeoff between the error probability and latency is provided for the AWGN channel. For efficiently evaluating EC in the FBL regime, we propose a Laplace's-method-based EC approximation approach to derive an analytical expression for EC. Based on the discussion on the fixed-power allocation scheme, we further extend our analysis to consider the EC approximation for certain kinds of channel-gain-based power allocation schemes, with a generalization to systems incorporating the simple retransmission mechanism.

The rest of this paper is organized as follows. Section II-A presents the system model, including the physical layer model and network layer model. Section II-B introduces the main results of this paper, which are summarized by four main theorems. In Section III A-C, we provide conclusions and corresponding analysis for the revised gain-conservation equations. The tradeoff between the error probability and latency in the AWGN channel is discussed in Section III-D. Further, the EC approximation approach is proposed in Section IV. Numerical results are presented in Section V. Finally, the conclusion is given in Section VI.

Throughout this paper, $u'(x)$ denotes the first-order derivative of $u(x)$ for $x$, while $u''(x)$ denotes the second-order derivative. $\mathbb{N}^+$ denotes the set of positive natural numbers. $\mathbb{E}_X\{X\}$ is the expectation of random variable $X$, while ${\rm Var}_X\{X\}$ is its variance. For $X[n]$, we omit index $n$ in the subscripts. $e$ denotes Euler's number. $\boldsymbol I_N$ denotes a $N$-dimensional identity matrix. For two random variables $X$ and $Y$, ${\rm Cov}(X,Y)$ denote the covariance between $X$ and $Y$.  The Gaussian $Q$-function is $Q(x)=\frac{1}{\sqrt{2 \pi}}\int_x^\infty e^{-\frac{t^2}{2}}{\rm d}t$, while $Q^{-1}(\cdot)$ represents the inverse Gaussian $Q$-function. $\Gamma(s)=\int_0^{\infty} t^{s-1}e^{-t}{\rm d}t$ denote the gamma function. Besides, define the incomplete gamma function as $\Gamma(s,x)=\int_{x}^{\infty}t^{s-1}e^{-t}{\rm d}t$, and the exponential integral function as ${\rm Ei}(x)=-\int_{-x}^{\infty}\frac{e^{-t}}{t}{\rm d}t$. $W(\cdot)$ denotes the Lambert W function \cite{lambertw}. $O(\cdot)$ is the big-O notation. For two functions $f(x)$ and $g(x)$, $f(x)=O(g(x))$ holds if there exists a positive real number $M$ and a real number $x_0$ such that $|f(x)|\leq M|g(x)|$ for all $x\geq x_0$.

\section{System Model and Main Results}

In this section, we present a system model with FBC, focusing on the statistical delay performance. To evaluate the latency performance, we adopt two key metrics: the queue-length-violation probability (QVP) and delay-violation probability (DVP). Specifically, we will first describe the system model in Section II-A, which includes physical layer and queueing models. Following that, we will introduce the main results of this paper, summarized in four main theorems in Section II-B.

\subsection{System Model}

As shown in Fig. 1, we focus on a point-to-point low-latency communication system. The arrival data generated by the source will be transmitted through the channel. Time is divided into time slots with equal length $T_0=\frac{N}{B}$ (in seconds), where $B$ is the bandwidth of this system and $N$ is the blocklength. Let $T_f$ (in seconds) denote the frame length. Each frame contains $T=\frac{T_f}{T_0}$ time slots, where $T\in\mathbb{N}^+$.\footnote{Note that, for the high-mobility or THz
communication system, the channel coherence time is significantly small \cite{thz}. Thus, $T=1$ also covers the low-latency scenarios. } Let $h[n]$ denote the channel coefficient of time slot $n$. Assume that the frame duration is equal to the channel coherence time. Thus, the channel coefficient remains constant within a frame, i.e., $h[iT+1]=h[iT+2]=\cdots=h[(i+1)T]$, $i=0,1,\cdots$, and varies in an independently and identically distributed ($i.i.d.$) manner across different frames.	With blocklength $N$, the channel output $\boldsymbol y[n]$ at the $n$-th time slot with the input $\boldsymbol x[n]$ can be expressed as 
\begin{equation}
\boldsymbol y[n]=h[n]\boldsymbol x[n]+\boldsymbol z[n],
\end{equation}
where $\boldsymbol z[n]\sim\mathcal{CN}(0,N_0 \boldsymbol {I}_N)$ denotes the  AWGN. For the AWGN channel, $\boldsymbol{x}[n]\in \mathbb{C}^N$ and $h[n]=1$. For the fading channel, the probability density function ($p.d.f.$) of $|h[n]|^2$ is denoted by $f(x)$, while its cumulative distribution function ($c.d.f.$) is denoted by $F(x)$. For a wireless system in a Nakagami-$m$ fading channel with $\mathbb{E}_{|h|^2}\left\{|h[n]|^2\right\}=\Omega$, we have
\begin{equation}
f(x)= \left(\frac{m}{\Omega}\right)^m\frac{x^{m-1}}{\Gamma(m)} e^{-\frac{m}{\Omega}x}, \quad x>0, \, m\geq 0.5. \label{nakadis}
\end{equation}
With $\Omega=1$ and $m=1$, the system is in a Rayleigh fading channel with $\mathbb{E}_{|h|^2}\left\{|h[n]|^2\right\}=1$. For a wireless system with spatial or frequency diversity $\kappa$ in Rayleigh fading channels, the equivalent channel power gain by adopting maximum-ratio combining can be represented by setting $m=\kappa$ and $\Omega=\kappa$.

\begin{figure*}[b]
	\normalsize
	\hrulefill
	\setcounter{equation}{6} 
	\begin{equation}
		\begin{aligned}
			R_N^{\epsilon} = N\log_2(1+\gamma)-\sqrt{N\left(1-\frac{1}{(1+\gamma)^2}\right)}Q^{-1}(\epsilon)\log_2 e  + \frac{\log_2 N}{2}+G(N,\gamma,\epsilon). \label{c1}
		\end{aligned}
	\end{equation}
	\setcounter{equation}{8} 
	\begin{equation}
		\begin{aligned}
			R^{\epsilon}_N[n]  = N\log_2\left(1+|h[n]|^2\gamma\right)-\sqrt{N\left(1-\frac{1}{\left(1+|h[n]|^2\gamma\right)^2}\right)}Q^{-1}(\epsilon)\log_2 e+\frac{\log_2 N}{2}+G\left(N,|h[n]|^2\gamma,\epsilon\right). \label{c2}
		\end{aligned}
	\end{equation}
\end{figure*}

Next, we introduce the notion of the channel code. In this paper, we assume that the channel state information (CSI) is perfectly known at both the receiver and the transmitter (CSIRT) through the channel estimation and feedback as shown in Fig. \ref{sys}.\footnote{To ensure the perfect CSI assumption, either the average power of pilot symbols should be high, or the pilot length should be large \cite{Tse}. In environments with large coherence bandwidth or coherence time, such as indoor scenarios with small delay and Doppler spread, a large pilot length is feasible \cite{gold,robert}. Additionally, achieving perfect CSIT requires a feedback path with high-resolution quantization and sufficient SNR to transmit the CSI accurately and reliably \cite{robert}. In latency-constrained systems, available time resources for transmission is limited. Therefore, using high SNR or considering systems with a large coherence bandwidth can help maintain the perfect CSI assumption while adhering to latency constraints.} $\Delta N= B\Delta T$ channel uses are utilized for the channel estimation in each frame, which we neglect in this paper.\footnote{Note that for short packet communication with extremely small blocklength, the assumption that perfect CSIRT may be unrealistic. Moreover, analyzing the system with extremely small blocklength requires more effort to derive a precise expression for the coding rate based on \cite{codingrate}.} With CSIRT, the transmitter can adaptively determine the coding rate. Specifically, with $h[n]=h$, an $(M_h,N,\gamma,\epsilon)$ code consists of \cite{Yang2014}
\begin{enumerate}
\item An encoder $f^h_{\rm tx}:\{1,\cdots,M_h\}\times h \to \mathbb{C}^N$ that maps the message $j \in \{1,\cdots,M_h\}$ and the channel $h$ to a codeword $\boldsymbol{x}=f^h_{\rm tx}(j,h)$ satisfying
\setcounter{equation}{2} 
\begin{equation}
\frac{\|\boldsymbol x[n]\|^2}{N_0}=N\gamma.
\end{equation}
\item A decoder $g_{\rm rx}^h: \mathbb{C}^{N}\times h\to \{1,\cdots,M_h\}$ satisfying
\begin{equation}
\sum_{j=1}^{M_h}\Pr\left\{g_{\rm rx}^h(\boldsymbol{x},h)\neq J|J=j\right\}\Pr\{J=j\}\leq \epsilon.
\end{equation}
\end{enumerate}

The maximum instantaneous coding rate in time slot $n$, which is defined as $R_N^{\epsilon}[n]$, with $(M_h,N,\gamma,\epsilon)$ code is thus defined as 
\begin{equation}
	\begin{aligned}
		R_N^{\epsilon}[n]=\sup&\left\{\log_2 M_h:\right. \\
		& \left.\forall(M_h,N,\gamma,\epsilon) \text{ code} \,\, | \,\,  h[n]=h\right\}.
	\end{aligned}
\end{equation}

The conclusion in \cite{codingrate, thirdorder} showed that for channels with capacity $C$, the instantaneous maximum coding rate (in bits per $N$ channel uses) with $\gamma$, $N$, and $\epsilon$, can be expressed as
\begin{equation}
R_N^\epsilon[n]=NC[n]-\sqrt{NV[n]}Q^{-1}(\epsilon)+\frac{\log N}{2}+O\left(1\right),\label{normal}
\end{equation}
where $V[n]$ is the channel dispersion, and $O(1)$ is the error term that is bounded for large $N$ and may be related to $\gamma$ and $\epsilon$ at the receiver \cite{codingrate}.\footnote{Note that there is no existing work presenting a general conclusion on whether the error term is bounded for $\gamma$. We will discuss this in Section II-B.} For the AWGN channel, we omit the index $n$. According to \cite{thirdorder}, the maximum coding rate (in bits per $N$ channel uses) for a given $\epsilon$, $N$, and $\gamma$ for the AWGN channel is presented in Eq. \eqref{c1}, which is located at the bottom of this page. In Eq. \eqref{c1}, we call $G(N,\gamma,\epsilon)$ error term function. For the given $\gamma$ and $\epsilon$, $G(N,\gamma,\epsilon)$ satisfies \cite{thirdorder}
\setcounter{equation}{7} 
\begin{equation}
\lim_{N\to+\infty} |G(N,\gamma, \epsilon)|<+\infty.
\end{equation}

Moreover, given the channel gain $|h[n]|^2$, the maximum coding rate (in bits per $N$ channel uses) with $\epsilon$, $N$, and $\gamma$ for the fading channel is given in Eq. \eqref{c2}, which is presented at the bottom of this page. For simplicity, we define $\tilde{R}^{\epsilon}_{N}[n]=R^{\epsilon}_{N}[n]-G(N,|h[n]|^2\gamma,\epsilon)$, which is called normal approximation in \cite{codingrate}.

To backlog the packets to be transmitted, an infinite-length buffer is assumed to be available at the transmitter. Let $a[n]$ denote the size of the arrival data at the beginning of time slot $n$. In this paper, we assume that $a[n]$ follows an arbitrary stochastic distribution with finite expectation and variance in this system, and $a[n]$ changes in an $i.i.d.$ manner across different time slots. The average size of the arrival data is denoted by $\mu_a$, while the variance is denoted by $\sigma_a^2$. Accordingly, the accumulated number of the arrival data until time slot $M$ can be expressed as $A[M]=\sum_{n=1}^{M}a[n]$. Similarly, $s[n]$ is defined as the size of the service data in time slot $n$. The average size of the service data is denoted by $\mu_s$, while the variance of $s[n]$ is denoted by $\sigma_s^2$. The accumulated number of service data until time slot $M$ can be expressed as $S[M]=\sum_{n=1}^{M}s[n]$.

As shown in Fig. \ref{sys}, we assume a first-in first-out (FIFO) policy for serving the backlogged packets. 
Let $Q[n]$ denote the length of the queue at the end of time slot $n$. Then, the queue length is updated as
\setcounter{equation}{9} 
\begin{equation}
Q[n]=\left(Q[n-1]+a[n]-s[n]\right)^+,
\end{equation}
where $(x)^+=\max\{0,x\}$.

\emph{Remark 1: In this paper, we primarily focus on the analysis with CSIRT. There are also some existing conclusions on the coding rate in the FBL regime with CSIR only. A rigorous analysis of $R_N^{\epsilon}$ in the block fading channel with CSIR for single-input and single-out (SISO) systems was carried out in \cite{CSIR1}. For real transmitted symbols and noise, an analytical expression for the dispersion was given in Eqs. (27) and (36) of \cite{CSIR1} under the assumption that each symbol experiences $i.i.d.$ fading. For simplicity, we omit the index $n$ in this remark. The dispersion for the real block fading channel is given by}
	\begin{equation}
		\begin{aligned}
			V_r={\rm Var}_{|h|^2}&\left(\frac{1}{2}\log_2\left(1+|h|^2\gamma \right)\right)+\\
			&\frac{\log_2 e}{2}\left(1-\mathbb{E}_{|h|^2}^2\left\{\frac{1}{1+|h|^2\gamma}\right\}\right).
		\end{aligned}
	\end{equation}
\emph{Moreover, the dispersion for the complex block fading channel is given by}
\begin{equation}
	\begin{aligned}
		V_c={\rm Var}_{|h|^2}&\left(\log_2\left(1+|h|^2\gamma \right)\right)+\\
		&\log_2 e\left(1-\mathbb{E}_{|h|^2}^2\left\{\frac{1}{1+|h|^2\gamma}\right\}\right).
	\end{aligned}
\end{equation}

	\emph{This conclusion was further generalized in \cite{fbl4}, which considered multiple-input and multiple-output (MIMO) systems with real transmitted symbols and noise. Additionally, they considered the case in which $T_d$ symbols experience the same fading. According to Eqs. (16) and (17) in \cite{fbl4}, the dispersion $V$ for a SISO system with a real block fading channel is represented by Eq. \eqref{vm}, which is presented at the bottom of this page. In Eq. \eqref{vm}, $\eta_1$, $\eta_2$, $v^*(1,T_d)$ are defined by Eqs. (14), (15), and (23) of \cite{fbl4}, respectively.}
	\begin{figure*}[b]
			\normalsize
		\hrulefill
		\setcounter{equation}{12} 
	\begin{equation}
		\begin{aligned}\label{vm}
				V_m=&T_d {\rm Var}_{|h|^2}\left(\frac{1}{2}\log_2\left(1+|h|^2\gamma \right)\right)+\frac{\log_2^2e}{2}\mathbb{E}_{|h|^2}\left\{1-\frac{1}{(1+|h|^2\gamma)^2}\right\}+\gamma^2\left(\eta_1-\frac{\eta_2}{T_d}v^*(1,T_d)\right)\\
				=&T_d{\rm Var}_{|h|^2}\left(\frac{1}{2}\log_2\left(1+|h|^2\gamma \right)\right)+\frac{\log_2^2e}{2}\left(1-\mathbb{E}_{|h|^2}^2\left\{\frac{1}{1+|h|^2 \gamma}\right\}\right).
			\end{aligned}
	\end{equation}
\end{figure*}

	\emph{A comparison between the results of \cite{CSIR1} and \cite{fbl4} reveals that the findings in \cite{fbl4} serve as a generalization of \cite{CSIR1} for the real block fading channel. When $T_d=1$, the results align with each other. However, for more general cases, such as $T_d>1$ in SISO systems or MIMO systems with real block fading channel, \cite{fbl4} offers additional insights.}

\subsection{Effective Bandwidth and Capacity}
In this subsection, we will introduce effective bandwidth (EB) and effective capacity (EC). 
The essence of EB and EC is to use large deviation theory (LDT) to characterize the tail distribution of the length and delay violation probability. By using the Gartner-Ellis theorem \cite{ldt} and given a length threshold $L$, QVP satisfies
\setcounter{equation}{13}
\begin{equation}\label{ttt1}
	\lim_{L \to +\infty} \frac{\ln\Big(\sup_{m}\big\{\Pr\{Q[m] \geq L\}\big\}\Big)}{L}=-\theta  ,
\end{equation}
where $\theta$ is called QoS exponent, indicating the decay rate of the tail distribution of the queue length. According to \cite{eb}, the EB of the arrival process is defined as 
\begin{equation}
	\alpha_A(\theta)=\lim_{M \to +\infty} \frac{1}{\theta M} \ln \mathbb{E}_{A} \left\{ e^{ \theta A[M]} \right\}.\label{eba}
\end{equation}
EB of the arrival process indicates the minimum constant transmission rate for guaranteeing certain QoS requirements as shown in Eq. \eqref{ttt1}. Similarly, according to \cite{ec}, the EC of the service process is defined as  
\begin{equation}
	\alpha_S(\theta)=\lim_{M \to +\infty}- \frac{1}{\theta M} \ln \mathbb{E}_S \left\{ e^{- \theta S[M]} \right\}.\label{ebs}
\end{equation}
EC of the service process refers to the maximum constant arrival rate that meets the QoS requirements shown in Eq. \eqref{ttt1}.

With a given $\gamma$, $s[n]$ is $i.i.d.$. Thus, for $T\geq 1$, EC is given by
\begin{equation}\label{multitec}
\begin{aligned}
\alpha_S(\theta)=&\lim_{M \to +\infty} -\frac{1}{\theta M} \ln \mathbb{E}_S \left\{ e^{- \theta S[M]} \right\}  \\
=&\lim_{M\to +\infty}-\frac{1}{\theta T}\cdot\frac{T}{M} \ln \mathbb{E}_s\left\{e^{-\theta \sum_{i=0}^{\frac{M}{T}-1}\sum_{j=1}^Ts[iT+j]}\right\}  \\
=&-\frac{1}{\theta T}\ln \mathbb{E}_{|h|^2}\left\{e^{-\theta T s[n]}\right\}.
\end{aligned}
\end{equation}

To simplify the derivation, we normalize the EC by the blocklength, which is also called effective rate \cite{rayleigh1}. We let $\Lambda(\gamma)$ denote the normalized EC, which is a function of $\gamma$ with a given $\theta$. According to Eq. \eqref{multitec},  $\Lambda(\gamma)$ of the $i.i.d.$ $s[n]$ is given by
\begin{equation}
	\begin{aligned}
		\Lambda(\gamma)&=-\frac{1}{\theta NT}\ln \mathbb{E}_{|h|^2}\left\{e^{-\theta T s[n]}\right\}\\
		&=-\frac{1}{\theta NT}\ln\mathbb{E}_{|h|^2}\left\{e^{-\theta N T \frac{R_{N}^{\epsilon}[n]}{N}}\right\} \quad \text{bits/s/Hz}. \label{normalizeddefine}
	\end{aligned}
\end{equation}  

Note that the effective rate defined in \cite{rayleigh1} is equal to $\Lambda(\gamma)$ when $T=1$. In this paper, we use $\Lambda(\gamma)$ to characterize the service-rate gain of the low-latency communication systems, which aligns with \cite{rayleigh1}.

Let $Q[\infty]$ denote the steady state of $Q[n]$. If $Q[\infty]$ exists, QVP is denoted by $\chi=\Pr\{Q[\infty]\geq L\} $.  With a large threshold $L$ (in bits), $\chi$ can be approximated according to Eq. \eqref{ttt1} and \cite{ec} as
\begin{equation}
\chi\simeq \eta e^{-\theta L}.\label{9}
\end{equation}
In Eq. \eqref{9}, $\eta$ is the probability that the queue is not empty, which can be approximated by the ratio of $\mu_a$ to $\mu_s$.\footnote{Note that by setting $\eta=1$ we obtain upper bounds of $\chi$ and $\delta$ defined below with arbitrary loads \cite{Whitt1996} Thus, the derived results in this paper can be seen as a converse analysis for the low-load scenario.} The notation $a(x)\simeq b(x)$ means that $\lim_{x\to \infty}\frac{a(x)}{b(x)}=1$. In this paper, we focus on the high-load scenario, i.e., $\eta\to 1$, which is more common in emerging applications with the increasing demand for information exchange \cite{demand} and is more likely to cause the occurrence of extreme events, e.g., queue length and delay violations. $\theta$ is the solution to the following equation \cite{chang2012performance}:
\begin{equation}
\alpha_A(\theta)-\alpha_S(\theta)=0.\label{eqso1}
\end{equation}

Let $D[m]$ denote the delay experienced by the data arriving at the start of time slot $m$.  The steady state of $D[n]$ is denoted by $D[\infty]$. If $D[\infty]$ exists, DVP is denoted by $\delta=\Pr\{D[\infty]\geq D_{\rm max}\}$.  With a large delay bound $D_{\rm max}$ (in slots), $\delta$ in the high load scenario is approximated as \cite{ec}, \cite{Whitt1996}
\begin{equation}
	\delta \simeq  e^{-\theta \alpha_A(\theta) D_{\rm max}}.  \label{delta}
\end{equation}

\begin{figure*}[t]
	\centerline{\includegraphics[width=18cm]{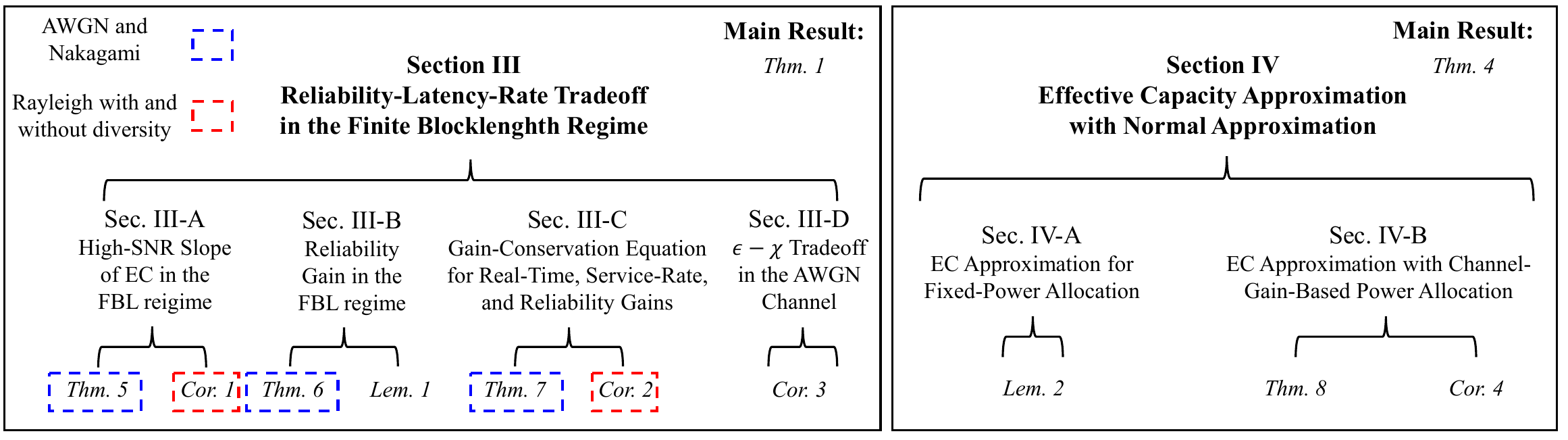}}
	\caption{Structures of Section III and Section IV.}
	\label{struc}
\end{figure*}

\subsection{Main Results}

In this subsection, we present the main results of this paper. The main results of this paper can be divided into two aspects, which refer to the revised gain-conservation equations with random arrivals in the high-SNR regime and a Laplace's-method-based approximation approach for EC. Detailed analysis and discussions are provided in Sections III and IV. The structures of Sections III and IV are illustrated in Fig. \ref{struc}. 

Before presenting the main results, we first introduce key definitions. We define $\zeta=\lim\limits_{\gamma\to+\infty}\frac{\partial{\Lambda(\gamma)}}{\partial{\log_2\gamma}}$ as the QoS-constrained service-rate gain, similar to the concept of the service-rate gain in \cite{YLTCOM}. $\zeta$ indicates the growth rate of throughput with respect to $\log_2\gamma$ in the high-SNR regime. Next, we define the reliability gain $\varpi=\sqrt{2\log_2e}\lim\limits_{\gamma\to+\infty}\frac{\partial\sqrt{-\log_2 \epsilon(\gamma)}}{\sqrt{N}\partial \log_2\gamma}$, which quantifies the decay rate of error probability as $\log_2\gamma$ increases in the high-SNR regime, consistent with the reliability gain proposed in \cite{YLTCOM}.\footnote{Note that the constant $\sqrt{2\log_2e}$ aligns with \cite{YLTCOM} since dispersion term is treated as a constant in \cite{YLTCOM}. Omitting this constant does not alter the conclusions but would modify the coefficient of $\varpi$ in the gain-conservation equations.} We also define the real-time gain $\tau=-\frac{1}{D_{\rm max}}\frac{\partial\log_2\delta}{\partial \log_2\gamma}$, which is analogous to the definition in \cite{YLTCOM}. $\tau$ captures the decay rate of DVP with increasing $\log_2\gamma$ in the high-SNR regime. We call $\tau$ the real-time gain since real-time usually refers to being controlled and deterministic \cite{6Gtime}. $\tau$ represents the decay rate of the probability of extreme events happening. Alternatively, $\tau$ can also be named differently, e.g., revised diversity gain \footnote{Diversity gain was initially proposed in \cite{Zheng2003}. A detailed comparison between $\tau$ and diversity gain is provided in Section III-C.}, low-latency gain, or timeliness gain. 

\begin{figure*}[t]
\centerline{\includegraphics[width=18cm]{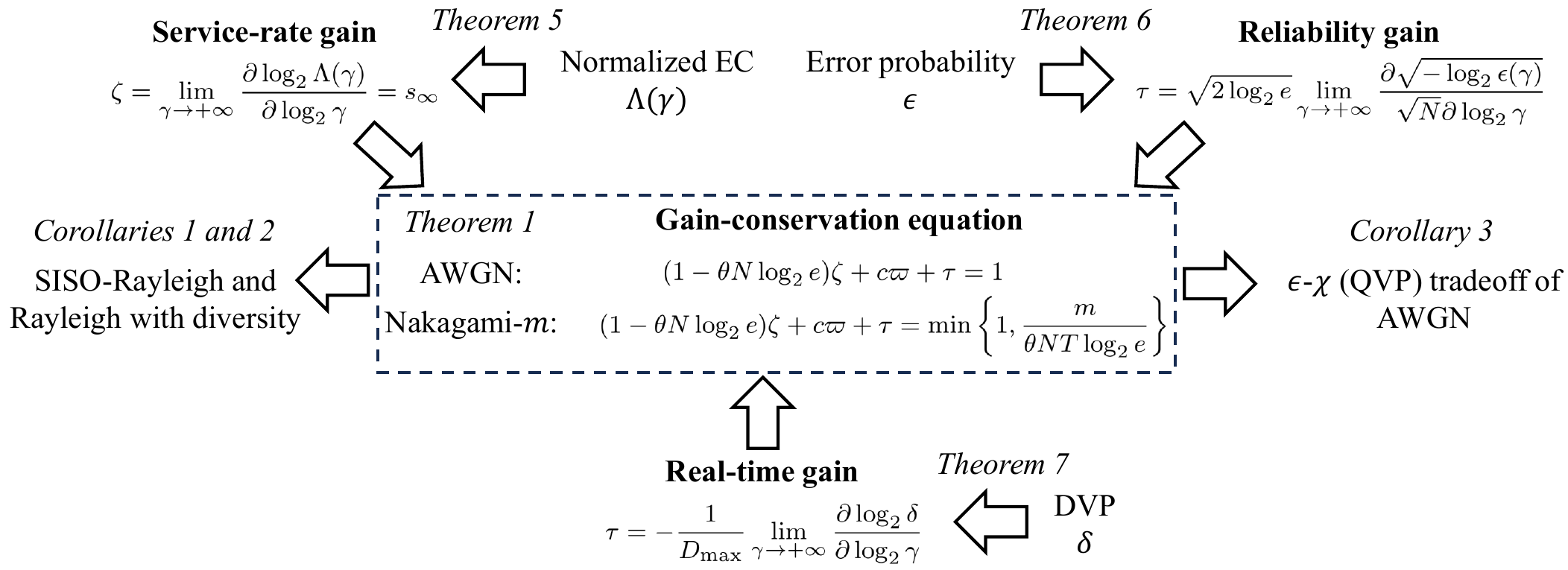}}
\caption{Main results for gain-conservation equations and relationships between different quantities.}
\label{diag}
\end{figure*}

Based on the above definitions of $\zeta$, $\varpi$, and $\tau$, Theorem 1 presents the high-SNR reliability-latency-rate tradeoff through revised gain-conservation equations for low-latency systems with FBC and random arrivals. For simplicity, let $\varrho$ denote an arbitrarily positive constant, which is named as revised QoS exponent. The main results and terms related to the gain-conservation equations are illustrated in Fig. \ref{diag}. 

\textbf{Theorem 1}. Let $\epsilon \in (0,0.5]$ be a function of $\gamma$ with $\lim\limits_{\gamma\to+\infty}\epsilon(\gamma)=0$. Additionally, assume that $\varpi$ is finite, $\theta N = \varrho$, and $N=\Psi(\gamma)$, where $\Psi(\gamma)$ satisfies the following condition:
\begin{equation}\label{newcod1}
\sup_{\gamma>0}\left|\frac{G\left(\Psi(\gamma),|h[n]|^2\gamma,\epsilon(\gamma)\right)}{\Psi(\gamma)}\right|\leq \nu, \quad \text{for all }|h[n]|^2.
\end{equation}
In Eq. \eqref{newcod1}, $\nu\geq 0$ is a finite constant.
\begin{itemize}
\item  For a SISO system over an AWGN channel, the following relations hold:
\begin{align}
&\zeta+c\varpi=1, \, \text{and}\label{tt0111}\\
&\tau=\left(\theta N \log_2 e\right)\zeta. \label{tt0121}
\end{align}

By combining Eqs. \eqref{tt0111} and \eqref{tt0121}, we obtain
\begin{equation}
(1-\theta N \log_2 e)\zeta+c\varpi+\tau=1; \label{awgncon1}
\end{equation}
\item For a SISO system over a Nakagami-$m$ fading channel, the following relations hold:
\begin{align}
&\zeta+c\varpi=\min\left\{1,\frac{m}{\theta NT\log_2e}\right\}, \, \text{and}\label{tt0091}\\
&\tau=\left(\theta N \log_2 e\right)\zeta. \label{tt0101}
\end{align}

By combining Eqs. \eqref{tt0091} and \eqref{tt0101}, we obtain
\begin{equation}
	\begin{aligned}
		(1-\theta N\log_2e)\zeta&+c\varpi+\tau\\
		& =\min\left\{1,\frac{m}{\theta NT\log_2e}\right\}.\label{gainequ33}
	\end{aligned}
\end{equation}
\end{itemize}

In Eqs. \eqref{tt0111}, \eqref{awgncon1}, \eqref{tt0091} and  \eqref{gainequ33}, $c\in[0,1]$ is given by
\begin{equation}
c=\lim_{\gamma\to+\infty}\frac{1}{1+\frac{\partial \sqrt{\Psi(\gamma)}}{\partial \log_2 \gamma}\frac{\log_2\gamma}{\sqrt{\Psi(\gamma)}}}.
\end{equation}

\begin{IEEEproof}
	This theorem summarizes the main results in Section III. The proof of this theorem is derived from the proofs of Theorem 5, Theorem 6, Lemma 1, and Theorem 7, as detailed in Appendix D, Appendix E, Appendix F, and Section III-C, respectively. 
\end{IEEEproof}

\begin{figure*}[b]
	\normalsize
	\hrulefill
	\setcounter{equation}{30}
	\begin{equation}\label{up1}
		\begin{aligned}
			R_{N}^{\epsilon}&\leq NC+\sqrt{NV\left(|h|^2\gamma\right)}Q^{-1}(1-\epsilon)+\frac{1}{2}\log_2 N+g_u\left(|h|^2\gamma,\epsilon\right)\\
			&=\tilde{R}_N^{\epsilon}+\frac{c_1}{V\left(|h|^2\gamma\right)}+\frac{3}{2}\log_2 V\left(|h|^2\gamma\right) -\log_2 K\left(|h|^2\gamma\right).
		\end{aligned}
	\end{equation}
\end{figure*}

From the definitions of $\zeta$, $\varpi$, and $\tau$, we find that these three gains effectively characterize the performance of communication systems in terms of the service rate, reliability, and latency under a maximum power constraint. Consequently, this theorem presents a fundamental tradeoff among the service rate, reliability, and latency in the high-SNR regime for low-latency communications. By combining Eqs. \eqref{tt0091} and \eqref{tt0101}, we derive \eqref{gainequ33}, which allows for a comparison with Eq. (4) from \cite{YLTCOM}, as they exhibit a similar form. In this work, we adopt EC to characterize the service-rate and real-time gains, which is different from the method utilized for the asymptotic analysis in \cite{YLTCOM}. The primary difference between Eqs. \eqref{awgncon1}, \eqref{gainequ33}, and Eq. (4) in \cite{YLTCOM} results from the influence of random arrivals. Therefore, Theorem 1 can be regarded as a generalization of Eq. (4) in \cite{YLTCOM} for systems with unbounded random arrivals under a large delay threshold. Furthermore, \cite{YLTCOM} lacks a discussion of the assumptions underlying the use of normal approximation in asymptotic analysis, which are provided in this paper.

The assumption presented in Eq. \eqref{newcod1} can be met with certain $G$ functions. We provide the discussion on the requirements for $G$ function in the following. For simplicity, we omit the index $n$ in the subsequent analysis. Additionally, we let $V\left(|h|^2\gamma\right)=\left(\log_2e\right)^2\left(1-\left(1+|h|^2\gamma\right)^{-2}\right)$ denote the dispersion term, which is a function of $|h|^2\gamma$. In \cite{codingrate}, Polyanskiy, Poor, and Verdú proposed the upper and lower bounds for $R_N^{\epsilon}$, which are given by \footnote{There is a tighter lower bound proposed in \cite{thirdorder}, which shows that $NC-\sqrt{NV\left(|h|^2\gamma\right)}Q^{-1}(\epsilon)+\frac{1}{2}\log_2N +O(1)\leq R_N^{\epsilon}$. However, we do not use this bound due to the complexity of analyzing the $O(1)$ term in this bound. Moreover, the lower bound proposed in \cite{fbl1} is sufficient for our analysis. } 
	\setcounter{equation}{29}
\begin{equation}\label{ualb}
	\begin{aligned}
		NC&-\sqrt{NV\left(|h|^2\gamma\right)}Q^{-1}(\epsilon)+O(1)\leq R_N^{\epsilon}\\
		&\leq  NC-\sqrt{NV\left(|h|^2\gamma\right)}Q^{-1}(\epsilon)+\frac{\log_2 N}{2} +O(1),
	\end{aligned}
\end{equation}
where the upper bound requires 
\begin{equation}
	N>\left(\frac{2K\left(|h|^2\gamma\right)}{(1-\epsilon)V^{\frac{3}{2}}\left(|h|^2\gamma\right)}\right)^2, \nonumber 
\end{equation}
and the lower bound requires
\begin{equation}
	N>\left(\frac{2K\left(|h|^2\gamma\right)}{\epsilon V^{\frac{3}{2}}\left(|h|^2\gamma\right)}\right)^2.\nonumber 
\end{equation}

Based on the upper and lower bounds of $R_N^{\epsilon}$, we can derive the upper and lower bounds of $G$ function, which are denoted by $g_u(|h|^2\gamma,\epsilon)$ and $g_l(N,|h|^2\gamma,\epsilon)$, respectively. Specifically, according to Eq. (632) in \cite{codingrate}, the analytical form of the upper bound of $R_N^{\epsilon}$ is given in Eq. \eqref{up1}, which is presented at the bottom of the last page.

In Eq. \eqref{up1}, $K(\cdot)$ is defined as
\setcounter{equation}{31}
\begin{equation}\label{kfunc}
	\begin{aligned}
	K\left(|h|^2\gamma\right)=&c_0\left(\frac{|h|^2\gamma}{1+|h|^2\gamma}\right)^3\cdot\\
	&\quad \quad \quad 	\mathbb{E}_z\left\{\left|z^2-\frac{2}{\sqrt{|h|^2\gamma}}z-1\right|^3\right\},
	\end{aligned}
\end{equation}
where $c_0$ is a positive constant and $z$ represents a standard normal variable. 

$c_1$ in Eq. \eqref{up1} is given by
\begin{equation}\label{lll1}
c_1=-2K\left(|h|^2\gamma\right)\cdot\min_{y\in\left[1-\epsilon-b_1,1-\epsilon\right]}\frac{{\rm d}Q^{-1}(y)}{{\rm d}y},
\end{equation}
where $b_1$ is given by
\begin{equation}
	b_1=\frac{2K\left(|h|^2\gamma\right)}{V^{\frac{3}{2}}\left(|h|^2\gamma\right)\sqrt{\left(\frac{2K\left(|h|^2\gamma\right)}{V^{\frac{3}{2}}\left(|h|^2\gamma\right)(1-\epsilon)}\right)^2+1}}.
\end{equation}

According to Eqs. (650)-(654) in \cite{codingrate}, the lower bound of $R_N^{\epsilon}$ is expressed in Eq. \eqref{low1} (located at the bottom of this page), where $c_2>0$ is a constant. Based on Eqs. \eqref{up1} and \eqref{low1}, we can derive the upper and lower bounds of $G(N,|h|^2\gamma,\epsilon)$, respectively. Consequently, the requirements for $\Psi(\gamma)$ to satisfy Eq. \eqref{newcod1} can be determined using Eqs. \eqref{up1} and \eqref{low1}. In Theorems 2 and 3, we present the sufficient conditions under which Eq. \eqref{newcod1} holds for different system assumptions.

\begin{figure*}[b]
	\normalsize
	\hrulefill
\setcounter{equation}{34}
\begin{equation}\label{low1}
	\begin{aligned}
		R_N^{\epsilon}&\geq NC+\sqrt{NV\left(|h|^2\gamma\right)}Q^{-1}(1-\epsilon)+\frac{1}{2}\log_2 N+g_l(N,\gamma,\epsilon)\\
		&=\tilde{R}_N^{\epsilon}-\frac{1}{2}\log_2 N+\log_2\frac{K\left(|h|^2\gamma\right)}{c_2 V^{\frac{3}{2}}\left(|h|^2\gamma\right)} -\log_2 \left[2\left(\frac{\ln 2}{\sqrt{2\pi}}+\frac{2K\left(|h|^2\gamma\right)}{V^{\frac{3}{2}}\left(|h|^2\gamma\right)}\right)\right]\\
		&\quad \quad \quad\quad\quad\quad +\sqrt{NV\left(|h|^2\gamma\right)}\left[Q^{-1}\left(1-\epsilon+\frac{2K\left(|h|^2\gamma\right)}{\sqrt{N}V^{\frac{3}{2}}\left(|h|^2\gamma\right)}\right)+Q^{-1}(\epsilon)\right].
	\end{aligned}
\end{equation}

\end{figure*}

\textbf{Theorem 2}. Given $\epsilon\in(0,0.5]$, the condition proposed in Eq. \eqref{newcod1} is satisfied if $\Psi(\gamma)$ follows
\setcounter{equation}{35}
\begin{equation}
	\Psi(\gamma)>\varsigma_1, \quad \text{for all } \gamma>0,
\end{equation} 
where $\Psi(\gamma)$ may either be bounded or satisfy $\lim_{\gamma\to+\infty}\Psi(\gamma)=+\infty$. $\varsigma_1$ is finite and is given by
\begin{equation}
	\varsigma_1=\max_{x\geq 0}\left\{\left(\frac{2K(x)}{(1-\epsilon)V^{\frac{3}{2}}(x)}\right)^2,\left(\frac{2K(x)}{\epsilon V^{\frac{3}{2}}(x)}\right)^2\right\}.
\end{equation}
\begin{IEEEproof}
	See Appendix A.
\end{IEEEproof}

From Theorem 2 we find that with a given $\epsilon$, $\Psi(\gamma)$ can be set as a constant larger than $\varsigma_1$. Moreover, it can also be chosen as $\Psi(\gamma)=\varsigma_1+U(\gamma)$, where $U(x)>0$ for $x>0$ and $\lim\limits_{x\to+\infty}U(x)=+\infty$. Theorem 2 is used to determine the forms of $\Psi(\gamma)$ in Theorem 5 and Corollary 1 in Section III-A.

Furthermore, the proof of Theorem 2 reveals that, for $N>\varsigma_1$ and a given $\epsilon$, the error term $G(N,\gamma, \epsilon)$ remains bounded with respect to $\gamma$, which provides additional insights complementary to those in \cite{highsnr}. An important future work is to further investigate the case where $N<\varsigma_1$.

 \textbf{Theorem 3}. Under the assumption that $\epsilon  \in (0,0.5]$ is a function of $\gamma$ and $\lim\limits_{\gamma\to+\infty}\epsilon(\gamma)=0$, the condition proposed in Eq. \eqref{newcod1} is satisfied if $\Psi(\gamma)$ follows
\begin{equation}\label{newthm3}
	\begin{cases}
		&\Psi(\gamma)\epsilon^{2}(\gamma)>\varsigma_2, \quad \gamma>0,\\
		&\lim_{\gamma\to+\infty}\frac{e^{\left [\frac{Q^{-1}(\epsilon(\gamma))}{2}\right]^2}}{\Psi(\gamma)}<+\infty,\\
		&\lim\limits_{\gamma\to+\infty}\frac{Q^{-1}(\epsilon(\gamma))}{\sqrt{\Psi(\gamma)}}<+\infty,
	\end{cases}
\end{equation} 
where $\varsigma_2$ is a finite and positive constant defined as 
\begin{equation}
	\varsigma_2=4\max_{x\geq 0}\left\{\frac{K^2(x)}{V^{3}(x)}\right\}.
\end{equation}
\begin{IEEEproof}
	See Appendix B.
\end{IEEEproof}

Theorem 3 outlines the specific requirements for $\Psi(\gamma)$ under the conditions that $\epsilon(\gamma)\in(0,0.5]$ and  $\lim\limits_{\gamma\to+\infty}\epsilon(\gamma)=0$. The form of $\Psi(\gamma)$ is related to $\epsilon(\gamma)$. Given a specific form of $\epsilon(\gamma)$, $\Psi(\gamma)$ can be formulated according to Eq. \eqref{newthm3}. For example, one can let $\Psi(\gamma)=e^{\left [\frac{Q^{-1}(\epsilon(\gamma))}{2}\right]^2}+\frac{\varsigma_2+1}{\epsilon^2(\gamma)}$, which satisfies Eq. \eqref{newthm3}. Theorem 3 serves as a foundation for determining the forms of $\Psi(\gamma)$ utilized in Theorems 1, 6, 7 and Corollary 2.

In this paper, we define the constant $\varrho$ as the revised QoS exponent. Different from the definition of $\theta$ in Eq. \eqref{ttt1}, $\varrho$ is associated with the value of the DVP. We can gain a clearer understanding of  the interpretation of $\varrho$ by presenting the expression of the DVP. By combining Eqs. \eqref{normalizeddefine}, \eqref{eqso1}, and \eqref{delta}, we obtain 
\begin{equation}\label{tt0001}
\begin{aligned}
	\delta\simeq \exp\left(\frac{D_{\rm max}}{T}\ln\mathbb{E}_{|h|^2}\left\{\exp\left(-\varrho T \frac{R_N^{\epsilon}[n]}{N}\right)\right\}\right).
\end{aligned}
\end{equation}

From Eq. \eqref{tt0001}, we observe that a larger $\varrho$ results in a smaller $\delta$ given $\gamma$, $\Psi(\cdot)$, and $D_{\rm max}$. Therefore, the value of $\varrho$ can represent the QoS requirements for the DVP.

EC is widely utilized in research on low-latency communications to assess the QoS-constrained throughput of wireless systems. However, it is usually challenging to obtain a closed-form expression of EC, which brings difficulty for the designs of resource allocation and scheduling strategies aimed at maximizing EC. In this paper, we propose a Laplace's-method-based EC approximation approach, in which normal approximation is employed to evaluate the maximum coding rate. Thus, this approximation is applicable when the normal approximation has satisfying accuracy. For the simplicity of expression, let  	
\begin{equation}\label{maaa1}
	\begin{aligned}
		r(x)=&\log_2\left(1+\Xi(x) x\right)-\\
		&\sqrt{\frac{1}{N}\left(1-\frac{1}{(1+\Xi(x) x)^{2}}\right)}Q^{-1}(\epsilon)\log_2 e+\frac{\log_2 N}{2N},
	\end{aligned}
\end{equation}
where $x$ denotes the channel power gain $|h[n]|^2$ in time slot $n$. Since $|h[n]|^2$ is $i.i.d.$ across time slots, we omit the argument for $x$. The function $\Xi(x)$ denotes the channel-power-gain-based power allocation scheme. Note that $\Xi(x)$ can also denote the fixed power allocation by setting $\Xi(x)=\gamma$. Besides, let $\ell =\left(\frac{Q^{-1}(\epsilon)}{\sqrt{N}}\right)^2$. 

\textbf{Theorem 4}. For a low-latency communication system that adopts a power allocation scheme $\Xi(x)$ satisfying $\Xi(x)+\frac{{\rm d}\Xi(x)}{{\rm d} x} x\geq 0$ with the channel distribution $f(\cdot)$, EC is approximated using Eq. \eqref{theo31}, which is presented at the bottom of this page. In Eq. \eqref{theo31}, $x^*$ is given by
\setcounter{equation}{42}
\begin{equation}
\Xi(x^*)x^*=\sqrt{\frac{1+\sqrt{1+4\ell}}{2}}-1. \label{theo32}
\end{equation}

\begin{figure*}[b]
	\normalsize
	\hrulefill
\setcounter{equation}{41}
\begin{equation}
	\alpha_S(\theta)= Nr(x^*)-\frac{\ln f(x^*)+\frac{1}{2}\ln 2\pi -\frac{1}{2}\ln |r''(x^*)|-\frac{1}{2}\ln \theta NT}{\theta T}+O\left(\frac{1}{\theta^2 T^2N}\right). \label{theo31}
\end{equation}
\end{figure*}

For fixed power allocation $\Xi(x)=\gamma$, $x^*$ is given by
\setcounter{equation}{43} 
\begin{equation}
	x^*=\frac{1}{\gamma}\left(\sqrt{\frac{1+\sqrt{1+4\ell}}{2}}-1\right).
\end{equation}

\begin{IEEEproof}
	Theorem 4 summarizes the main results in Section IV. The proof of this theorem follows from the proofs of Lemma 2 and Theorem 8 (in Appendix H and I, respectively).
\end{IEEEproof}

The approximation proposed in Theorem 4 provides a method to obtain a closed-form expression for EC, which facilitates further analysis of the QoS-constrained performance of low-latency communication systems and allows for theoretical comparison of different scheduling schemes. The condition stated in Theorem 4 can be satisfied by various classical allocation schemes. More detailed discussions on $\Xi(x)$ are presented in Section IV-B.

\section{Reliability-latency-rate tradeoff in the finite blocklength regime}

In this section, we mainly focus on characterizing the reliability-latency-rate tradeoff in the FBL regime for low-latency communication systems. In Section III-A, we begin by deriving the high-SNR slope of EC in the FBL regime with a fixed error probability, which indicates the service-rate gain of low-latency communication systems. Based on the results of the high-SNR slope of EC in the FBL regime, we further generalize the conclusions to characterize the reliability gain by representing error probability as a function of SNR in Section III-B. Finally, in Section III-C, we propose the gain-conservation equation, which reveals the tradeoff among the service-rate gain, reliability gain, and real-time gain in the FBL and high-SNR regimes.

\subsection{High-SNR Slope of EC in the FBL regime}

In this subsection, we focus on characterizing the high-SNR slope of EC in the FBL regime. Specifically, we derive the high-SNR slope of EC with a given $\epsilon$ in both the AWGN and Nakagami-$m$ fading channels with FBC. The derived conclusions can be viewed as a generalization of the conclusions in \cite{rayleigh1}. We then provide corollaries for systems in Rayleigh fading channels, considering both scenarios with and without frequency or spatial diversity.

We start our discussion with the FBL-SISO system in an AWGN channel. Specifically, the high-SNR slope of EC in the FBL regime is defined as \cite{rayleigh1}
\begin{equation}\label{defi1}
s_{\infty}=\lim_{\gamma\to+\infty}\frac{\Lambda(\gamma)}{\log_2 \gamma}.
\end{equation}

We then provide the discussion for the FBL-SISO system in the AWGN and Nakagami-$m$ fading channels, encompassing wired communications, satellite communications, and other typical scenarios in wireless communications. In Theorem 5, we derive the high-SNR slope of the normalized EC for the FBL-SISO system in both the AWGN and Nakagami-$m$ fading channel. 

\textbf{Theorem 5}. Under the condition that $N=\Psi(\gamma)$ satisfies Theorem 2 for a given $\epsilon$,
\begin{itemize}
\item the high-SNR slope of the normalized EC for the FBL-SISO system in the AWGN channel is $s_{\infty}=1$;

\item the high-SNR slope of the normalized EC for the FBL-SISO system in the Nakagami-$m$ fading channel with $m\in[0.5,+\infty)$ is given by
\begin{equation}\label{thm1_1}
\begin{aligned}
		\zeta=\begin{cases}
			1, \quad &1\leq \frac{m}{\theta N T\log_2 e},\\
			\frac{m}{\theta NT \log_2 e}, \quad &1>\frac{m}{\theta N T\log_2 e}.
		\end{cases}
\end{aligned}
\end{equation}
\end{itemize}

\begin{IEEEproof}
See Appendix D.
\end{IEEEproof}

In Theorem 5, we derive the value of $s_{\infty}$ for the FBL-SISO system over AWGN and Nakagami-$m$ fading channels. We can compare this result with the well-known multiplexing gain \cite{Zheng2003}. Multiplexing gain is defined as the ratio of the maximum coding rate to $\log_2 \gamma$ as $\gamma\to+\infty$, which is equal to 1 for the SISO system. However, the definition of multiplexing gain does not account for latency requirements. As we mentioned in Section II-B, $\varrho=\theta N$ represents the constraint imposed by DVP, which is one of the key performance indicators of QoS. A larger $\theta N$ indicates a stricter QoS constraint, as outlined in Eq. \eqref{tt0001}. Consequently, $s_{\infty}$ can be interpreted as a revised multiplexing gain, which takes the requirements of DVP into account. Since $s_{\infty}$ considers the gain under the QoS constraint, it can not be larger than the multiplexing gain, as validated in Theorem 5. 

The results in Theorem 5 demonstrate that for small values of $\theta N$, $s_{\infty}$ of the fading channel is independent of QoS constraints, which aligns with the analysis in \cite{rayleigh1} for the infinite-blocklength case in the Rayleigh fading channel. This result indicates that when the requirement of DVP is loose, it does not influence the increasing rate of the QoS-constrained throughput. Moreover, we find that $s_{\infty}$ of the fading channel can be smaller than 1, which is also consistent with the common understanding that the randomness of fading has a negative influence on the QoS-constrained throughput in queueing systems.

Based on Theorem 5, we can obtain $s_{\infty}$ for several common cases. In Corollary 1, we present the $s_{\infty}$ of Rayleigh fading channels with and without frequency or spatial diversity.

\textbf{Corollary 1.} Under the same assumption used in Theorem 5, 
\begin{itemize}
\item the high-SNR slope of the normalized EC in the Rayleigh fading channel is given by
\begin{equation}
\begin{aligned}
		s_{\infty}=
		\begin{cases}
			1, \quad &1 \leq  \frac{1}{\theta N T\log_2 e},\\
			\frac{1}{\theta N T \log_2 e}, \quad & 1 > \frac{1}{\theta N T\log_2 e}. 
		\end{cases}\label{newlm2}
\end{aligned}
\end{equation}
\item the high-SNR slope of the normalized EC in Rayleigh fading channels with frequency or spatial diversity $\kappa$ by adopting maximum-ratio combining is given by
\begin{equation}
\begin{aligned}
		s_{\infty}=\begin{cases}
			1, \quad &1 \leq \frac{\kappa}{\theta N T \log_2 e},\\
			\frac{\kappa}{\theta  N T \log_2 e}, \quad & 1 > \frac{\kappa}{\theta NT \log_2 e}.
		\end{cases}
\end{aligned}
\end{equation}
\end{itemize}
\begin{IEEEproof}
The results can be obtained directly by setting $m=\Omega=1$ and $m=\Omega=\kappa$ from Eq. \eqref{thm1_1}. 
\end{IEEEproof}

In Theorem 5 and Corollary 1, we characterize the high-SNR slope of the normalized EC in the FBL regime with a given $\epsilon$. Comparing the results in Theorem 5 and Corollary 1, we find that in the FBL regime, $s_{\infty}$ of the systems with frequency or spatial diversity is at least as large as that of the SISO system over a Rayleigh fading channel. For the SISO system over a Nakagami-$m$ fading channel, $s_{\infty}$ is no smaller than that of the SISO system over a Rayleigh fading channel with $m\geq 1$. For $m\in[0.5,1)$, $s_{\infty}$ is not larger than that of the SISO system over a Rayleigh fading channel. Furthermore, we observe that as $\kappa$ or $m$ increases, the length of the $\theta N$ interval, where $s_{\infty}$ is irrelevant to $\theta N$, is larger. According to \cite{gold}, $m$ represents the ratio of line-of-sight signal power to multipath power, with $m<1$ indicating more severe fading than Rayleigh fading and $m>1$ indicating less severe fading than the Rayleigh fading. In summary, the severity of fading and the degree of spatial or frequency diversity both influence $s_{\infty}$. Less serve fading and larger spatial or frequency diversity lead to larger $s_{\infty}$ under a given QoS constraint and ensure that $s_{\infty}$ remains irrelevant to QoS constraints over a broader range of $\theta N$ values.

\emph{Remark 2: We can also define $s_0$ as the low-SNR slope of the normalized EC in the FBL regime. However, identifying the appropriate denominator in the definition of $s_0$ is crucial. As far as we know, $\gamma$ is not a proper choice since it can result in $\frac{\Lambda(\gamma)}{\gamma}\to+\infty$ when $\gamma$ approaches 0 by adopting a kind of orthogonal code proposed in Chapter 8.5 of \cite{galla}. Once a proper denominator is determined, we can obtain the upper bound of $s_0$ by considering the infinite blocklength regime by leveraging Shannon's formula. Performing the above analysis is not trivial, which makes it an important future work.}

\subsection{Reliability Gain in the FBL regime}

In this subsection, we extend the analysis to the case where $\epsilon$ is a function of $\gamma$. Under this assumption, we will refine the conclusions in Theorem 5. The updated results are summarized in Theorem 6.

\textbf{Theorem 6}. Assume $\epsilon  \in (0,0.5]$ is a function of $\gamma$ and $\lim\limits_{\gamma\to+\infty}\epsilon(\gamma)=0$. Under the assumption that $N=\Psi(\gamma)$ satisfies Theorem 3,

\begin{itemize}
\item for an FBL-SISO system over an AWGN channel, we have
\begin{equation}\label{awgn1}
\lim_{\gamma\to+\infty}\frac{\Lambda(\gamma)}{\log_2\gamma}+\log_2 e\lim_{\gamma\to+\infty}\frac{Q^{-1}(\epsilon(\gamma))}{\sqrt{N}\log_2\gamma}=1.
\end{equation}
\item for an FBL-SISO system over a Nakagami-$m$ fading channel, we have
\begin{equation}\label{fading1}
	\begin{aligned}
		\lim_{\gamma\to+\infty}\frac{\Lambda(\gamma)}{\log_2\gamma}+\log_2 e&\lim_{\gamma\to+\infty}\frac{Q^{-1}(\epsilon(\gamma))}{\sqrt{N}\log_2\gamma}\\
		&= \min\left\{1,\frac{m}{\theta NT\log_2e}\right\}.
	\end{aligned}
\end{equation} 
\end{itemize}

\begin{IEEEproof}
See Appendix E.
\end{IEEEproof}

In Theorem 6, we proposed the result of the high-SNR asymptotic analysis of the EC, which is related to the error probability. Before characterizing the tradeoff between the service-rate gain, reliability gain, and real-time gain, we need to first explore the concept of reliability gain to establish the link between these metrics. We define the reliability gain as  $\varpi=\sqrt{2\log_2e}\lim\limits_{\gamma\to+\infty}\frac{\partial\sqrt{-\log_2 \epsilon(\gamma)}}{\sqrt{N}\partial \log_2\gamma}$. $\varpi$ is related to the decay rate of the error probability with respect to $\log_2\gamma$ in the high-SNR regime, which is consistent with \cite{YLTCOM}. Using this definition, we can relate the reliability gain to the second term on the left-hand side of Eqs. \eqref{awgn1} and \eqref{fading1}, which is summarized in Lemma 1.

\textbf{Lemma 1}. For a finite reliability gain $\varpi$, the following holds:
\begin{equation}
\lim\limits_{\gamma\to+\infty}\frac{Q^{-1}(\epsilon(\gamma))}{\sqrt{N} \log_2\gamma}=\frac{c_4}{\log_2e}\varpi, \label{newlemma2}
\end{equation}
where $c_4\in[0,1]$ is given by
\begin{equation}\label{tttt1}
c_4=\lim_{\gamma\to+\infty}\frac{1}{1+\frac{\partial \sqrt{\Psi(\gamma)}}{\partial \log_2 \gamma}\frac{\log_2\gamma}{\sqrt{\Psi(\gamma)}}}.
\end{equation}
\begin{IEEEproof}
See Appendix F.
\end{IEEEproof}

Through Lemma 1, we establish a link between $s_{\infty}$ and the reliability gain in Eqs. \eqref{awgn1} and  \eqref{fading1}. In Lemma 1, we mention that $c_4\in [0,1]$. The value of $c_4$ depends on the form of $\Psi(\gamma)$. Specifically, when $\lim\limits_{\gamma\to+\infty}\frac{\log_2\gamma}{\sqrt{\Psi(\gamma)}}$ exists and is positive, we have $c_4=\frac{1}{2}$. This is due to the fact that in this case, $\lim\limits_{\gamma\to+\infty}\frac{\Psi(\gamma)}{\log_2\gamma}=\lim\limits_{\gamma\to+\infty}\frac{\partial \sqrt{\Psi(\gamma)}}{\partial \log_2 \gamma}$ is finite. Thus, we have
\begin{equation}
\begin{aligned}
	&\lim_{\gamma\to+\infty}\frac{\partial \sqrt{\Psi(\gamma)}}{\partial \log_2 \gamma}\frac{\log_2\gamma}{\sqrt{\Psi(\gamma)}} \\
	=&\lim_{\gamma\to+\infty}\frac{\partial \sqrt{\Psi(\gamma)}}{\partial \log_2 \gamma}\lim_{\gamma\to+\infty}\frac{\log_2\gamma}{\sqrt{\Psi(\gamma)}}\\
	=&\lim_{\gamma\to+\infty}\frac{\Psi(\gamma)}{\log_2\gamma}\lim_{\gamma\to+\infty}\frac{\log_2\gamma}{\sqrt{\Psi(\gamma)}}\\
	=&1.
\end{aligned}
\end{equation}

For $\lim\limits_{\gamma\to+\infty}\frac{\log_2\gamma}{\sqrt{\Psi(\gamma)}}=0$ or $\lim\limits_{\gamma\to+\infty}\frac{\log_2\gamma}{\sqrt{\Psi(\gamma)}}=+\infty$, $\lim\limits_{\gamma\to+\infty}\frac{\partial \sqrt{\Psi(\gamma)}}{\partial \log_2 \gamma}\frac{\log_2\gamma}{\sqrt{\Psi(\gamma)}}$ can be 0, $+\infty$, or lie in $(0,+\infty)$.

\subsection{Gain-Conservation Equation for Reliability, Real-time, and Service-Rate Gains}

In this subsection, we will propose revised gain-conservation equations for low-latency communication systems. In contrast to \cite{YLTCOM}, the revised terms in the gain-conservation equation account for the influence of random arrivals. Before presenting revised gain-conservation equations, we first provide the definitions of service-rate and real-time gains.  

The service-rate gain is defined as $\zeta=\lim\limits_{\gamma\to+\infty}\frac{\partial{\Lambda(\gamma)}}{\partial{\log_2\gamma}}$, which is similar to the definition of the service-rate gain presented in \cite{YLTCOM}. Note that EC represents the maximum constant source rate that can be supported by the given service process to satisfy the statistical queueing requirements \cite{ec}.  Thus, $\zeta$ indicates the increasing rate of throughput with respect to $\log_2\gamma$ in the high-SNR regime. According to L'Hospital's rule \cite{f1}, we have
\begin{equation}
	\begin{aligned}
		s_{\infty}&=\lim\limits_{\gamma\to+\infty}\frac{\Lambda(\gamma)}{\log_2\gamma}\\
		&=\lim\limits_{\gamma\to+\infty}\frac{\partial{\Lambda(\gamma)}}{\partial{\log_2\gamma}}\\
		&=\zeta.
	\end{aligned}
\end{equation}
Therefore, we can use the conclusions from Theorem 5 to characterize the service-rate gain $\zeta$.

For the real-time gain, we adopt the same definition as proposed in \cite{YLTCOM}, which is given by $\tau=-\frac{1}{D_{\rm max}}\frac{\partial\log_2\delta}{\partial \log_2\gamma}$. $\tau$ represents the decay rate of DVP as $\log_2\gamma$ increases in the high-SNR regime.

Based on Theorem 5, Theorem 6, and Lemma 1, we can then derive revised gain-conversation equations for low-latency communication systems with random arrivals in the FBL regime. The results are summarized in Theorem 7.

\textbf{Theorem 7}. In the FBL regime, under the same assumptions used in Theorem 6,
\begin{itemize}
\item  for a SISO system over an AWGN channel, we have
\setcounter{equation}{54}
\begin{align}
&\zeta+c_4\varpi=1, \, \text{and}\label{tt011}\\
&\tau=\left(\theta N \log_2 e\right)\zeta. \label{tt012}
\end{align}
By combining Eqs. \eqref{tt011} and \eqref{tt012}, we obtain
\begin{equation}\label{tttt2}
(1-\theta N \log_2 e)\zeta+c_4\varpi+\tau=1;
\end{equation}
\item for a SISO system over a Nakagami-$m$ fading channel with a finite $\varpi$, we have
\begin{align}
&\zeta+c_4\varpi=\min\left\{1,\frac{m}{\theta NT\log_2e}\right\}, \, \text{and}\label{tt009}\\
&\tau=\left(\theta N \log_2 e\right)\zeta. \label{tt010}
\end{align}
By combining Eqs. \eqref{tt009} and \eqref{tt010}, we obtain
\begin{equation}\label{tttt3}
	\begin{aligned}
		(1-\theta N\log_2e)\zeta+&c_4\varpi+\tau\\
		&= \min\left\{1,\frac{m}{\theta NT\log_2e}\right\}.
	\end{aligned}
\end{equation}

In Eqs. \eqref{tt011}, \eqref{tttt2}, \eqref{tt009}, and \eqref{tttt3}, $c_4$ are defined in Eq. \eqref{tttt1}.
\end{itemize}

\begin{IEEEproof}
Based on Eqs. \eqref{eqso1} and \eqref{delta}, we obtain 
\begin{equation}
\ln\delta=-\theta N \Lambda(\gamma)D_{\rm max}.\label{newtheo31}
\end{equation}

Next, by taking the derivative on both sides of Eq. \eqref{newtheo31}, we have
\begin{equation}
	\begin{aligned}
		-\frac{1}{D_{\rm max}}\lim_{\gamma\to+\infty}\frac{\partial\log_2\delta}{\partial\log_2\gamma}&=\theta N \log_2 e \lim_{\gamma\to+\infty}\frac{\partial \Lambda(\gamma)}{\partial \log_2\gamma}\\
		&=\left(\theta N \log_2 e \right)\zeta.
	\end{aligned}
\label{newtheo32}
\end{equation}

By substituting Eqs. \eqref{newlemma2} and \eqref{newtheo32} into Eqs. \eqref{awgn1}, and \eqref{fading1}, we obtain the results shown in Theorem 7.
\end{IEEEproof}

In Theorem 7, we present revised gain-conservation equations for low-latency communication systems. There are some differences between Eq. \eqref{tttt3} and Eq. (4) in \cite{YLTCOM}. First, the parameter $c_4$ equals 1 in \cite{YLTCOM}. This is mainly because \cite{YLTCOM} lacks a discussion on the conditions under which normal approximation can be applied for high-SNR asymptotic analysis. Second, Eq. \eqref{tttt3} includes the term $\theta N$, which is absent in \cite{YLTCOM}. This discrepancy arises because the arrival in \cite{YLTCOM} is deterministic. The randomness of the arrival causes a deterioration in the performance of queueing systems with respect to DVP and QoS-constrained throughput.  

Moreover, we can derive the real-time gain based on Theorem 5 and Eq. \eqref{tt010} for Nakagami-$m$ fading channels as
\setcounter{equation}{62}
\begin{equation}\label{mmm1}
\begin{aligned}
	\tau=\begin{cases}
		\theta N \log_2 e, \quad &1\leq \frac{m}{\theta N T \log_2 e},\\
		\frac{m}{T}, \quad &1>\frac{m}{\theta NT\log_2 e}.
	\end{cases}
\end{aligned}
\end{equation}

Considering the case in which $T=1$, we can interpret $\tau$ as a revised version of diversity gain defined in \cite{Zheng2003}. The diversity gain is defined as the ratio of the outage probability to $\log_2 \gamma$ as $\gamma\to+\infty$, which equals $m$ for the Nakagami-$m$ fading channel. According to L'Hospital's rule, $\tau$ can be treated as the ratio of DVP to $\log_2\gamma $ as $\gamma \to +\infty$. From Eq. \eqref{mmm1}, if $ 1\leq \frac{m}{\theta N T \log_2 e}$, then $\theta N \log_2 e\leq m$. Consequently, we have $\tau\leq m$. The diversity gain proposed in \cite{Zheng2003} is an upper bound of $\tau$, as demonstrated in Eq. \eqref{ine1}, which is presented at the bottom of the next page. Thus, we conclude that $-\frac{1}{D_{\rm max}}\frac{\log_2\delta}{\log_2\gamma}$ is not larger than the diversity gain proposed in \cite{Zheng2003}.

Based on Theorem 7, we provide the conclusions for the Rayleigh fading channel with and without frequency or spatial diversity without proof in Corollary 2.

\textbf{Corollary 2}. In the FBL regime, under the assumptions used in Theorem 6,
\begin{itemize}
\item  for a SISO system over a Rayleigh fading channel, we have
\setcounter{equation}{64}
\begin{align}
&\zeta+c_4\varpi=\min\left\{1,\frac{1}{\theta NT\log_2e}\right\}, \, \text{and}\label{kkk1} \\ 
&\tau=\left(\theta N \log_2 e\right)\zeta. \label{kkk2}
\end{align}
By combining Eqs. \eqref{kkk1} and \eqref{kkk2}, we have
\begin{equation}\label{kkk3}
	\begin{aligned}
		(1-\theta N\log_2e)\zeta+&c_4\varpi+\tau\\
		&= \min\left\{1,\frac{1}{\theta NT\log_2e}\right\}.
	\end{aligned}
\end{equation}
\item for a system over Rayleigh fading channels with frequency or spatial diversity $\kappa$, we have
\begin{align}
&\zeta+c_4\varpi=\min\left\{1,\frac{\kappa}{\theta NT\log_2e}\right\}, \, \text{and}\label{kkk4} \\ 
&\tau=\left(\theta N \log_2 e\right)\zeta. \label{kkk5}
\end{align}
By combining Eqs. \eqref{kkk4} and \eqref{kkk5}, we have
\begin{equation}\label{kkk6}
	\begin{aligned}
		(1-\theta N\log_2e)\zeta+&c_4\varpi+\tau\\
		&= \min\left\{1,\frac{\kappa}{\theta NT\log_2e}\right\}.
	\end{aligned}
\end{equation}

In Eqs. \eqref{kkk1}, \eqref{kkk3}, \eqref{kkk4}, and \eqref{kkk6}, $c_4$ is the same as in Eq. \eqref{tttt1}.
\end{itemize}

\begin{figure*}[b]
	\hrulefill
	\setcounter{equation}{63}
	\begin{equation}\label{ine1}
		\begin{aligned}
			\log_2 \Pr\{D[n]>D_{\rm max}\}&\geq \log_2 \left(\Pr\left\{R^{\epsilon}_N[n]<\frac{a[n]}{D_{\rm max}}\right\}\right)^{D_{\rm max}}\\
			&\geq  D_{\rm max}\log_2 \left(\Pr\left\{C[n]<\frac{a[n]}{N D_{\rm max}}\right\}\right).
		\end{aligned}
	\end{equation}
\end{figure*}

\subsection{$\epsilon-\chi$ Tradeoff in the AWGN Channel}

In this subsection, we will take AWGN as an example to characterize the $\epsilon-\chi$ tradeoff through an inequality. The AWGN channel model is widely employed in wired communications and satellite communications. With the rapid development of deep learning technologies, communication between processing units, which usually belongs to wired communication, has also raised much attention from both academia and industry. Therefore, characterizing the performance limits of the AWGN channel in the FBL regime is important. Different from the discussion in Subsections III-A, III-B, and III-C, we will not adopt the assumptions presented in Theorem 1 in this subsection.

For the AWGN channel, the DVP is equivalent to QVP since the service rate is fixed given $N$, $\epsilon$, and $\gamma$. Therefore, we can choose to use QVP representing the delay performance of the communication system in the AWGN channel. Similarly, we use $\epsilon$ to evaluate the reliability of the system. Based on these considerations, we derive the $\epsilon-\chi$ tradeoff as an inequality in Corollary 3. With some abuse of notations, we let $C=\log_2(1+\gamma)$ and $V=\log_2^2e \left(1-\frac{1}{(1+\gamma)^2}\right)$ in Corollary 3.

\textbf{Corollary 3}. Given $\gamma$, $N$, a large $L$, and the requirement $\chi\leq \chi_{\rm th}$, we have
\setcounter{equation}{70}
\begin{equation}\label{nn002}
\epsilon\geq Q\left(\frac{NC+\frac{\log_2 N}{2}+G(N,\gamma,\epsilon)-\alpha_A\left(-\frac{1}{L}\ln \chi_{\rm th}\right)}{\sqrt{NV}}\right).
\end{equation}

\begin{IEEEproof}
See Appendix G.
\end{IEEEproof}

Corollary 3 presents the relationships between $\epsilon$ and $\chi$. In practical applications, we can use these relationships to determine the appropriate choice of $\epsilon$ for a given $\chi_{\rm th}$. For the fading channel, a similar analysis can be performed. However, it is not trivial, as obtaining the analytical expression for $\theta$ is challenging. With the help of approximation methods and under certain assumptions, one may derive an analytical or even closed-form approximation for $\theta$ \cite{li2}. Subsequently, a similar expression to Eq. \eqref{nn002} can be derived, which is beyond the scope of this paper.

\begin{figure*}[b]
	\hrulefill
	\setcounter{equation}{71}
	\begin{equation}
		\alpha_S(\theta)= Nr(x_1^*)-\frac{\ln f(x_1^*)+\frac{1}{2}\ln 2\pi -\frac{1}{2}\ln |r''(x_1^*)|-\frac{1}{2}\ln \theta NT}{\theta T}+O\left(\frac{1}{\theta^2 T^2N}\right). \label{lemma31}
	\end{equation}
	\setcounter{equation}{73}
\begin{equation}\label{appx2}
	\alpha_S(\theta)= Nr(x_2^*)-\frac{\ln f(x_2^*)+\frac{1}{2}\ln 2\pi -\frac{1}{2}\ln |r''(x_2^*)|-\frac{1}{2}\ln \theta NT}{\theta T}+O\left(\frac{1}{\theta^2 T^2N}\right). 
\end{equation}
\setcounter{equation}{76}
\begin{equation}\label{add1}
	\begin{aligned}
		\Xi_1(x)+\frac{{\rm d}\Xi_1(x)}{{\rm d}x}x&=\frac{1}{a_1^{\frac{1}{a_2+1}}(\gamma x)^{\frac{a_2}{a_2+1}}}-\frac{1}{\gamma x}+x\left(-\frac{a_2}{a_2+1}\frac{1}{a_1^{\frac{1}{a_2+1}}\gamma^{\frac{a_2}{a_2+1}}}x^{-\frac{a_2}{a_2+1}-1}+\frac{1}{\gamma}\frac{1}{x^2}\right)=\frac{1-\frac{a_2}{a_2+1}}{a_1^{\frac{1}{a_2+1}}\gamma^{\frac{a_2}{a_2+1}}}x^{-\frac{a_2}{a_2+1}}.
	\end{aligned}
\end{equation}

\end{figure*}

\section{Effective Capacity Approximation with Normal Approximation}

In this section, we present an EC approximation method for low-latency communication systems. The closed-form expression of this approximation characterizes the influence of the channel distribution, error probability, and blocklength on EC.	As shown in Eqs. \eqref{9} and \eqref{delta}, there are limitations in using LDT to determine the QVP and DVP in URLLC scenarios. However, EC has been widely adopted in various studies of URLLC to evaluate the QoS-constrained throughput \cite{zxjsac,8466036,ecresouce1,thirdorder2}. In these studies, the value of the QoS exponent $\theta$ is typically given to represent the QoS requirement, while EC is employed to quantify the throughput subject to these QoS requirements. Several scheduling schemes for URLLC have been proposed to maximize EC in these works. Therefore, deriving a closed-form expression of EC is crucial for further system optimization. However, it is challenging to derive such a closed-form expression for EC, which complicates throughput analysis and the comparison of different scheduling schemes. To address this issue, we propose a Laplace's-method-based approximation for EC in this section. In this subsection, we set $T\geq1$, which covers the general low-latency scenarios. Note that we use normal approximation to approximate the maximum coding rate in this section. Thus, the proposed approximation can serve as an efficient EC approximation method for scenarios where normal approximation achieves satisfactory accuracy.

\subsection{EC Approximation for Fixed-Power Allocation}
In this subsection, we start with a simple case in which the transmitted power remains constant in each time slot. For low-latency scenarios, the thresholds for queue length $L$ and latency $D_{\rm max}$ are usually small due to the stringent QoS requirements \cite{fbleb1}. Thus, to achieve extremely small QVP or DVP, the value of the QoS exponent $\theta$ or the revised QoS exponent $\theta N$ needs to be large. Consequently, $\theta NT$ should have a large value. With this consideration, we begin our analysis by the  fixed-power allocation in Lemma 2. Note that $\ell=\left(\frac{Q^{-1}(\epsilon)}{\sqrt{N}}\right)^2$, as defined before Theorem 4.

\textbf{Lemma 2}. For a low-latency communication system adopting fixed-power allocation, EC is approximated using Eq. \eqref{lemma31}, which is shown at the bottom of this page. In Eq. \eqref{lemma31}, $r(x)$ is defined in Eq. \eqref{maaa1}, and $x_1^*$ is given by 
\setcounter{equation}{72}
\begin{equation}
x_1^*=\frac{1}{\gamma}\left(\sqrt{\frac{1+\sqrt{1+4\ell}}{2}}-1\right).
\end{equation}
\begin{IEEEproof}
See Appendix H.
\end{IEEEproof}

In Lemma 2, we provide an approximation of EC for the fixed-power allocation, which provides a closed-form expression of EC applicable to arbitrary fading channels. Additionally, the approximation error is $O\left(\frac{1}{\theta^2 T^2N}\right)$. Thus, as mentioned earlier, this approximation performs better in a low-latency communication system, which we will validate through numerical calculation in Section V. 

\subsection{EC Approximation with Channel-Gain-Based Power Allocation}
In this subsection, we will generalize the proposed EC approximation method for the fixed-power allocation scheme to certain types of channel-gain-based power allocation schemes. The conditions for using the proposed approximation are provided. Moreover, we discuss the applicability of the proposed approximation for several classical channel-gain-based power allocation schemes.

Building on Lemma 2, we extend the approximation to more general power allocation schemes, which we summarize in Theorem 8.

\textbf{Theorem 8}. For a low-latency communication system which adopts a channel-gain-based power allocation scheme $\Xi(x)$ satisfying $\Xi(x)+\frac{{\rm d}\Xi(x)}{{\rm d} x} x\geq 0$ with a channel distribution $f(\cdot)$, EC is approximated using Eq. \eqref{appx2}, which is presented at the bottom of this page. In Eq. \eqref{appx2}, $r(x)$ is defined in Eq. \eqref{maaa1}, and $x_2^*$ is given by
\setcounter{equation}{74}
\begin{equation}\label{mm2}
\Xi(x_2^*)x_2^*=\sqrt{\frac{1+\sqrt{1+4\ell}}{2}}-1. 
\end{equation}

In Eq. \eqref{mm2}, $\ell=\left(\frac{Q^{-1}(\epsilon)}{\sqrt{N}}\right)^2$.
\begin{IEEEproof}
See Appendix I.
\end{IEEEproof}

In Theorem 8, we present an EC approximation for a specific kind of power allocation schemes. Specifically, for the policies satisfying $\frac{{\rm d}\Xi(x)}{{\rm d} x}\geq 0$, the condition $\Xi(x)+\frac{{\rm d}\Xi(x)}{{\rm d} x} x\geq 0$ is satisfied. The requirement $\frac{{\rm d}\Xi(x)}{{\rm d} x}\geq 0$ indicates that the allocated power does not decrease as the channel power gain increases. This condition holds for certain power allocation schemes, including the well-established water-filling scheme \cite{wf}. Furthermore, when $\frac{{\rm d}\Xi(x)}{{\rm d} x}\geq 0$, $\Xi(x)x$ is a monotonic function of $x$. Consequently, the solution to Eq. \eqref{theo32} can be efficiently obtained via the binary search algorithm. 

A power-control policy designed to maximize EC while adhering to the average power constraint is proposed in \cite{pc1}. This policy also complies with the constraints outlined in Theorem 8. Let $\Xi_1(x)$ denote the power-control policy introduced in \cite{pc1}, which can be expressed as
\begin{equation}
\begin{aligned}
	\Xi_1(x)=\begin{cases}
		\frac{1}{a_1^{\frac{1}{a_2+1}}(\gamma x)^{\frac{a_2}{a_2+1}}}-\frac{1}{\gamma x}, \quad &x\geq \frac{a_1}{\gamma}, \\
		0, \quad & x< \frac{a_1}{\gamma},
	\end{cases}
\end{aligned}
\end{equation} 
where $a_1$ and $a_2$ are positive constants. $a_1$ is related to the average power constraint, while $a_2$ is a linear and increasing function of $\theta$. For $x\geq \frac{a_1}{\gamma}$, we have Eq. \eqref{add1}, which is provided at the bottom of this page. Since $\frac{a_2}{a_2+1}<1$, it follows that $\Xi_1(x)+\frac{{\rm d}\Xi_1(x)}{{\rm d}x}x\geq 0$ for $x>0$, indicating that the power-control policy proposed in \cite{pc1} satisfies the condition of Theorem 8. There are two special cases to consider. When $a_2$ approaches 0, the power-control policy proposed in \cite{pc1} simplifies to the water-filling policy, which has been discussed above. When $a_2$ approaches infinity, the power-control policy proposed in \cite{pc1} becomes channel inversion policy, with which the rate remains constant over all channel states. In this case, EC is trivial, as the rate remains fixed.

\begin{figure*}[t]
	\begin{minipage}[t]{0.48\linewidth}
		\centering
		\includegraphics[width=9cm]{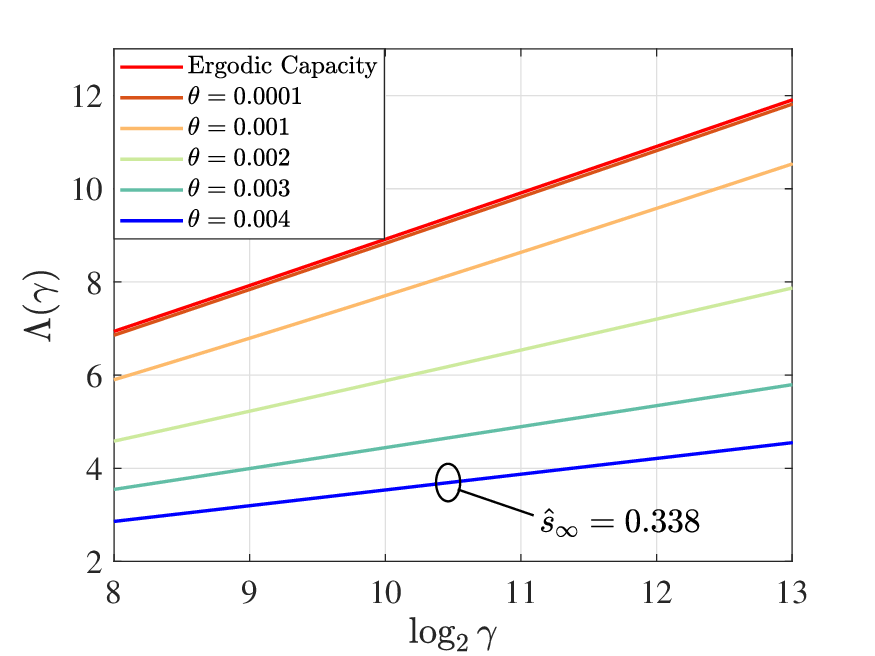}
		\caption{$s_{\infty}$ in the Rayleigh fading channel.}
		\label{titadd_rayleigh}
	\end{minipage}
	\hspace{3mm}
	\begin{minipage}[t]{0.48\linewidth}
		\centering
		\includegraphics[width=9cm]{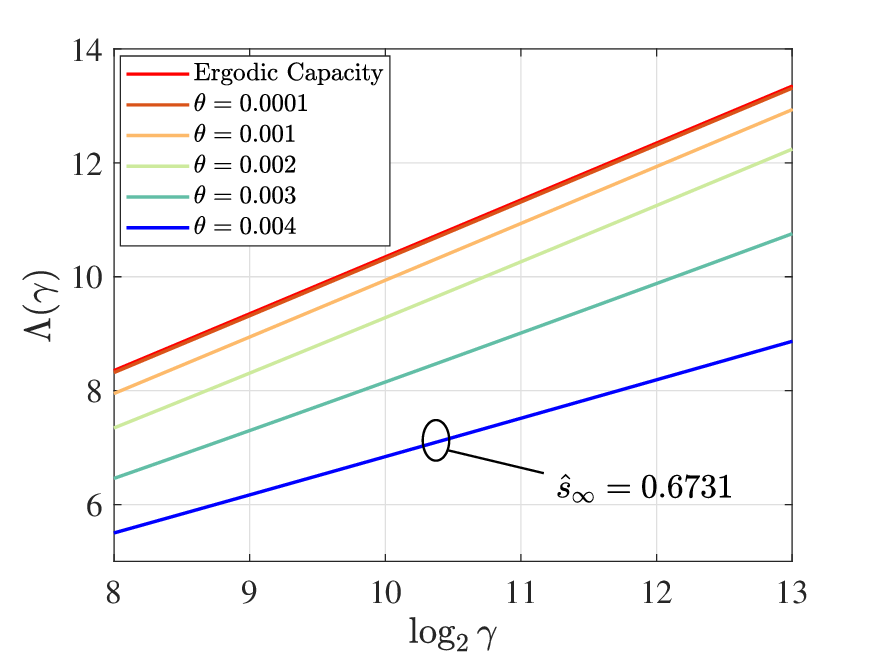}
		\caption{$s_{\infty}$ in Rayleigh fading channels with $\kappa=2$.}
		\centering
		\label{titadd_erlang}
	\end{minipage}
\end{figure*}

Although the approximation provided in Theorem 8 can be applied to policies satisfying $\Xi(x)+\frac{{\rm d}\Xi(x)}{{\rm d} x} x\geq 0$, its accuracy can be further enhanced through additional optimization tailored to the specific properties of each policy. For example, in the water-filling scheme, there exists a channel-gain threshold $h_{\rm th}$. If the channel gain falls below this threshold, the allocated power becomes 0, implying that the transmission rate remains constant in the FBL regime when $h[n]<h_{\rm th}$.\footnote{The actual rate should be zero when $\gamma=0$. However, since we use the normal approximation as mentioned at the beginning of this section, we let the rate be equal to $\frac{\log_2N}{2}$ when $\gamma=0$, in line with Eq. \eqref{c2}. This setting can be easily adjusted by setting the rate to 0 when $\gamma=0$, which does not affect the accuracy of this approximation.} Let $\xi(\theta)=e^{-\theta T \frac{\log_2(N)}{2}}\int_{0}^{h_{\rm th}} f(x)   {\rm d}x$. Based on the above analysis and Theorem 8, EC can be approximated using Eq. \eqref{trunc}, which is presented at the bottom of this page. For those policies exhibiting truncated characteristics,  Eq. \eqref{trunc} provides a more accurate approximation of their EC.

\begin{figure*}[b]
	\hrulefill
	\setcounter{equation}{77}
	\begin{equation}
		\alpha_S(\theta)= -\frac{1}{\theta N} \ln\left( \xi(\theta) + \sqrt{\frac{2\pi}{\theta NT |r''(x^*)|}}f(x^*)e^{-\theta NT r(x^*)}\left(1+O\left((\theta NT)^{-1}\right)\right)\right). \label{trunc}
	\end{equation}
	\setcounter{equation}{78}
	\begin{equation}
		\alpha_S(\theta)=-\frac{1}{\theta T}\ln\left( \epsilon+(1-\epsilon)\sqrt{\frac{2\pi}{\theta NT |r''(x^*)|}}f(x^*)e^{-\theta NT r(x^*)}\left(1+O\left((\theta NT)^{-1}\right)\right)\right). \label{coro31}
	\end{equation}
\end{figure*} 

For the low-latency system incorporating the retransmission mechanism, EC is discussed in \cite{eur}, which differs from the analysis provided above where no retransmission mechanism is considered. In \cite{eur}, a simple ARQ mechanism is assumed. Under this mechanism, the receiver sends a negative acknowledgment to request the retransmission of the message in case of an erroneous reception. In Corollary 4, we generalize the EC approximation to the low-latency communication system incorporating this ARQ mechanism.

\textbf{Corollary 4.} For a low-latency system utilizing the above simple ARQ mechanism and a channel-gain-based power allocation scheme $\Xi(x)$ that satisfies $\Xi(x)+\frac{{\rm d}\Xi(x)}{{\rm d} x} x\geq 0$, with channel distribution $f(\cdot)$, EC is approximated using Eq. \eqref{coro31}, which is located at the bottom of this page. In Eq. \eqref{coro31}, $r(x)$ is defined by Eq. \eqref{maaa1}, and $x^*$ is given by Eq. \eqref{theo32}.
\begin{IEEEproof}
See Appendix J. 
\end{IEEEproof}

\section{Numerical Results}

In this section, we perform numerical calculations to validate the theoretical results and conclusions presented in this paper. Moreover, the numerical results provide a more visual explanation of some characteristics of the statistical-delay performance in low-latency systems operating in the FBL regime.

\begin{figure*}[t]
	\centering
	\subfigure[$s_{\infty}$ in the Nakagami-$m$ channel with $m=0.5$.]{\includegraphics[width=9cm]{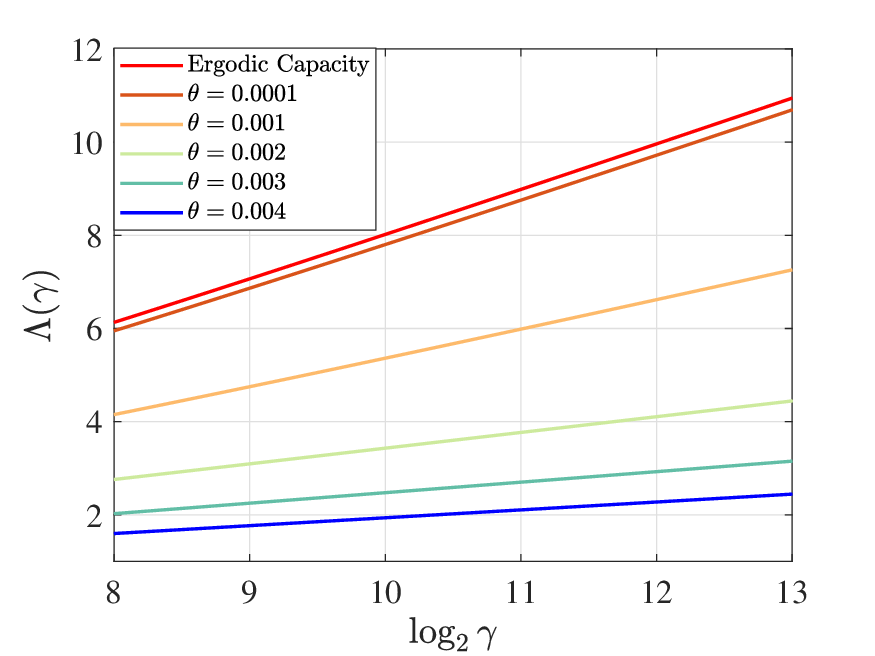}}
	\subfigure[$s_{\infty}$ in the Nakagami-$m$ channel with $m=2$.]{\includegraphics[width=9cm]{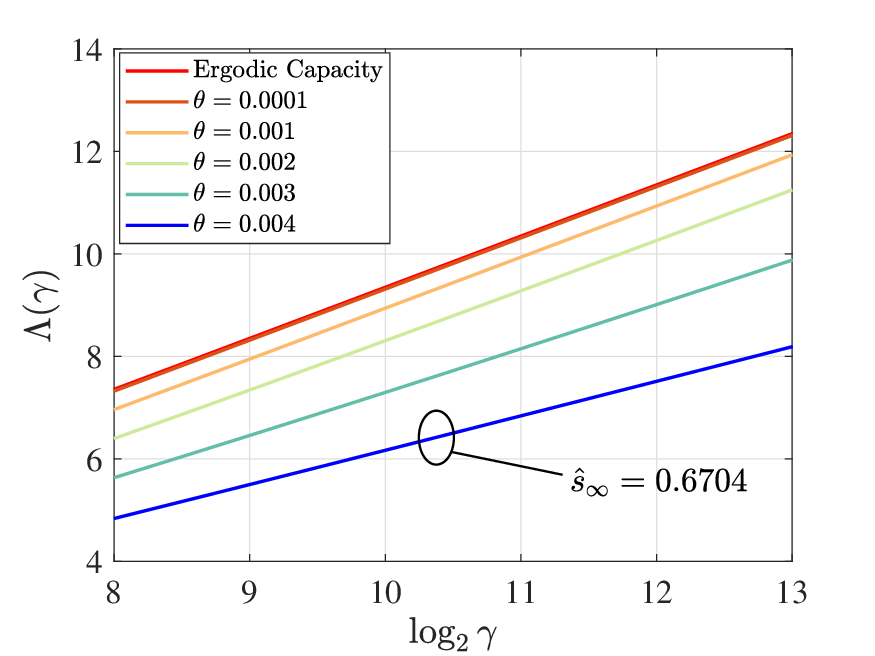}}
	\caption{$s_{\infty}$ in the Nakagami-$m$ fading channel.}
	\label{titadd_naka}
\end{figure*}

First, we validate the conclusions regrading the high-SNR slopes of EC in the FBL regime, shown in Figs. \ref{titadd_rayleigh}, \ref{titadd_erlang}, and \ref{titadd_naka}. In these figures, we set $N=512$, $\epsilon=10^{-5}$, and $T=1$. In Fig. \ref{titadd_rayleigh}, $\mathbb{E}_{|h|^2}\{|h[n]|^2\}=1$. In Fig. \ref{titadd_erlang}, $f(x)$ is an Erlang distribution with $\kappa=2$ and $\mathbb{E}_{|h|^2}\{|h[n]|^2\}=2$. In Fig. \ref{titadd_naka}, we set $\Omega=1$. For the Rayleigh fading case, $\frac{1}{T\log_2e}\approx0.6931$. From Fig. \ref{titadd_rayleigh}, we find that the slopes of $\Lambda(\gamma)$ for $\theta N\leq 0.6931$, i.e. $\theta\leq 0.0014$, are almost identical and close to 1, while the slopes of $\Lambda(\gamma)$ for $\theta>0.0014$ are smaller than 1. Moreover, for $\theta>0.0014$, a larger $\theta$ leads to a smaller slope of $\Lambda(\gamma)$. Specifically, the slope of $\Lambda(\gamma)$ for $\theta=0.004$ is $0.338$ in Fig. \ref{titadd_rayleigh}, while $\frac{1}{\theta N T \log_2 e}$ with $\theta=0.004$ equals 0.3385. These results validate the accuracy of Corollary 1. 

Next, we discuss the numerical results of the case in which $|h[n]|^2$ follows an Erlang distribution. In Fig. \ref{titadd_erlang}, $\frac{\kappa}{T\log_2e}\approx1.3863$. We observe that for $\theta=0.002$, $\theta N=1.024$ is larger than $\frac{1}{T\log_2e}\approx0.6931$ in Fig. \ref{titadd_rayleigh}, but smaller than $\frac{\kappa}{T\log_2e}\approx1.3863$ in Fig. \ref{titadd_erlang}. Accordingly, the slope of $\Lambda(\gamma)$ for $\theta=0.002$ in Fig. \ref{titadd_rayleigh} differs from the slope of ergodic capacity, while the slope of $\Lambda(\gamma)$ for $\theta=0.002$ is nearly identical to the slope of ergodic capacity in Fig. \ref{titadd_erlang}. This result highlights the differences in $s_{\infty}$ performance with and without spatial or frequency diversity in Rayleigh fading channels, further validating Corollary 1.

\begin{figure*}[t]
	\centerline{\includegraphics[width=15cm]{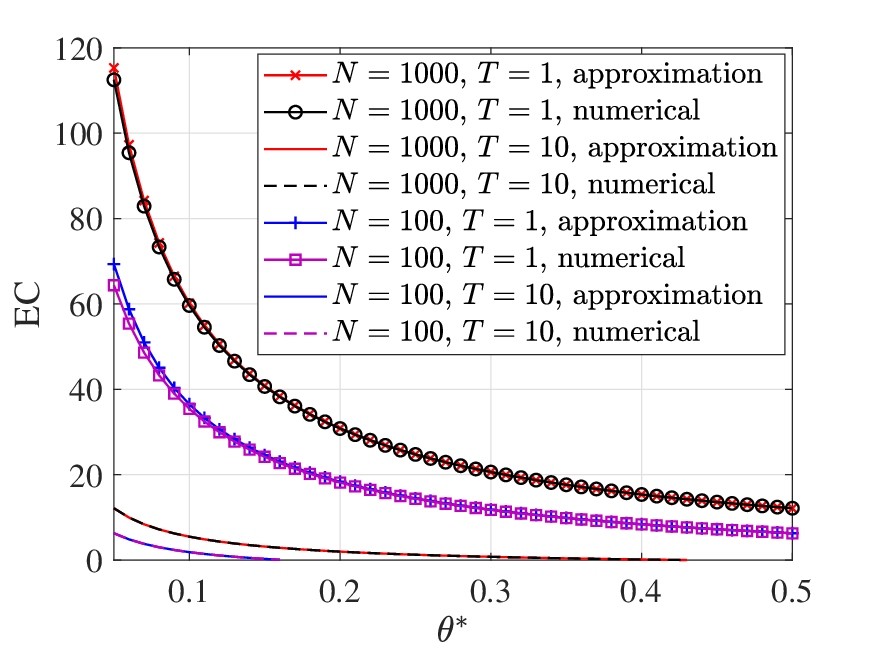}}
	\caption{EC approximation for the fixed-power allocation scheme.}
	\label{fig6}
\end{figure*}

In Fig. \ref{titadd_naka}, we present the numerical results for $s_{\infty}$ in the Nakagami-$m$ fading channel. In Fig. \ref{titadd_naka}(a), $\frac{m}{TN\log_2e}\approx0.3466$, while in Fig. \ref{titadd_naka}(b), $\frac{m}{TN\log_2e}\approx1.3836$. The slopes of $\Lambda(\gamma)$ in Figs. \ref{titadd_naka}(a) and \ref{titadd_naka}(b) validate the accuracy of Theorem 5. Besides, by comparing the results in Figs. \ref{titadd_erlang} and \ref{titadd_naka}(b), we find that the slopes of $\Lambda(\gamma)$ are nearly identical for the same value of $\theta$. For example, when $\theta=0.004$, i.e., $\theta N=2.0480$, the slope of $\Lambda(\gamma)$ in Fig. \ref{titadd_erlang} is $0.6731$, while it is $0.6704$ in Fig. \ref{titadd_naka}(b). This finding further validates the accuracy of Theorem 5 and Corollary 1, as $s_{\infty}$ with $\kappa = 2$ and $m = 2$ are shown to be identical in both Theorem 5 and Corollary 1.

\begin{figure*}[t]
	\centerline{\includegraphics[width=15cm]{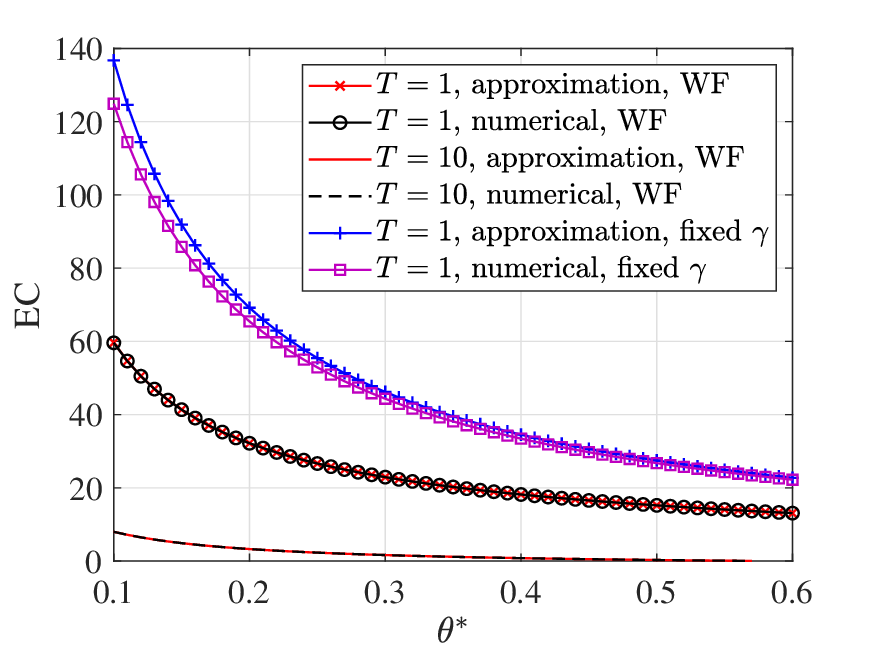}}
	\caption{EC approximation for the water-filling policy.}
	\label{fig7}
\end{figure*}

Additionally, we validate the accuracy of the EC approximation through numerical calculations. In Fig. \ref{fig6}, we present the results of the EC approximation for the fixed-power allocation, where $\gamma=10$ and $\epsilon=10^{-3}$. The channel is modeled as a Rayleigh fading channel with $\mathbb{E}_{|h|^2}\{|h[n]|^2\}=1$. The results with the legend ``numerical'' are calculated based on the definition of the EC in Eq. \eqref{ebs}, while the results with the legend ``approximation'' are obtained from Eq. \eqref{lemma31}. As shown in Fig. \ref{fig6}, we find that the approximation results match well with the numerical results. However, the difference between the numerical and approximation results with $N=100$ and $T=1$ is slightly more evident when $\theta$ is small compared to the other three pairs. This is reasonable since the approximation error of Laplace's method is $O\left(\frac{1}{\theta^2 T^2N}\right)$. From these results, we see that the approximation in Lemma 2 satisfactorily approximates the EC in the low-latency scenarios.

Fig. \ref{fig7} illustrates the results of EC approximation for the water-filling (WF) policy \cite{wf}. To evaluate the performance of the EC approximation in different channel models, the channel gain in Fig. \ref{fig7} is set as $|h[n]|^2=|h_1[n]|^2+|h_2[n]|^2$, where $|h_i[n]|^2$ follows an exponential distribution with expectation 1. The approximation for the WF policy used here is based on Eq. \eqref{trunc}. As seen in Fig. \ref{fig7}, the approximation results closely matches the numerical results, which validates our analysis in Section III-C. Note that, for the WF policy with the average SNR $\mathbb{E}_{|h|^2}\{\gamma\}=10$,\footnote{For simplicity, we use $\gamma$ to denote the SNR of the WF policy, which should be a random variable related to $|h[n]|^2$.} the average transmission rate $\mathbb{E}_{{|h|^2}}\{s[n]\}\approx 3932$, while for the fixed-power allocation policy with $\gamma=10$, $\mathbb{E}_{|h|^2}\{s[n]\}\approx3923$. This result indicates that the transmission optimization under the infinite blocklength assumption can improve the average transmission rate in the FBL regime under certain parameter settings. However, the result in Fig. \ref{fig7} also shows that, although the WF policy leads to a larger average transmission rate compared to the fixed-power allocation scheme, it does not improve the EC. Since the EC represents the maximum throughput under the QVP or DVP constraints, the relatively poorer performance of the WF policy on the EC highlights the distinction between maximizing the average transmission rate and maximizing throughput under statistical QoS constraints.

\section{Conclusion}

In this paper, we mainly focused on characterizing the fundamental tradeoff between reliability, latency, and rate for low-latency systems operating under FBC. Based on the previous conclusions on the gain-conservation properties for low-latency systems with deterministic arrival, we extended the definitions of the reliability gain, real-time gain, and service-rate gain for the low-latency systems with random arrivals. By employing EB and EC, we established the connection between the gain-conservation equation and the influence of random arrivals. Our analysis, conducted over both AWGN and Nakagami-$m$ fading channels, provided insights into service-rate gains for both wired and wireless systems. Then, the revised gain-conservation equations were derived by incorporating service-rate, reliability, and real-time gains. Additionally, an EC approximation method was proposed to derive an analytical expression of EC in systems with channel-gain-based scheduling policies in the FBL regime.  The conclusions proposed in this paper offer a comprehensive characterization of the fundamental tradeoff between the reliability, latency, and throughput for low-latency systems with FBC, which will provide instructions for designing efficient scheduling schemes for the low-latency systems. Future research directions include refining the approximations of QVP and DVP with small thresholds, relaxing the assumptions of the proposed analysis through tighter bounds or exact expression of the FBC rate, and developing advanced scheduling schemes based on the revised gain-conservation equations and the proposed EC approximation method for low-latency communication systems.

\appendices

\section{Proof of Theorem 2}
According to Eqs. \eqref{up1} and \eqref{low1}, we derive Eqs. \eqref{upper} and \eqref{lower}, which are provided at the bottom of the next page.

\begin{figure*}[b]
	\hrulefill
	\setcounter{equation}{79}
	\begin{equation}\label{upper}
		g_u(|h|^2 \gamma,\epsilon)=\frac{c_1}{V\left(|h|^2\gamma\right)}+\frac{3}{2}\log_2 V\left(|h|^2\gamma\right) -\log_2 K\left(|h|^2\gamma\right).
	\end{equation}
	\begin{equation}\label{lower}
		\begin{aligned}
			g_l(N,|h|^2\gamma,\epsilon)&=-\frac{1}{2}\log_2 N+\log_2\frac{K\left(|h|^2\gamma\right)}{c_2 V^{\frac{3}{2}}\left(|h|^2\gamma\right)} -\log_2 \left[2\left(\frac{\ln 2}{\sqrt{2\pi}}+\frac{2K\left(|h|^2\gamma\right)}{V^{\frac{3}{2}}\left(|h|^2\gamma\right)}\right)\right]\\
			&\quad\quad\quad +\sqrt{NV\left(|h|^2\gamma\right)}\left[Q^{-1}\left(1-\epsilon+\frac{2K\left(|h|^2\gamma\right)}{\sqrt{N}V^{\frac{3}{2}}\left(|h|^2\gamma\right)}\right)+Q^{-1}(\epsilon)\right].
		\end{aligned}
	\end{equation}
\end{figure*}

Since $|h|^2$ and $\gamma$ are independent, the product $|h|^2\gamma$ can be zero, finite, or approach positive infinity. Hence, we have to consider all possible cases of $|h|^2\gamma$. The essence of this proof is to demonstrate that $G(N,|h|^2\gamma,\epsilon)$ remains bounded regardless of the value of $|h|^2\gamma$.
The proof is divided into four parts. We let $\vartheta>0$ denote an arbitrarily small constant. The first part is the proof for $|h|^2\gamma \to 0$, i.e., $|h|^2\gamma\in[0,\vartheta)$.  The second part is the proof for a bounded $\Psi(\gamma)$ and $|h|^2\gamma\in[\vartheta,+\infty)$. The third part is the proof for $\lim\limits_{\gamma\to+\infty} \Psi(\gamma)=+\infty$ and $|h|^2\gamma\in[\vartheta,+\infty)$. Finally, we will summarize the proof in the fourth subsection. To apply the bounds shown in Eqs. \eqref{upper} and \eqref{lower}, we assume that
\setcounter{equation}{81}
\begin{equation}\label{condN}
N>\max\left\{\left(\frac{2K(|h|^2 \gamma)}{(1-\epsilon)V^{\frac{3}{2}}(|h|^2 \gamma)}\right)^2, \left(\frac{2K(|h|^2 \gamma)}{\epsilon V^{\frac{3}{2}}(|h|^2 \gamma)}\right)^2 \right\}.
\end{equation}
The validity of Eq. \eqref{condN} is established in the fourth part.

\subsection{Proof for $|h|^2\gamma \to 0$, i.e., $|h|^2\gamma\in[0,\vartheta)$}

The analysis of the case where $|h|^2\gamma \to 0$ is grounded in its physical significance. The proof in this subsection is to demonstrate that $G(N,|h|^2\gamma,\epsilon)$ remains bounded as $|h|^2\gamma\to0$. When $|h|^2\gamma\in \to0$, it is clear that $\tilde{R}^{\epsilon}_N$ is bounded according to Eq. \eqref{c2}. Since we can not obtain an unbounded rate with a finite received SNR, the coding rate  $\tilde{R}^{\epsilon}_N+G(N,|h|^2 \gamma, \epsilon)$ must also be bounded as $|h|^2\gamma\to 0$. Consequently, $G(N,|h|^2 \gamma, \epsilon)$ must remain bounded when $|h|^2\gamma\to0$. This result implies that $|G(N,|h|^2 \gamma, \epsilon)|\leq u_1$ when $|h|^2\gamma\to 0$, where $u_1$ is a positive constant.

\subsection{Proof for a Bounded $\Psi(\gamma)$ and $|h|^2\gamma\in[\vartheta,+\infty)$}

We use the upper and lower bounds to demonstrate that $G(N,|h|^2\gamma,\epsilon)$ remains bounded as $|h|^2 \gamma\to +\infty$. We start the analysis with the upper bound $g_u(|h|^2 \gamma,\epsilon)$.  According to Eq. \eqref{kfunc}, $K\left(|h|^2\gamma\right)$ is finite for a given finite $|h|^2\gamma\in[\vartheta,+\infty)$. As $|h|^2\gamma \to +\infty$, $\frac{\gamma}{1+\gamma}\to1$ and
\begin{equation}
	\lim_{|h|^2\gamma\to+\infty}\mathbb{E}_z\left\{\left|z^2-\frac{2}{\sqrt{|h|^2\gamma}}z-1\right|^3\right\}<+\infty.
\end{equation} 
Thus, $K\left(|h|^2\gamma\right)$ remains bounded when $|h|^2\gamma\in[\vartheta,+\infty)$.

Moreover, we have 
\begin{equation}
	\left(1-\frac{1}{(1+\vartheta)^2} \right) \left(\log_2 e\right)^2\leq  V\left(|h|^2\gamma\right)\leq \left(\log_2 e\right)^2, \nonumber 
\end{equation}
 for $|h|^2\gamma\in[\vartheta,+\infty)$. This implies that both $K\left(|h|^2\gamma\right)$ and $V\left(|h|^2\gamma\right)$ are bounded for $|h|^2\gamma\in[\vartheta,+\infty)$.

Next, we analyze $b_1$. First, we want to show that $b_1<1-\epsilon$, i.e.,
\begin{equation}\label{appp1}
	\frac{2K\left(|h|^2\gamma\right)}{V^{\frac{3}{2}}\left(|h|^2\gamma\right)\sqrt{\left(\frac{2K\left(|h|^2\gamma\right)}{V^{\frac{3}{2}}\left(|h|^2\gamma\right)(1-\epsilon)}\right)^2+1}}< 1-\epsilon.
\end{equation}
By manipulating Eq. \eqref{appp1}, it is equivalent to show that the following inequality holds.
\begin{equation}\label{appp2}
	\frac{2K\left(|h|^2\gamma\right)}{\sqrt{(2K\left(|h|^2\gamma\right))^2+V^3\left(|h|^2\gamma\right)(1-\epsilon)^2}}< 1.
\end{equation}

Eq. \eqref{appp2} holds since $\left(1-\frac{1}{(1+\vartheta)^2} \right) \left(\log_2 e\right)^2\leq  V\left(|h|^2\gamma\right)$ with $|h|^2\gamma\in[\vartheta,+\infty)$. Hence, $1-\epsilon-b_1>0$ indicating that $\min_{y\in\left[1-\epsilon-b_1,1-\epsilon\right]}\frac{{\rm d}Q^{-1}(y)}{{\rm d}y}$ is finite. Therefore, $c_1$ is bounded for $|h|^2\gamma\in[\vartheta,+\infty)$ according to Eq. \eqref{lll1}. 

Thus, we have proved that for $|h|^2\gamma\in[\vartheta,+\infty)$, every term of $g_u\left(|h|^2\gamma,\epsilon\right)$ shown in Eq. \eqref{upper} is finite. Consequently,  $|g_u\left(|h|^2\gamma,\epsilon\right)|\leq u_2$ for $|h|^2\gamma\in[\vartheta,+\infty)$, where $u_2$ is a positive constant.

\begin{figure*}[b]
	\hrulefill
	\setcounter{equation}{87}
	\begin{equation}\label{add2}
		\begin{aligned}
			&\lim_{\gamma\to+\infty} \frac{\sqrt{\Psi(\gamma)V\left(|h|^2\gamma\right)}\left[Q^{-1}\left(1-\epsilon+\frac{2K\left(|h|^2\gamma\right)}{\sqrt{\Psi(\gamma)}V^{\frac{3}{2}}\left(|h|^2\gamma\right)}\right)+Q^{-1}(\epsilon)\right]}{\Psi(\gamma)}\\
			=&\lim_{\gamma\to+\infty}\frac{\sqrt{\Psi(\gamma)V\left(|h|^2\gamma\right)}\left[Q^{-1}\left(1-\epsilon\right)+Q^{-1}(\epsilon)+O\left(\frac{2K\left(|h|^2\gamma\right)}{\sqrt{\Psi(\gamma)}V^{\frac{3}{2}}\left(|h|^2\gamma\right)}\right)\right]}{\Psi(\gamma)}\\
			=&\lim_{\gamma\to+\infty} \frac{\sqrt{\Psi(\gamma)V\left(|h|^2\gamma\right)}O\left(\frac{2K\left(|h|^2\gamma\right)}{\sqrt{\Psi(\gamma)}V^{\frac{3}{2}}\left(|h|^2\gamma\right)}\right)}{\Psi(\gamma)}\\
			=&0.
		\end{aligned}
	\end{equation}

\end{figure*}

Then, we will analyze $g_l(N,|h|^2\gamma,\epsilon)$. We consider two cases. Following a similar approach to the analysis of $g_u\left(|h|^2\gamma,\epsilon\right)$, every term in $g_l(N,|h|^2\gamma,\epsilon)$ is finite for $|h|^2\gamma\in[\vartheta,+\infty)$. This result implies that $\left|g_l(N,|h|^2\gamma,\epsilon)\right|\leq u_3$, where $u_3$ is a positive constant.

In summary, we have proved that both $|g_u\left(|h|^2\gamma,\epsilon\right)|$ and $|g_l(N,|h|^2\gamma,\epsilon)|$ are bounded for $|h|^2\gamma\in[\vartheta,+\infty)$ and a bounded $\Psi(\gamma)$. Since $g_l(N,|h|^2\gamma,\epsilon) \leq  G(N,|h|^2 \gamma, \epsilon) \leq  g_u(|h|^2\gamma,\epsilon)$, it follows that $|G(N,|h|^2 \gamma, \epsilon)|\leq \max\{u_2,u_3\}$ for $|h|^2\gamma\in[\vartheta,+\infty)$.

Combining the results from the last subsection with those of this subsection, we have proved that  
\setcounter{equation}{85}
\begin{equation}\label{bo1}
	|G(N,|h|^2\gamma,\epsilon)|\leq \max\{u_1,u_2,u_3\}, \quad  |h|^2\gamma\in[0,+\infty),
\end{equation}
holds for a given $\epsilon$, a bounded $\Psi(\gamma)$, and an $N$ satisfying Eq. \eqref{condN}. With a bounded $\Psi(\gamma)$, Eq. \eqref{bo1} ensures that Eq. \eqref{newcod1} is satisfied. Therefore, to satisfy the condition in Eq. \eqref{newcod1} with a given $\epsilon$ and a bounded $\Psi(\gamma)$, it is sufficient to ensure that Eq. \eqref{condN} is fulfilled.

 \subsection{Proof for $\lim\limits_{\gamma\to+\infty} \Psi(\gamma)=+\infty$ and $|h|^2\gamma\in[\vartheta,+\infty)$}

Since the upper bound is independent of $N$, the only difference between this subsection and the last subsection lies in the analysis of the lower bound $g_l(N,|h|^2\gamma,\epsilon)$. For $|h|^2\gamma \in[0,\vartheta)$, the proof is the same as that presented in Subsection A. Therefore, the focus is placed on $|h|^2\gamma\in[\vartheta,+\infty)$. When $\lim\limits_{\gamma\to+\infty} \Psi(\gamma)=+\infty$, two terms in $g_l(N,|h|^2\gamma,\epsilon)$ are unbounded. The unbounded terms are $-\frac{1}{2}\log_2 \Psi(\gamma)$ and
\begin{equation}\label{unbterm}
	\begin{aligned}
		\sqrt{\Psi(\gamma)V\left(|h|^2\gamma\right)}&\Bigg[Q^{-1}(\epsilon(\gamma))+\\ Q^{-1}&\left(1-\epsilon(\gamma)+\frac{2K\left(|h|^2\gamma\right)}{\sqrt{\Psi(\gamma)}V^{\frac{3}{2}}\left(|h|^2\gamma\right)}\right)\Bigg]. \nonumber
	\end{aligned}
\end{equation}
 Since $\lim\limits_{\gamma\to+\infty}\Psi(\gamma)=+\infty$, it follows that 
\begin{equation}\label{appae1}
	\lim_{\gamma\to+\infty}\frac{-\frac{1}{2}\log_2\Psi(\gamma)}{\Psi(\gamma)}=0.
\end{equation}

Next, we consider the term shown in Eq. \eqref{unbterm}. By performing a Taylor-series expansion of the inverse Gaussian Q-function around $1-\epsilon$, we obtain Eq. \eqref{add2}, which appears at the bottom of this page. Consequently, for the lower bound $g_l(\Psi(\gamma),|h|^2\gamma,\epsilon(\gamma))$, any $\Psi(\gamma)$ that satisfies $\lim\limits_{\gamma\to+\infty}\Psi(\gamma)=+\infty$ can ensure $	\lim\limits_{\gamma\to+\infty}\frac{g_l(\Psi(\gamma),|h|^2\gamma,\epsilon(\gamma))}{\Psi(\gamma)}=0$. Given that the upper bound is also bounded, we have $\lim\limits_{\gamma\to+\infty}\frac{g_u(|h|^2\gamma,\epsilon)}{\Psi(\gamma)}=0$. Since we use the upper and lower bounds, it is essential to verify that Eq. \eqref{condN} holds. Thus, we have demonstrated that, with a $\Psi(\gamma)$ satisfying Eq. \eqref{condN} and $\lim\limits_{\gamma\to+\infty}\Psi(\gamma)=+\infty$, Eq. \eqref{newcod1} holds.

\subsection{Summary of the Proof}

From the above proof, we find that whether $\Psi(\gamma)$ is bounded or $\lim\limits_{\gamma\to+\infty}\Psi(\gamma)=+\infty$, Eq. \eqref{condN} is sufficient to ensure that Eq. \eqref{newcod1} holds for a given $\epsilon$. For the convenience of expression, we define 
	\setcounter{equation}{88}
	\begin{equation}
		\varsigma_1=\max_{x\geq 0}\left\{\left(\frac{2K(x)}{(1-\epsilon)V^{\frac{3}{2}}(x)}\right)^2,\left(\frac{2K(x)}{\epsilon V^{\frac{3}{2}}(x)}\right)^2\right\}.
	\end{equation}
 Therefore, to ensure that Eq. \eqref{condN} holds for a given $\epsilon$, we can choose a $\Psi(\gamma)$ satisfying
\begin{equation}\label{Ncond1}
	\Psi(\gamma)>\varsigma_1.
\end{equation}

To establish the validity of Eq. \eqref{Ncond1}, we have to prove that $\varsigma_1$ is bounded. It is equivalent to proving that  $\left(\frac{2K(x)}{V^{\frac{3}{2}}(x)}\right)^2$ is bounded on the interval $[0,+\infty)$.  For $\left(\frac{2K(x)}{V^{\frac{3}{2}}(x)}\right)^2$, we have shown that its value is bounded for both $x\to+\infty$ and finite $x$. As $x\to 0$, we have 
\begin{equation}\label{appxa1}
	\begin{aligned}
		\lim_{x\to0}\frac{2K(x)}{V^{\frac{3}{2}}(x)}=&\frac{2c_0}{(\log_2e)^3}\lim_{x\to0}\frac{x^3\mathbb{E}_z\left\{\left|z^2-\frac{2}{\sqrt{x}}z-1\right|^3\right\}}{(1+x)^3\left(1-\frac{1}{(1+x)^2}\right)^{\frac{3}{2}}}\\
		=&\frac{2c_0}{(\log_2e)^3}\lim_{x\to0}\frac{x^\frac{3}{2}}{(x+2)^\frac{3}{2}}\left(u_4 x^{-\frac{3}{2}}+o\left(x^{-\frac{3}{2}}\right)\right)\\
		=&u_5,
	\end{aligned}
\end{equation}
where $u_4$ and $u_5$ are finite constant, and $o\left(x^{-\frac{3}{2}}\right)$ satisfies $\lim\limits_{x\to0}\frac{o\left(x^{-\frac{3}{2}}\right)}{x^{-\frac{3}{2}}}=0$. Thus, we have proved that $\varsigma_1$ is bounded. The proof is completed.

\section{Proof of Theorem 3}
For finite $|h|^2\gamma$, the proof is the same as that in Appendix A, where both $g_u(|h|^2\gamma,\epsilon(\gamma))$ and $g_l(\Psi(\gamma),|h|^2\gamma,\epsilon(\gamma))$ are proved to be bounded. Therefore, in the following proof, we mainly focus on the analysis for $|h|^2\gamma\to+\infty$. 

Note that we use the upper and lower bounds in the following proof. Therefore, this result is based on the assumption shown in Eq. \eqref{condN2}, which is located at the bottom of this page.

\begin{figure*}[b]
	\hrulefill
	\setcounter{equation}{91}
	\begin{equation}\label{condN2}
		\begin{aligned}
			\Psi(\gamma)>\max\left\{\left(\frac{2K(|h|^2 \gamma)}{(1-\epsilon(\gamma))V^{\frac{3}{2}}(|h|^2 \gamma)}\right)^2, \left(\frac{2K(|h|^2 \gamma)}{\epsilon(\gamma) V^{\frac{3}{2}}(|h|^2 \gamma)}\right)^2 \right\}.
		\end{aligned}
	\end{equation}
\setcounter{equation}{96}
\begin{equation}\label{add3}
	H(\gamma,|h|^2\gamma,\epsilon(\gamma))=\sqrt{\Psi(\gamma)V\left(|h|^2\gamma\right)}\left[Q^{-1}\left(1-\epsilon(\gamma)+\frac{2K\left(|h|^2\gamma\right)}{\sqrt{\Psi(\gamma)}V^{\frac{3}{2}}\left(|h|^2\gamma\right)}\right)+Q^{-1}(\epsilon(\gamma))\right].
\end{equation}
\begin{equation}\label{appb1}
	Q^{-1}\left(1-\epsilon(\gamma)+\frac{2K\left(|h|^2\gamma\right)}{\sqrt{\Psi(\gamma)}V^{\frac{3}{2}}\left(|h|^2\gamma\right)}\right)+Q^{-1}(\epsilon(\gamma))\leq Q^{-1}\left(1-\epsilon(\gamma)\right)+Q^{-1}(\epsilon(\gamma))=0.
\end{equation}
\end{figure*}

 Since $\epsilon(\gamma)\in(0,0.5]$, it follows that $1-\epsilon(\gamma)\geq \epsilon(\gamma)$. Thus, to ensure Eq. \eqref{condN2} holds, it is sufficient to ensure the following condition: 
 	\setcounter{equation}{92}
 \begin{equation}
 	\Psi(\gamma)>\left(\frac{2K(|h|^2 \gamma)}{\epsilon(\gamma) V^{\frac{3}{2}}(|h|^2 \gamma)}\right)^2.
 \end{equation}
 
  We define
  \begin{equation}
  	\varsigma_2=4\max_{x\geq 0}\left\{\frac{K^2(x)}{V^{3}(x)}\right\}. \nonumber 
  \end{equation}
Similar to the analysis in Eq. \eqref{appxa1}, $\varsigma_2$ is bounded. Thus, we can impose the condition  $\Psi(\gamma)\epsilon^{2}(\gamma)>\varsigma_2$ to ensure Eq. \eqref{condN2} is satisfied. We let $\Psi(\gamma)\epsilon^{2}(\gamma)=(1+\omega)\varsigma_2$, where $\omega>0$ is an arbitrarily positive constant. Then we have
  \begin{equation}\label{condN3}
  	\epsilon(\gamma)=\sqrt{\frac{(1+\omega)\varsigma_2}{\Psi(\gamma)}}.
  \end{equation}
Eq. \eqref{condN3} will be used in the subsequent analysis.

\subsection{Analysis of $g_u(|h|^2\gamma,\epsilon(\gamma))$ for $\lim\limits_{\gamma\to+\infty}\epsilon(\gamma)=0$ and $\lim\limits_{\gamma\to+\infty}\Psi(\gamma)\to+\infty$}

	Given that $\lim\limits_{\gamma\to+\infty}\epsilon(\gamma)=0$, we have $1-\epsilon(\gamma) \to 1$ as $\gamma\to+\infty$. Thus, according to Eq. \eqref{lll1}, $c_1\to+\infty$ for $\gamma\to+\infty$ since  $\frac{dQ^{-1}(y)}{dy}\Big|_{y=1-\epsilon(\gamma)} \to -\infty$. Except for $c_1$, the other terms in $g_u(|h|^2\gamma, \epsilon(\gamma))$ remain bounded.  Therefore, to ensure that there exists a finite $\nu$ such that $\lim\limits_{\gamma\to+\infty}\left|\frac{g_u(|h|^2\gamma,\epsilon(\gamma))}{\Psi(\gamma)}\right|\leq \nu$ holds, it is equivalent to requiring
\begin{equation}\label{appbe1}
	\lim_{\gamma\to+\infty}\frac{\frac{dQ^{-1}(1-\epsilon(\gamma))}{dy}}{\Psi(\gamma)}< +\infty.
\end{equation} 

By substituting the derivative of inverse Gaussian Q-function, which is given by
\begin{equation}
	\frac{{\rm d}Q^{-1}(y)}{{\rm d}y}=\left[-\frac{1}{\sqrt{2\pi}}\exp\left(-\left(\frac{Q^{-1}(y)}{2}\right)^2\right)\right]^{-1}, \nonumber 
\end{equation} 
into Eq. \eqref{appbe1}, we arrive at the requirement that
\begin{equation}\label{varepcond1}
	\begin{aligned}
		&\lim_{\gamma\to+\infty}\frac{e^{\left(\frac{Q^{-1}(1-\epsilon(\gamma))}{2}\right)^2}}{\Psi(\gamma)}\\
		=& \lim_{\gamma\to+\infty}\frac{e^{\left(\frac{Q^{-1}(\epsilon(\gamma))}{2}\right)^2}}{\Psi(\gamma)}<+\infty.
	\end{aligned}
\end{equation}

\begin{figure*}[b]
	\hrulefill

	\setcounter{equation}{99}
\begin{equation}\label{add4}
	\begin{aligned}
		Q^{-1}\left(1-\epsilon(\gamma)+\frac{2K\left(|h|^2\gamma\right)}{\sqrt{\Psi(\gamma)}V^{\frac{3}{2}}\left(|h|^2\gamma\right)}\right)&\geq  Q^{-1}\left(1-\sqrt{\frac{\varsigma_2}{\Psi(\gamma)}}\left(\sqrt{1+\omega}-1\right)\right)\\
		&=-Q^{-1}\left(\sqrt{\frac{\varsigma_2}{\Psi(\gamma)}}\left(\sqrt{1+\omega}-1\right)\right).
	\end{aligned}
\end{equation}
\begin{equation}\label{add5}
	\begin{aligned}
		&\lim_{\gamma\to+\infty}\frac{\sqrt{\Psi(\gamma)V\left(|h|^2\gamma\right)}\left[Q^{-1}\left(1-\epsilon(\gamma)+\frac{2K\left(|h|^2\gamma\right)}{\sqrt{\Psi(\gamma)}V^{\frac{3}{2}}\left(|h|^2\gamma\right)}\right)+Q^{-1}(\epsilon(\gamma))\right]}{\Psi(\gamma)}
		\\
		\geq &\lim_{\gamma\to+\infty} \frac{-Q^{-1}\left(\sqrt{\frac{\varsigma_2}{\Psi(\gamma)}}\left(\sqrt{1+\omega}-1\right)\right)+Q^{-1}(\epsilon(\gamma))}{\sqrt{\Psi(\gamma)}}.
	\end{aligned}
\end{equation}

	\setcounter{equation}{106}
\begin{subequations}\label{add6}
	\begin{align}
		\lim_{\gamma\to+\infty}\frac{Q^{-1}\left(\frac{\omega_1}{\sqrt{\Psi(\gamma)}}\right)}{\sqrt{\Psi(\gamma)}}&\leq \lim_{\gamma\to+\infty}\frac{\sqrt{W\left(\frac{\Psi(\gamma)}{2\pi \omega_1^{2}}\right)}}{\sqrt{\Psi(\gamma)}} \nonumber \\
		&=\frac{1}{\sqrt{\log_2 e}}\lim_{\gamma\to+\infty}\frac{\sqrt{\log_2\Psi(\gamma)-\log_2 2\pi \omega_1^2}}{\sqrt{\Psi(\gamma)}} \label{tttt}\\
		&=\frac{1}{\sqrt{\log_2 e}}\lim_{\gamma\to+\infty}\sqrt{\frac{\log_2\Psi(\gamma)}{\Psi(\gamma)}} \nonumber \\
		&=0. \nonumber 
	\end{align}
\end{subequations}
\end{figure*}

\subsection{Analysis of $g_l(\Psi(\gamma),|h|^2\gamma,\epsilon(\gamma))$ for $\lim\limits_{\gamma\to+\infty}\epsilon(\gamma)=0$ and $\lim\limits_{\gamma\to+\infty}\Psi(\gamma)\to+\infty$}

Given that $\lim\limits_{\gamma\to+\infty}\epsilon(\gamma)=0$ and $\lim\limits_{\gamma\to+\infty}\Psi(\gamma)\to+\infty$, the two terms in Eq. \eqref{lower} are unbounded, which are the same as those mentioned in Appendix A. The term $-\frac{1}{2}\log_2\Psi(\gamma)$ follows Eq. \eqref{appae1} without changes. For simplicity, we define $H(\gamma,|h|^2\gamma,\epsilon(\gamma))$ in Eq. \eqref{add3}, which is located at the bottom of this page, to represent another unbounded term. Since $Q^{-1}(x)$ is a decreasing function and $\frac{2K\left(|h|^2\gamma\right)}{\sqrt{\Psi(\gamma)}V^{\frac{3}{2}}\left(|h|^2\gamma\right)}>0$, we obtain Eq. \eqref{appb1}, which is also located at the bottom of this page. From Eq. \eqref{appb1}, we have 
\setcounter{equation}{98}
\begin{equation}
	\lim_{\gamma\to+\infty}\frac{H(\gamma,|h|^2\gamma,\epsilon(\gamma))}{\Psi(\gamma)}\leq 0.
\end{equation}

We have proved that the upper bound of $\lim\limits_{\gamma\to+\infty}\frac{H(\gamma,|h|^2\gamma,\epsilon(\gamma))}{\Psi(\gamma)}$ is finite. Next, we demonstrate the condition under which the lower bound of $\lim\limits_{\gamma\to+\infty}\frac{H(\gamma,|h|^2\gamma,\epsilon(\gamma))}{\Psi(\gamma)}$ is finite.
According to Eq. \eqref{condN3} and the definition of $\varsigma_2$, we derive Eq. \eqref{add4} at the bottom of the next page. Based on Eq. \eqref{add4}, we obtain Eq. \eqref{add5}, which is also located at the bottom of the next page.

We will provide the proof of $\lim\limits_{\gamma\to+\infty}\frac{Q^{-1}\left(\sqrt{\frac{\varsigma_2}{\Psi(\gamma)}}\left(\sqrt{1+\omega}-1\right)\right)}{\sqrt{\Psi(\gamma)}}=0$ in Appendix C. Thus, $\lim\limits_{\gamma\to+\infty}\frac{H(\gamma,|h|^2\gamma,\epsilon(\gamma))}{\Psi(\gamma)}<+\infty$ holds if  
	\setcounter{equation}{101}
\begin{equation}\label{varepcond2}
	\lim_{\gamma\to+\infty}\frac{Q^{-1}(\epsilon(\gamma))}{\sqrt{\Psi(\gamma)}}<+\infty.
\end{equation}
Thus, if Eq. \eqref{varepcond2} holds, we have 
\begin{equation}
	\lim\limits_{\gamma\to+\infty}\left|\frac{g_l(\Psi(\gamma),|h|^2\gamma,\epsilon(\gamma))}{\Psi(\gamma)}\right|\leq \nu. \nonumber
\end{equation}

Therefore, if Eqs. \eqref{varepcond1}, \eqref{varepcond2}, and $\Psi(\gamma)\epsilon^{2}(\gamma)>\varsigma_2$ hold, the condition stated in Eq. \eqref{newcod1} is satisfied. The proof is completed.

\section{Proof of 
	 $\lim\limits_{\gamma\to+\infty}\frac{Q^{-1}\left(\sqrt{\frac{\varsigma_2}{\Psi(\gamma)}}\left(\sqrt{1+\omega}-1\right)\right)}{\sqrt{\Psi(\gamma)}}=0$}

Since $\Psi(\gamma)\epsilon^2(\gamma)=(1+\omega)\varsigma_2$, we have $\sqrt{\frac{\varsigma_2}{\Psi(\gamma)}}\left(\sqrt{1+\omega}-1\right)<\epsilon(\gamma)<\frac{1}{2}$. Thus, we obtain $\lim\limits_{\gamma\to+\infty}\frac{Q^{-1}\left(\sqrt{\frac{\varsigma_2}{\Psi(\gamma)}}\left(\sqrt{1+\omega}-1\right)\right)}{\sqrt{\Psi(\gamma)}}\geq 0$. To apply the squeeze theorem, we then need to demonstrate that $\lim\limits_{\gamma\to+\infty}\frac{Q^{-1}\left(\sqrt{\frac{\varsigma_2}{\Psi(\gamma)}}\left(\sqrt{1+\omega}-1\right)\right)}{\sqrt{\Psi(\gamma)}}\leq 0$. For simplicity, we define $\omega_1=\sqrt{\varsigma_2}\left(\sqrt{1+\omega}-1\right)$ in the following proof. 
Let $Q^{-1}\left(v\right)=x$, which implies $v=Q(x)$.  Based on Mills' ratio, for $x>0$, we obtain that
\begin{equation}
	\frac{1}{\sqrt{2\pi}}\frac{x}{1+x^2}e^{-\frac{x^2}{2}}\leq v \leq \frac{1}{\sqrt{2\pi}}\frac{1}{x}e^{-\frac{x^2}{2}}. \label{gain1}
\end{equation}

Then, we have
\begin{equation}
	x^2+\ln x^2\leq-2\ln\left(\sqrt{2\pi}v\right). \label{ttt2}
\end{equation}

Let us introduce the Lambert W function $W(\cdot)$. $W(v)$ is the solution of $ye^y=v$. Based on $W(\cdot)$, we can transform Eq. \eqref{ttt2} into
\begin{equation}
	-\sqrt{W\left(\frac{v^{-2}}{2\pi}\right)} \leq x\leq  \sqrt{W\left(\frac{v^{-2}}{2\pi}\right)}.\label{newgain11}
\end{equation}

Since $\frac{\omega_1}{\sqrt{\Psi(\gamma)}}<\frac{1}{2}$, we obtain the following relationships from Eq. \eqref{newgain11}:
\begin{equation}\label{ttt}
0<Q^{-1}\left(\frac{\omega_1}{\sqrt{\Psi(\gamma)}}\right)<\sqrt{W\left(\frac{\Psi(\gamma)}{2\pi \omega_1^{2}}\right)}.
\end{equation}

Based on Eq. \eqref{ttt}, we obtain Eq. \eqref{add6}, which is presented at the bottom of this page. Eq. \eqref{tttt} holds because, according to \cite{lambertw}, for $y>e$, the Lambert W function satisfies the following inequalities: 
\setcounter{equation}{107}
\begin{equation}
		W(y)\geq \ln y-\ln\ln y+\frac{\ln \ln y}{2\ln y} ,
	\label{gain3}
\end{equation}
and 
\begin{equation}
	W(y)\leq \ln y-\ln \ln y +\frac{e}{e-1}\frac{\ln\ln y}{\ln y}.\label{mm1}
\end{equation}

Based on Eqs. \eqref{gain3} and \eqref{mm1}, we find that for $y>e$, the following equation holds:
\begin{equation}\label{tt20}
	\lim_{y\to+\infty}\frac{\sqrt{W(y)}}{\sqrt{\log_2 y}}=\frac{1}{\sqrt{\log_2e}}.
\end{equation}

Since $\frac{\Psi(\gamma)}{2\pi \omega_1^{2}}$ approaches $+\infty$ with $\gamma\to+\infty$, we can apply Eq. \eqref{tt20} into Eq. \eqref{tttt}. The proof is completed.

\section{Proof of Theorem 5}

\begin{figure*}[b]
	\hrulefill
	
	\setcounter{equation}{110}
	\begin{equation}\label{add7}
		\begin{aligned}
			s_{\infty}&=-\frac{1}{\theta NT}\lim_{\gamma\to+\infty}\frac{\ln\mathbb{E}_{|h|^2}\left\{e^{-T\theta R_N^{\epsilon}[n]}\right\}}{\log_2\gamma}\\
			&=-\frac{1}{\theta NT}\lim_{\gamma\to+\infty}\frac{\ln \mathbb{E}_{|h|^2}\left\{e^{-T\theta \left(R_N^{\epsilon}[n]-\tilde{R}_N^{\epsilon}[n]\right)}e^{-T\theta \tilde{R}_N^{\epsilon}[n]}\right\}}{\log_2\gamma}\\
			&=-\frac{1}{\theta NT} \lim_{\gamma\to+\infty}\frac{\ln \mathbb{E}_{|h|^2}\left\{e^{-T\theta \tilde{R}_N^{\epsilon}[n]}e^{-T\theta G(\Psi(\gamma),|h[n]|^2\gamma)}\right\}}{\log_2\gamma}.
		\end{aligned}
	\end{equation}
	
	\setcounter{equation}{113}
	\begin{equation}\label{add8}
		\begin{aligned}
			s_{\infty}&=-\frac{1}{\theta NT}\lim_{\gamma\to+\infty}\frac{  \ln\int_{0}^{\infty}(1+\gamma x)^{-\theta NT \log_2 e}e^{\theta T\sqrt{N\left(1-\frac{1}{(1+\gamma x)^2}\right)}Q^{-1}(\epsilon)\log_2 e -\theta T \frac{\log_2N}{2}}\left(\frac{m}{\Omega}\right)^m\frac{x^{m-1}}{\Gamma(m)}e^{-\frac{m}{\Omega}x}{\rm d}x  }{\log_2 \gamma}\\
			&=-\frac{1}{\theta NT}\lim_{\gamma\to+\infty}\frac{  \ln\int_{0}^{\infty}(1+\gamma x)^{-\theta NT \log_2 e}e^{\theta T\sqrt{N\left(1-\frac{1}{(1+\gamma x)^2}\right)}Q^{-1}(\epsilon)\log_2 e -\theta T \frac{\log_2N}{2}}x^{m-1}e^{-\frac{m}{\Omega}x}{\rm d}x  }{\log_2 \gamma}.
		\end{aligned}
	\end{equation}
	\setcounter{equation}{114}
\begin{equation}
	\begin{aligned}
		s_{\infty}&\leq -\frac{1}{\theta NT}  \lim_{\gamma\to+\infty}  \frac{\ln\left[\int_{0}^{\infty} (1+\gamma x)^{-\theta NT\log_2 e}\left(\frac{m}{\Omega}\right)^m e^{-\theta T\frac{\log_2N}{2}}x^{m-1}e^{-\frac{m}{\Omega}x}{\rm d}x\right]}{\log_2 \gamma} \\
		&=-\frac{1}{\theta NT}  \lim_{\gamma\to+\infty}  \frac{\ln \left[e^{-\theta T\frac{\log_2N}{2}}\gamma^{-\theta NT \log_2 e}\int_{0}^{\infty} \left(\frac{1}{\gamma}+x\right)^{-\theta NT \log_2 e}x^{m-1} e^{-\frac{m}{\Omega}x}{\rm d}x\right]}{\log_2\gamma}.
		\label{nakaupp}
	\end{aligned}
\end{equation}

\setcounter{equation}{115}
\begin{equation}\label{tt007}
	\begin{aligned}
		\int_{0}^{\infty}\left(\frac{1}{\gamma}+x\right)^{-\theta NT \log_2 e}x^{m-1}e^{-\frac{m}{\Omega}x}{\rm d}x&\leq \int_{0}^{\infty} x^{m-\theta NT \log_2 e-1}e^{-\frac{m}{\Omega}x}{\rm d}x\\
		&=\left(\frac{\Omega}{m}\right)^{m-\theta NT \log_2e}\int_0^{\infty}t^{m-\theta NT\log_2 e-1}e^{-t}{\rm d}t.
	\end{aligned}
\end{equation}
\end{figure*}

  Based on the definition of $s_{\infty}$ proposed in Eq. \eqref{defi1} and $\tilde{R}^{\epsilon}_{N}[n]=R^{\epsilon}_{N}[n]-G(N,|h[n]|^2\gamma)$, we obtain Eq. \eqref{add7} at the bottom of this page.

Since $N=\Psi(\gamma)$ and $N\theta=\varrho$, we have $\theta=\frac{\varrho}{\Psi(\gamma)}$. As discussed in Section II, if $\Psi(\gamma)$ satisfies the conditions of Theorem 2, Eq. \eqref{newcod1} holds for a given $\epsilon$. Thus, the following inequality holds:
\setcounter{equation}{111}
\begin{equation}\label{nnn1}
\left|\theta G(N,|h[n]|^2\gamma)\right|=\left|\frac{\varrho G(N,|h[n]|^2\gamma)}{\Psi(\gamma)}\right|\leq \varrho\nu.
\end{equation}

From Eq. \eqref{nnn1}, we obtain that $c_5\leq e^{-T\theta G(N,|h[n]|^2\gamma)}\leq c_6$, where $c_5$ and $c_6$ are positive constants. Under the above conditions, we obtain
\setcounter{equation}{112}
\begin{equation}\label{ttmm1}
s_{\infty}=-\frac{1}{\theta NT}\lim_{\gamma\to+\infty}\frac{\ln\mathbb{E}_{|h|^2}\left\{e^{-T\theta \tilde{R}_N^{\epsilon}[n]}\right\}}{\log_2\gamma}.
\end{equation}

From Eq. \eqref{ttmm1}, it is obvious that $s_{\infty}=1$ for the AWGN channel. For the Nakagami-$m$ fading channel, we obtain Eq. \eqref{add8}, which is presented at the bottom of this page.

We will utilize Eq. \eqref{add8} to carry out the following proof. The following proof is divided into four parts. We analyze the upper bound of $s_{\infty}$ for $m\geq 1$ in the first part, followed by its lower bound in the second part. Next, we present the upper and lower bounds of $s_{\infty}$ for $m\in[0.5,1)$ in the third and fourth part, respectively. For simplicity, we let $\beta= \sqrt{N} Q^{-1}(\epsilon)\log_2 e$ in the proofs of this paper.

\subsection{Upper bound of $s_{\infty}$ for $m\geq 1$}

For $m\geq 1$, we first derive the upper bound of $s_{\infty}$. Since $R_\epsilon[n]\leq N\log_2\left(1+|h[n]|^2\gamma\right)$, we derive Eq. \eqref{nakaupp}, which is presented at the bottom of this page.

\begin{figure*}[b]
	\hrulefill

	\setcounter{equation}{117}
	\begin{subequations}\label{tt11}
		\begin{align}
			&\int_0^{\infty}\left(\frac{1}{\gamma}+x\right)^{-\theta NT \log_2 e}x^{m-1}e^{-\frac{m}{\Omega}x}{\rm d}x \nonumber \\ =&\int_{0}^{\infty}\left(\frac{1}{\gamma}+x\right)^{-\theta NT \log_2 e+m -1}\cdot \left(\frac{x}{\frac{1}{\gamma}+x}\right)^{m-1} e^{-\frac{m}{\Omega}x}{\rm d}x \nonumber \\
			\geq & \int_{\frac{1}{\gamma}}^{\infty}\left(\frac{1}{\gamma}+x\right)^{-\theta NT \log_2 e+m-1}\left(\frac{1}{2}\right)^{m-1}e^{-\frac{m}{\Omega}x}{\rm d}x   \label{tt2} \\
			=&\left(\frac{1}{2}\right)^{m-1}\left(\frac{m}{\Omega}\right)^{\theta NT\log_2 e-m}e^{\frac{m}{\gamma\Omega}}\Gamma\left(m-\theta NT \log_2 e,\frac{2m}{\Omega\gamma}\right) . \label{tt3}
		\end{align}
	\end{subequations}
		\setcounter{equation}{118}
	\begin{subequations}\label{tt16}
		\begin{align}
			s_{\infty}&\leq-\frac{1}{\theta NT} \lim_{\gamma\to+\infty} \frac{\ln\left[e^{-\theta T\frac{\log_2N}{2}}\gamma^{-\theta NT \log_2 e}\left(\frac{1}{2}\right)^{m-1}\left(\frac{m}{\Omega}\right)^{\theta NT\log_2 e-m}e^{\frac{m}{\gamma\Omega}}\Gamma\left(m-\theta NT \log_2 e,\frac{2m}{\Omega\gamma}\right)\right]}{\log_2 \gamma} \nonumber \\
			&=-\frac{1}{\theta NT }\lim_{\gamma \to+\infty} \frac{\ln\left[e^{-\theta T\frac{\log_2N}{2}}\gamma^{-m}\left(\frac{1}{2}\right)^{\theta NT \log_2 e-1}e^{\frac{m}{\gamma\Omega}}\frac{\Gamma\left(m-\theta NT \log_2 e,\frac{2m}{\Omega\gamma}\right)}{\left(\frac{2m}{\Omega\gamma}\right)^{m-\theta NT \log_2 e}}\right]}{\log_2 \gamma} \nonumber \\
			&=-\frac{1}{\theta NT}\left(\lim_{\gamma\to+\infty}\frac{-m\ln \gamma}{\log_2 \gamma}+\lim_{\gamma \to+\infty} \frac{\ln\left[\frac{\Gamma\left(m-\theta NT \log_2 e,\frac{2m}{\Omega\gamma}\right)}{\left(\frac{2m}{\Omega\gamma}\right)^{m-\theta NT \log_2 e}}\right]}{\log_2 \gamma}+\lim_{\gamma \to+\infty}\frac{m}{\Omega\gamma\log_2\gamma}\right) +\lim_{\gamma\to+\infty}\frac{\log_2N}{2N}\nonumber \\
			&=-\frac{1}{\theta NT}\left(\lim_{\gamma\to+\infty}\frac{-m\ln \gamma}{\log_2 \gamma}+\lim_{\gamma \to+\infty}\frac{m}{\Omega\gamma\log_2\gamma}\right)\label{tt4}\\
			&=\frac{m}{\theta NT\log_2 e}.
		\end{align}
	\end{subequations}

	\setcounter{equation}{120}
\begin{subequations}\label{tt15}
	\begin{align}
		s_{\infty}&\leq -\frac{1}{\theta NT}\lim_{\gamma\to+\infty}\frac{\ln\left[e^{-\theta T\frac{\log_2N}{2}}\gamma^{-\theta NT \log_2 e}\left(\frac{1}{2}\right)^{m-1}\left(\frac{m}{\Omega}\right)^{\theta NT\log_2 e-m}e^{\frac{m}{\gamma\Omega}}E_1\left(\frac{2m}{\Omega\gamma}\right)\right]}{\log_2 \gamma} \nonumber \\
		&\leq -\frac{1}{\theta NT}\lim_{\gamma \to+\infty}\frac{\ln\left[e^{-\theta T\frac{\log_2N}{2}}\gamma^{-\theta NT \log_2 e}\left(\frac{1}{2}\right)^{m}e^{-\frac{m}{\Omega\gamma}}\ln\left(1+\frac{\Omega\gamma}{m}\right)\right]}{\log_2 \gamma} \label{tt5} \\
		&=-\frac{1}{\theta NT}\left(\lim_{\gamma\to+\infty}\frac{-\theta NT \log_2 e\ln\gamma}{\log_2\gamma}-\lim_{\gamma\to+\infty}\frac{m}{\Omega\gamma\log_2\gamma}+\lim_{\gamma\to+\infty}\frac{\ln\ln\left(1+\frac{\Omega\gamma}{m}\right)}{\log_2\gamma}\right)+\lim_{\gamma\to+\infty}\frac{\log_2N}{2N} \nonumber \\
		&=1.\nonumber 
	\end{align}
\end{subequations}
\end{figure*}

\begin{figure*}[b]
	\hrulefill

	\setcounter{equation}{122}
	\begin{subequations}\label{lownakaadd1}
		\begin{align}
			s_{\infty}&\geq-\frac{1}{\theta NT}  \lim_{\gamma\to+\infty}  \frac{\ln\left[e^{-\theta T\frac{\log_2N}{2}}\int_{0}^{\infty} (1+\gamma x)^{-\theta NT\log_2 e}e^{\theta T \beta-\theta T \frac{\log_2 N}{2}} x^{m-1}e^{-\frac{m}{\Omega}x}{\rm d}x\right]}{\log_2 \gamma} \nonumber \\
			&=-\frac{1}{\theta NT}  \lim_{\gamma\to+\infty}  \frac{\ln \left[e^{-\theta T\frac{\log_2N}{2}}\gamma^{-\theta NT \log_2 e}e^{\theta T \beta}\int_{0}^{\infty} \left(\frac{1}{\gamma}+x\right)^{-\theta NT \log_2 e}x^{m-1} e^{-\frac{m}{\Omega}x}{\rm d}x\right]}{\log_2\gamma} \label{tt006} \\
			&\geq -\frac{1}{\theta NT}\lim_{\gamma \to+\infty} \frac{\ln\left[e^{-\theta T\frac{\log_2N}{2}}\gamma^{-\theta NT \log_2 e}e^{\theta T\beta}\int_0^{\infty}\left(\frac{1}{\gamma}+x\right)^{-\theta NT \log_2 e+m-1}e^{-\frac{m}{\Omega}x}{\rm d}x\right]}{\log_2 \gamma}  \label{tt6} \\
			&=-\frac{1}{\theta NT}\lim_{\gamma \to+\infty}\frac{\ln\left[e^{-\theta T\frac{\log_2N}{2}}\gamma^{-\theta NT \log_2 e}e^{\theta T\beta}\left(\frac{m}{\Omega}\right)^{\theta NT\log_2 e-m}e^{\frac{m}{\gamma\Omega}}\Gamma\left(m-\theta NT \log_2 e,\frac{m}{\Omega\gamma}\right)\right]}{\log_2 \gamma}. \nonumber 
		\end{align}
	\end{subequations}
		\setcounter{equation}{124}
	\begin{subequations}\label{slb2}
		\begin{align}
			s_{\infty}&\geq  -\frac{1}{\theta NT}\lim_{\gamma \to+\infty} \frac{\ln\left[e^{-\theta T\frac{\log_2N}{2}}\gamma^{-\theta NT \log_2 e}e^{\theta T\beta}\left(\frac{m}{\Omega}\right)^{\theta NT\log_2 e-m}\frac{\left(\frac{m}{\Omega\gamma}+c_7\right)^{m-\theta NT \log_2 e}-\left(\frac{m}{\Omega\gamma}\right)^{m-\theta NT \log_2 e}}{c_7(m-\theta NT \log_2 e)}\right]}{\log_2 \gamma}\nonumber \\
			&=-\frac{1}{\theta NT}\left(\lim_{\gamma \to+\infty} \frac{\ln\left[\gamma^{-m}e^{\theta T\beta}\left((1+\frac{c_7\Omega}{m}\gamma)^{m-\theta NT \log_2 e}-1\right)\right]}{\log_2 \gamma}-\lim_{\gamma\to+\infty}\frac{\ln\left[c_7\left(m-\theta NT\log_2e\right)\right]}{\log_2 \gamma}\right) \nonumber\\
			&=-\frac{1}{\theta NT}\lim_{\gamma \to+\infty}\frac{-m\ln\gamma}{\log_2 \gamma}-\lim_{\gamma \to+\infty}\frac{\beta}{N\log_2 \gamma}-\frac{m-\theta NT \log_2 e}{\theta NT}\lim_{\gamma \to+\infty}\frac{\ln\left(1+\frac{c_7\Omega}{m}\gamma\right)}{\log_2 \gamma}\label{lb3}\\
			&=1 \label{lbres}.
		\end{align}
	\end{subequations}
	\setcounter{equation}{126}
\begin{equation}\label{add9}
	\begin{aligned}
		s_{\infty}&\geq -\frac{1}{\theta NT}\lim_{\gamma\to+\infty}\frac{-\theta N T \log_2 e \ln \gamma}{\log_2\gamma}-\lim_{\gamma\to+\infty}\frac{\beta}{N\log_2\gamma}+\lim_{\gamma\to+\infty}\frac{\log_2 N}{2N}\\
		&=1.
	\end{aligned}
\end{equation}
	\setcounter{equation}{127}
\begin{subequations}\label{tt17}
	\begin{align}
		s_{\infty}&\geq -\frac{1}{\theta NT}\lim_{\gamma \to+\infty}\frac{  \ln\left[e^{-\theta T\frac{\log_2N}{2}}\gamma^{-m}e^{\theta T\beta}e^{\frac{m}{\Omega \gamma}}\frac{\Gamma\left(m-\theta NT \log_2 e,\frac{m}{\Omega\gamma}\right)}{\left(\frac{m}{\Omega \gamma}\right)^{m-\theta NT\log_2 e}}\right]    }{\log_2 \gamma} \nonumber \\
		&=-\frac{1}{\theta NT}\left(-\lim_{\gamma\to+\infty}\frac{m\ln\gamma}{\log_2 \gamma}+\lim_{\gamma\to+\infty}\frac{\theta T\beta}{\log_2 \gamma}+\lim_{\gamma\to+\infty}\frac{m}{\Omega\gamma\log_2\gamma}\right) +\lim_{\gamma\to+\infty}\frac{\log_2N}{2N}\label{tt7}\\
		&=\frac{m}{\theta NT \log_2 e}.\nonumber 
	\end{align}
\end{subequations}
\end{figure*}

\subsubsection{$ 1< \frac{m}{\theta NT\log_2 e}$}
\ 
\newline \indent   
For $1< \frac{m}{\theta  NT\log_2 e}$, we find that Eq. \eqref{tt007}, which is presented at the bottom of this page, holds.

According to \cite{rayleigh1}, $0<\int_0^{\infty}z^{a}e^{-x}{\rm d}x<\infty$ if $a>-1$. Since in this case $m>\theta NT \log_2 e$, we obtain $m-\theta NT \log_2 e-1>-1$. Therefore, the integral $\int_{0}^{\infty}\left(\frac{1}{\gamma}+x\right)^{-\theta NT \log_2 e}x^{m-1}e^{-\frac{m}{\Omega}x}{\rm d}x$ is bounded by a finite constant. Thus, under the condition $1< \frac{m}{\theta  N T\log_2 e}$, Eq. \eqref{nakaupp} can be further expressed as
\setcounter{equation}{116}
\begin{equation}\label{tt14}
\begin{aligned}
	s_{\infty}&\leq \lim_{\gamma\to+\infty}\frac{\log_2N}{2N} -\frac{1}{\theta NT }\lim_{\gamma\to+\infty}\frac{-\theta NT \log_2 e\ln\gamma}{\log_2 \gamma}\\
	&=1.
\end{aligned}
\end{equation}
Eq. \eqref{tt14} holds because \begin{equation}
	\lim\limits_{\gamma\to+\infty}\frac{\log_2N}{N}=\lim\limits_{\gamma\to+\infty}\frac{\log_2\Psi(\gamma)}{2\Psi(\gamma)}=0. \nonumber 
\end{equation}
is valid under the condition that $\lim\limits_{\gamma\to+\infty}\Psi(\gamma)=+\infty$.

\subsubsection{$ 1> \frac{m}{\theta NT\log_2 e}$}
\ 
\newline \indent  
For $1> \frac{m}{\theta NT\log_2 e}$, we focus on the integral 
\begin{equation}
	\int_0^{\infty}\left(\frac{1}{\gamma}+x\right)^{-\theta NT \log_2 e}x^{m-1}e^{-\frac{m}{\Omega}x}{\rm d}x. \nonumber 
\end{equation}
 This yields Eq. \eqref{tt11}, which is shown at the bottom of this page. Eq. \eqref{tt2} holds because $\frac{x}{\frac{1}{\gamma}+x}$ an increasing function for $x\in \big[\frac{1}{\gamma},\infty\big)$.

By substituting Eq. \eqref{tt3} into Eq. \eqref{nakaupp}, we arrive at Eq. \eqref{tt16}, which is presented at the bottom of this page. Eq. \eqref{tt4} holds because according to \cite{incgamma2}, the following inequality holds for $s<0$:
\setcounter{equation}{119}
\begin{equation}
\lim_{x\to0}\frac{\Gamma(s,x)}{x^s}=-\frac{1}{s}. \label{limitgamma}
\end{equation}

\subsubsection{$1=\frac{m}{\theta NT\log_2e}$}
\ 
\newline \indent  
For $1=\frac{m}{\theta NT\log_2e}$, we have
\begin{equation}
	\Gamma\left(m-\theta NT \log_2 e,\frac{2m}{\Omega\gamma}\right)=E_1\left(\frac{2m}{\Omega\gamma}\right). \nonumber
\end{equation} 
By substituting this equation and Eq. \eqref{tt3} into Eq. \eqref{nakaupp}, we arrive at Eq. \eqref{tt15}, which is shown at the bottom of this page. Eq. \eqref{tt5} holds because, as demonstrated in \cite{e1}, the following inequality holds:
\setcounter{equation}{121}
\begin{equation}
E_1(x)>\frac{1}{2}e^{-x}\ln\left(1+\frac{2}{x}\right), \quad x>0.
\end{equation}

Thus, we have completed the proof for the upper bound of $s_{\infty}$ for $m\geq 1$.

\begin{figure*}[b]
	\hrulefill

		\setcounter{equation}{128}
	\begin{subequations}\label{tt18}
		\begin{align}
			s_{\infty}&\geq -\frac{1}{\theta NT}\lim_{\gamma \to+\infty}\frac{\ln\left[e^{-\theta T\frac{\log_2N}{2}}\gamma^{-\theta NT \log_2 e}e^{\theta T\beta}\left(\frac{m}{\Omega}\right)^{\theta NT\log_2 e-m}e^{\frac{m}{\gamma\Omega}}E_1\left(\frac{m}{\Omega\gamma}\right)\right]}{\log_2 \gamma}\nonumber \\
			&\geq -\frac{1}{\theta NT}\lim_{\gamma \to+\infty}\frac{\ln\left[e^{-\theta T\frac{\log_2N}{2}}\gamma^{-\theta NT \log_2 e}e^{\theta T\beta}\left(\frac{m}{\Omega}\right)^{\theta NT\log_2 e-m}\ln\left(1+\frac{\Omega\gamma}{m}\right)\right]}{\log_2 \gamma} \label{tt10} \\
			&=-\frac{1}{\theta NT}\left(\lim_{\gamma\to+\infty}\frac{-\theta NT\log_2 e\ln\gamma}{\log_2\gamma}+\lim_{\gamma\to+\infty}\frac{\theta T \beta}{\log_2\gamma}+\lim_{\gamma\to+\infty}\frac{\ln\ln\left(1+\frac{\Omega\gamma}{m}\right)}{\log_2 \gamma}\right)+\lim_{\gamma\to+\infty}\frac{\log_2N}{2N} \label{tt002}\\
			&=1.\nonumber 
		\end{align}
	\end{subequations}
		\setcounter{equation}{131}
	\begin{subequations}\label{add10}
		\begin{align}
			s_{\infty}&\leq -\frac{1}{\theta NT}\lim_{\gamma\to+\infty}\frac{\ln\left[e^{-\theta T\frac{\log_2N}{2}}\gamma^{-\theta NT\log_2e}\left(\frac{m}{\Omega}\right)^{\theta NT\log_2 e-m}e^{\frac{m}{\Omega\gamma}}\Gamma\left(m-\theta NT \log_2 e,\frac{m}{\Omega\gamma}\right)\right]}{\log_2\gamma}\nonumber \\
			& =-\frac{1}{\theta NT}\lim_{\gamma \to+\infty} \frac{\ln\left[e^{-\theta T\frac{\log_2N}{2}}\gamma^{-m}e^{\frac{m}{\Omega\gamma}}\frac{\Gamma\left(m-\theta NT \log_2 e,\frac{m}{\Omega\gamma}\right)}{\left(\frac{m}{\Omega\gamma}\right)^{m-\theta NT\log_2 e}}\right]}{\log_2\gamma} \nonumber \\
			&=-\frac{1}{\theta NT}\left(\lim_{\gamma\to+\infty}\frac{-m\ln\gamma}{\log_2\gamma}+\lim_{\gamma\to+\infty}\frac{m}{\Omega \gamma \log_2 \gamma}\right)+\lim_{\gamma\to+\infty}\frac{\log_2N}{2N}\label{tt13}\\
			&=\frac{m}{\theta NT \log_2 e}. \nonumber 
		\end{align}
	\end{subequations}
	
	\setcounter{equation}{132}
	\begin{equation}
		\begin{aligned}
			s_{\infty}\geq-\frac{1}{\theta NT}\lim_{\gamma \to+\infty} \frac{\ln\left[e^{-\theta T\frac{\log_2N}{2}}\gamma^{-\theta NT \log_2 e}e^{\theta T\beta}\int_0^{\infty}\left(\frac{1}{\gamma}+x\right)^{-\theta NT \log_2 e}x^{m-1}e^{-\frac{m}{\Omega}x}{\rm d}x\right]}{\log_2 \gamma} .
		\end{aligned}\label{nakalow}
	\end{equation}

\end{figure*}

\subsection{Lower bound of $s_{\infty}$ for $m\geq 1$}

Since $\sqrt{1-x}\leq1$ holds for $x\in(0,1)$, the lower bound of $s_{\infty}$ in the Nakagami-$m$ fading channel is derived in Eq. \eqref{lownakaadd1}, shown at the bottom of this page. Eq. \eqref{tt6} holds since $m\geq 1$ and $\frac{1}{\gamma}+x>x$ for $x>0$. Since $\frac{m}{\Omega}$ is a finite constant, we have 
\begin{equation}
	\lim\limits_{\gamma\to+\infty}\frac{m}{\Omega\gamma}=0. \nonumber 
\end{equation}
Based on Eq. \eqref{lownakaadd1}, we will divide the following discussion into two parts regrading the value of $\frac{m}{\theta N T \log_2e}$.

\subsubsection{$1< \frac{m}{\theta NT\log_2 e}$}
\ 
\newline \indent  
According to \cite{incomplete}, for $s\in(0,1]\cup [2,+\infty)$, $\Gamma(s,x)$ satisfies
\setcounter{equation}{123}
\begin{equation}
\Gamma(s,x)<\frac{(x+v_s)^s-x^s}{v_s s }e^{-x}, \label{addcon1}
\end{equation}
where $v_s=\Gamma(s+1)^{\frac{1}{s-1}}>0$. For simplicity, we define 
\begin{equation}
	c_7=\Gamma(m+1-\theta NT\log_2 e)^{\frac{1}{m-1-\theta NT \log_2e}}. \nonumber 
\end{equation}
 Based on Eq. \eqref{addcon1}, for $m-\theta NT \log_2e\in(0,1]\cup[2,\infty)$, we obtain Eq. \eqref{slb2}, which is presented at the bottom of this page. Eq. \eqref{lb3} holds because  
\setcounter{equation}{125}
\begin{equation}
	\lim_{\gamma\to+\infty} \frac{\ln\left[\left(1+\frac{c_7\Omega}{m}\gamma\right)^{m-\theta NT \log_2 e}-1\right]}{(m-\theta NT \log_2 e)\ln\left(1+\frac{c_7\Omega}{m} \gamma\right)}=1,
\end{equation}
 where $c_7>0$ and $m-\theta NT \log_2 e>0$. Eq. \eqref{lbres} holds because $\lim\limits_{\gamma\to+\infty}\frac{\ln\left(1+\frac{c_7\Omega}{m}\gamma\right)}{\ln\gamma}=1$.

For $m-\theta NT\log_2e \in(1,2)$, we focus on the integral in Eq. \eqref{tt006}. When $m-\theta NT\log_2e \in(1,2)$,  the condition $\theta NT\log_2 e < m$ holds. Therefore, we find that the integral $	\int_{0}^{\infty}\left(\frac{1}{\gamma}+x\right)^{-\theta NT \log_2 e}x^{m-1}e^{-\frac{m}{\Omega}x}{\rm d}x$ is bounded by a finite constant according to Eq. \eqref{tt007}. Thus, we obtain Eq. \eqref{add9}, which is presented at the bottom of this page.

\subsubsection{$1\geq \frac{m}{\theta NT\log_2e}$}
\ 
\newline \indent  
For $m-\theta NT\log_2 e< 0$, we obtain Eq. \eqref{tt17}, which is presented at the bottom of the last page. Eq. \eqref{tt7} holds according to Eq. \eqref{limitgamma}.

\subsubsection{$1=\frac{m}{\theta NT\log_2 e}$}
\ 
\newline \indent  
For $m-\theta NT \log_2 e=0$, we have
\begin{equation}
	\Gamma\left(m-\theta NT \log_2 e,\frac{m}{\Omega\gamma}\right)=E_1\left(\frac{m}{\Omega\gamma}\right). \nonumber 
\end{equation}
 Based on Eq. \eqref{lownakaadd1}, we obtain Eq. \eqref{tt18}, which is presented at the bottom of this page. Eq. \eqref{tt10} holds because according to \cite{e1}, the following inequality holds:
\setcounter{equation}{129}
\begin{equation}
E_1(x)<e^{-x}\ln\left(1+\frac{1}{x}\right), \quad x>0.
\end{equation}

Therefore, according to the squeeze theorem, we have completed the proof for $m\geq 1$ based on the results shown in Eqs. \eqref{tt14}, \eqref{tt16}, \eqref{tt15}, \eqref{slb2}, \eqref{tt17}, and \eqref{tt18}. 

\subsection{Upper bound of $s_{\infty}$ for $m\in[0.5,1)$}

For $0.5\leq m<1$, there are some differences from the previous proofs for $m\geq 1$. The differences mainly lie in the proof for the upper bound when $1>\frac{m}{\theta NT\log_2e}$ and the lower bound. 

For the upper bound when $ 1\leq \frac{m}{\theta NT \log_2 e}$, the proof follows the same reasoning as presented in Eqs. \eqref{tt14} and \eqref{tt15}. In contrast, for the upper bound with $1>\frac{m}{\theta NT\log_2e}$, Eq. \eqref{tt11} is changed into
\begin{subequations}\label{nakaupp1}
\begin{align}
	&\int_0^{\infty}\left(\frac{1}{\gamma}+x\right)^{-\theta NT \log_2 e}x^{m-1}e^{-\frac{m}{\Omega}x}{\rm d}x \nonumber  \\
	=&\int_{0}^{\infty}\left(\frac{1}{\gamma}+x\right)^{-\theta NT \log_2 e+m -1} \left(\frac{1}{\gamma x}+1\right)^{1-m} e^{-\frac{m}{\Omega}x}{\rm d}x  \nonumber \\
	\geq & \int_{0}^{\infty}\left(\frac{1}{\gamma}+x\right)^{-\theta NT \log_2 e+m-1}e^{-\frac{m}{\Omega}x}{\rm d}x \label{tt12}  \\
	=&\left(\frac{m}{\Omega}\right)^{\theta NT\log_2 e-m}e^{\frac{m}{\Omega\gamma}}\Gamma\left(m-\theta NT \log_2 e,\frac{m}{\Omega\gamma}\right). \nonumber
\end{align}
\end{subequations}
Eq. \eqref{tt12} holds because $1-m>0$ and $1+\frac{1}{\gamma x}>1$ for $x>0$.

Then, by substituting Eq. \eqref{nakaupp1} into Eq. \eqref{nakaupp}, we derive Eq. \eqref{add10}, which is presented at the bottom of this page. Eq. \eqref{tt13} holds because $m-\theta NT\log_2 e<0$, which allows the application of Eq. \eqref{limitgamma}.

\begin{figure*}[b]
	\hrulefill

	\setcounter{equation}{133}
	\begin{equation}
		s_{\infty}\geq -\frac{1}{\theta NT}\lim_{\gamma \to+\infty}\frac{\ln\left[e^{-\theta T\frac{\log_2N}{2}}\gamma^{-\theta NT \log_2 e}e^{\theta T\beta}\int_0^{\infty}x^{-\theta NT \log_2 e+m-1}e^{-\frac{m}{\Omega}x}{\rm d}x\right]}{\log_2 \gamma}.\label{nakalow1}
	\end{equation}
	\setcounter{equation}{135}
	\begin{subequations}\label{nakalow2}
		\begin{align}
			&\int_0^{\infty}\left(\frac{1}{\gamma}+x\right)^{-\theta NT \log_2 e}x^{m-1}e^{-\frac{m}{\Omega}x}{\rm d}x \nonumber \\
			=& \int_0^{\infty}\left(\frac{1}{\gamma}+x\right)^{-\theta NT \log_2 e+m}\left(\frac{x}{\frac{1}{\gamma}+x}\right)^{m-\frac{1}{2}}\frac{x^{-\frac{1}{2}}}{\left(\frac{1}{\gamma}+x\right)^\frac{1}{2}}e^{-\frac{m}{\Omega}x}{\rm d}x  \nonumber\\
			\leq& \left(\frac{1}{\gamma}\right)^{-\theta NT \log_2 e+m}\int_0^{\infty}\frac{e^{-\frac{m}{\Omega}x}}{\sqrt{x\left(\frac{1}{\gamma}+x\right)}}{\rm d}x \label{nakalowprocess}\\
			=&\gamma^{\theta NT\log_2 e-m}e^{\frac{m}{2\Omega\gamma}}K_0\left(\frac{m}{2\Omega\gamma}\right). \nonumber
		\end{align}	
	\end{subequations}
	\setcounter{equation}{137}
	\begin{equation}\label{nakalow3}
		\begin{aligned}
			\int_0^{\infty}\left(\frac{1}{\gamma}+x\right)^{-\theta NT \log_2 e}x^{m-1}e^{-\frac{m}{\Omega}x}{\rm d}x		\leq \gamma^{\theta NT\log_2 e-m}e^{\frac{m}{2\Omega\gamma}}\ln\left(\frac{2\Omega\gamma}{m}\right)\frac{K_0\left(\frac{m}{2\Omega\gamma}\right)}{-\ln\left(\frac{m}{2\Omega\gamma}\right)}.
		\end{aligned}
	\end{equation}
\setcounter{equation}{138}
\begin{subequations}\label{add11}
	\begin{align}
		s_{\infty}&\geq-\frac{1}{\theta NT}\lim_{\gamma \to+\infty} \frac{\ln\left[e^{-\theta T\frac{\log_2N}{2}}\gamma^{-m}e^{\theta T\beta}e^{\frac{m}{2\Omega\gamma}}\ln\left(\frac{2\Omega\gamma}{m}\right)\frac{K_0\left(\frac{m}{2\Omega\gamma}\right)}{-\ln\left(\frac{m}{2\Omega\gamma}\right)}\right]}{\log_2 \gamma} \nonumber \\
		&=-\frac{1}{\theta NT}\lim_{\gamma\to+\infty}\left(\frac{-m\ln\gamma}{\log_2\gamma}+\frac{\theta T\beta}{\log_2\gamma}+\frac{\ln\ln\gamma}{\log_2\gamma}+\frac{\ln\frac{K_0\left(\frac{m}{2\Omega\gamma}\right)}{-\ln\left(\frac{m}{2\Omega\gamma}\right)}}{\log_2\gamma}\right)+\lim_{\gamma\to+\infty}\frac{\log_2N}{2N}\label{nakalow12}\\
		&=\frac{m}{\theta NT \log_2e}\label{nakalow4}.
	\end{align}
\end{subequations}
\end{figure*}

\subsection{Lower Bound of $s_{\infty}$ for $m\in[0.5,1)$}

Next, we focus on the lower bound of $s_{\infty}$ for $0.5\leq m<1$. The essence of this part of the proof remains the scaling of the integral  $\int_0^{\infty}\left(\frac{1}{\gamma}+x\right)^{-\theta NT \log_2 e}x^{m-1}e^{-\frac{m}{\Omega}x}{\rm d}x$. According to Eq. \eqref{lownakaadd1}, the lower bound of $s_{\infty}$ is given in Eq. \eqref{nakalow}, located at the bottom of this page.

\subsubsection{$1<\frac{m}{\theta NT \log_2 e}$}
\ 
\newline \indent
For $1<\frac{m}{\theta NT \log_2 e}$, we transform Eq. \eqref{nakalow} into Eq. \eqref{nakalow1}, which is presented at the bottom of the next page. Eq. \eqref{nakalow1} holds because $-\theta NT \log_2 e<0$ and $\frac{1}{\gamma}+x>x$. As mentioned previously, $\int_0^{\infty}z^{a}e^{-x}{\rm d}x<\infty$ holds when $a>-1$. Since in this case $1<\frac{m}{\theta NT \log_2 e}$, we have $-\theta NT \log_2 e+m-1>-1$. Thus, in this case the following integral is a finite constant:
\begin{equation}
	\begin{aligned}
		&\int_0^{\infty}x^{-\theta NT \log_2 e+m-1}e^{-\frac{m}{\Omega}x}{\rm d}x\\
		=&\left(\frac{\Omega}{m}\right)^{m-\theta NT \log_2e}\int_0^{\infty}t^{m-\theta NT\log_2 e-1}e^{-t}{\rm d}t.
	\end{aligned} \nonumber 
\end{equation}
Consequently, we obtain 
\setcounter{equation}{134}
\begin{equation}
	\begin{aligned}
		s_{\infty}&\geq -\frac{1}{\theta NT }\lim_{\gamma\to+\infty}\frac{-\theta NT \log_2 e\ln \gamma}{\log_2\gamma}-\\
		&\quad \quad \quad \quad \frac{1}{N}\lim_{\gamma\to+\infty}\frac{\beta}{\log_2\gamma}+\lim_{\gamma\to+\infty}\frac{\log_2N}{2N}\\
		&=1.\label{nakalow11}
	\end{aligned}
\end{equation}

\begin{figure*}[b]
	\hrulefill

\setcounter{equation}{140}
\begin{equation}
	\lim_{\gamma\to+\infty}\frac{\Lambda(\gamma)}{\log_2\gamma}+\log_2 e\lim_{\gamma\to+\infty}\frac{Q^{-1}(\epsilon(\gamma))}{\sqrt{N}\log_2\gamma}\geq \min\left\{1,\frac{m}{\theta NT\log_2e}\right\}.\label{newcoro13}
\end{equation}
\begin{equation}
	\lim_{\gamma\to+\infty}\frac{\Lambda(\gamma)}{\log_2\gamma}+(1-c_3)\log_2 e\lim_{\gamma\to+\infty}\frac{Q^{-1}(\epsilon(\gamma))}{\sqrt{N}\log_2\gamma}\leq \min\left\{1,\frac{m}{\theta NT\log_2e}\right\}.\label{newcorou3} 
\end{equation}

\setcounter{equation}{142}
\begin{equation}\label{newupp30}
	\begin{aligned}
		s_{\infty}&\leq -\frac{1}{\theta NT}\lim_{\gamma\to +\infty}\frac{e^{-\theta T\frac{\log_2N}{2}} \ln \left[\int_{0}^{\infty} e^{-\theta T\left( N\log_2(1+\gamma x)-\beta\left(1-\frac{1}{1+\gamma x}\right) \right)}\left(\frac{m}{\Omega}\right)^m\frac{x^{m-1}}{\Gamma(m)}e^{-\frac{m}{\Omega}x}{\rm d}x\right] }{\log_2 \gamma}\\
		&=-\frac{1}{\theta NT}\lim_{\gamma\to +\infty}\frac{ \ln \left[\gamma^{-\theta NT \log_2 e} e^{\theta T\beta}\int_{0}^{\infty} \left(\frac{1}{\gamma}+x\right)^{-\theta NT \log_2 e}x^{m-1}e^{-\left(\frac{m}{\Omega}x+\frac{\theta T\beta}{1+\gamma x}\right)}{\rm d}x\right] }{\log_2 \gamma} .
	\end{aligned}
\end{equation}

\setcounter{equation}{143}
\begin{equation}\label{add12}
	s_{\infty}\leq -\frac{1}{\theta NT}\lim_{\gamma\to +\infty}\frac{ \ln \left[\gamma^{-\theta NT \log_2 e} e^{\theta T\beta}\int_{0}^{\infty} \left(\frac{1}{\gamma}+x\right)^{-\theta NT \log_2 e}\left(1+\frac{1}{\gamma x}\right)^{-\theta T \beta}x^{m-1}e^{-\frac{m}{\Omega}x}{\rm d}x\right] }{\log_2 \gamma}.
\end{equation}
\setcounter{equation}{144}
\begin{equation}\label{add13}
	\begin{aligned}
		&\int_0^{\infty}\left(\frac{1}{\gamma}+x\right)^{-\theta NT \log_2 e}\left(1+\frac{1}{\gamma x}\right)^{-\theta T \beta}x^{m-1}e^{- \frac{m}{\Omega}x}{\rm d}x \\
		=&\int_{0}^{\infty}\left(\frac{1}{\gamma}+x\right)^{-\theta NT \log_2 e+m -1} \left(\frac{1}{\gamma x}+1\right)^{1-m-\theta T \beta} e^{-\frac{m}{\Omega}x}{\rm d}x . \\
	\end{aligned}
\end{equation}

\setcounter{equation}{145}
\begin{equation}\label{add14}
	\begin{aligned}
		&\int_{0}^{\infty}\left(\frac{1}{\gamma}+x\right)^{-\theta NT \log_2 e+m -1} \left(\frac{1}{\gamma x}+1\right)^{1-m-\theta T \beta} e^{-\frac{m}{\Omega}x}{\rm d}x\\
		\geq & \left(\frac{c_9}{c_9+1}\right)^{\theta T \beta+m-1}\int_{\frac{c_9}{\gamma}}^{\infty}\left(\frac{1}{\gamma}+x\right)^{-\theta NT \log_2 e+m-1} e^{-\frac{m}{\Omega}x}{\rm d}x \\
		=& \left(\frac{c_9}{c_9+1}\right)^{\theta T \beta+m-1}\left(\frac{m}{\Omega}\right)^{\theta N \log_2 e -m}e^{\frac{m}{\Omega\gamma}}\Gamma\left(m-\theta NT \log_2 e,\frac{m(1+c_9)}{\Omega\gamma}\right).
	\end{aligned}
\end{equation}
\setcounter{equation}{146}
\begin{equation}\label{tt004}
	s_{\infty}\leq -\frac{1}{\theta NT}\lim_{\gamma\to+\infty}\frac{ \ln \left[\gamma^{-\theta NT \log_2 e} e^{\theta T\beta}\left(\frac{c_9}{c_9+1}\right)^{\theta T \beta+m-1} \left(\frac{m}{\Omega}\right)^{\theta N \log_2 e -m}e^{\frac{m}{\Omega\gamma}}\Gamma\left(m-\theta NT\log_2 e,\frac{m(1+c_9)}{\Omega\gamma}\right)\right] }{\log_2 \gamma}.
\end{equation}
\setcounter{equation}{147}
\begin{equation}\label{newupp34}
	\begin{aligned}
		s_{\infty}&\leq -\frac{1}{\theta NT}\lim_{\gamma\to+\infty}\frac{-m\ln \gamma}{\log_2\gamma}-\left(1+\ln\frac{c_9}{c_9+1}\right)\lim_{\gamma\to+\infty}\frac{\beta}{N\log_2\gamma}\\
		&=\frac{m}{\theta NT\log_2 e}-\log_2e\left(1-\ln\left(1+\frac{1}{c_9}\right)\right)\lim_{\gamma\to+\infty}\frac{Q^{-1}(\epsilon(\gamma))}{\sqrt{N}\log_2\gamma}.
	\end{aligned}
\end{equation}
\end{figure*}

\subsubsection{$ 1\geq \frac{m}{\theta NT\log_2 e}$}
\ 
\newline \indent
For $1\geq \frac{m}{\theta NT\log_2 e}$, we obtain Eq. \eqref{nakalow2}, which is presented at the bottom of this page. In Eq. \eqref{nakalow2}, $K_v(z)$ denotes the modified Bessel function of the second kind \cite{e1}. Eq. \eqref{nakalowprocess} holds because in this case, we have $m-\frac{1}{2}\geq 0$ and $\frac{x}{\frac{1}{\gamma}+x}\leq 1$. According to \cite{e1}, for the modified Bessel function of the second kind $K_{v}(z)$ with $v=0$, we obtain
\setcounter{equation}{136}
\begin{equation}
\lim_{z\to0}\frac{K_0(z)}{-\ln(z)}=c_8, \label{bess1}
\end{equation} 
where $c_8>0$ is a constant. 

Based on this conclusion, we transformed Eq. \eqref{nakalow2} into Eq. \eqref{nakalow3}, which is presented at the bottom of this page. By substituting Eq. \eqref{nakalow3} into Eq. \eqref{nakalow}, we obtain Eq. \eqref{add11}, located at the bottom of this page. Eq. \eqref{nakalow4} holds according to Eq. \eqref{bess1}. By applying the squeeze theorem, we have completed the proof for $m\in[0.5,1)$. Thus, we have finished the proof.

\section{Proof of Theorem 6}
With the $\Psi(\gamma)$ satisfying Theorem 3, Eq. \eqref{ttmm1} still holds. For the AWGN channel, we can easily obtain 
\setcounter{equation}{139}
\begin{subequations}
\begin{align}
	s_{\infty}
	&=\lim_{\gamma\to+\infty}\frac{\tilde{R}_N^{\epsilon}}{N\log_2\gamma} \label{tt001}\\
	&=1-\log_2 e\lim_{\gamma\to+\infty}\frac{Q^{-1}\left(\epsilon(\gamma)\right)}{\sqrt{N}\log_2\gamma}. \nonumber 
\end{align}
\end{subequations}

For an FBL-SISO system in a Nakagami-$m$ fading channel, we will prove two inequalities shown in Eqs. \eqref{newcoro13} and \eqref{newcorou3}, which are located at the bottom of this page. In Eq. \eqref{newcorou3}, $c_3>0$ is an arbitrary constant. Since $c_3>0$ can be an arbitrarily small constant, we obtain Eq. \eqref{fading1} based on Eqs. \eqref{newcoro13} and \eqref{newcorou3}.

The derivation of Eq. \eqref{newcoro13} is based on the derivation of the lower bounds of $s_{\infty}$ in the proof of Theorem 5. For $m\geq 1$, we can directly obtain Eq. \eqref{newcoro13} according to Eqs. \eqref{lb3}, \eqref{tt7}, and \eqref{tt002}. For $m\in[0.5,1)$, we can derive Eq. \eqref{newcoro13} according to Eqs. \eqref{nakalow11} and \eqref{nakalow12}. Thus, we omit the details here.

The derivation of Eq. \eqref{newcorou3} is based on the derivation of the upper bounds of $s_{\infty}$. Different from the proof of Theorem 5, we start the proof with the help of $\sqrt{1-x}\geq1-\sqrt{x}$ with $x\in(0,1)$. We obtain Eq. \eqref{newupp30}, which is presented at the bottom of this page.

Since $\frac{x}{x+1}\leq \ln(1+x)$ holds for $x>0$, we have
\begin{equation}
	\frac{1}{1+\gamma x}\leq\ln\left(1+\frac{1}{\gamma x}\right). \nonumber 
\end{equation}
Therefore, we can transform Eq. \eqref{newupp30} into Eq. \eqref{add12}, which is presented at the bottom of this page. The integral in Eq. \eqref{add12} can be further manipulated to obtain Eq. \eqref{add13}, which is also presented at the bottom of this page.

\begin{figure*}[b]
	\hrulefill

\setcounter{equation}{149}
\begin{equation}\label{newupp35}
	\begin{aligned}
		s_{\infty}&\leq -\frac{1}{\theta NT}\lim_{\gamma\to+\infty}\frac{ \ln \left[\gamma^{-\theta NT \log_2 e} e^{\theta T\beta}\left(\frac{c_9}{c_9+1}\right)^{\theta T \beta} e^{-\frac{c_9m}{\Omega\gamma}}\left(1+\frac{m(1+c_9)}{\Omega\gamma}\right)^{m-\theta NT \log_2 e-1}\right] }{\log_2 \gamma}\\
		&=-\frac{1}{\theta NT}\lim_{\gamma\to+\infty}\left(\frac{-\theta NT\log_2 e \ln \gamma}{\log_2\gamma}+\left(1+\ln\frac{c_9}{c_9+1}\right)\frac{\theta T\beta}{\log_2\gamma}+(m-\theta NT-1) \log_2 e\frac{m(1+c_9)}{\gamma \log_2\gamma}\right)\\
		&=1-\log_2e\left(1-\ln\left(1+\frac{1}{c_9}\right)\right)\lim_{\gamma\to+\infty}\frac{Q^{-1}(\epsilon(\gamma))}{\sqrt{N}\log_2\gamma}.
	\end{aligned}
\end{equation}
\setcounter{equation}{151}
\begin{equation}\label{tt008}
	\begin{aligned}
		s_{\infty}
		&\leq-\frac{1}{\theta NT}\lim_{\gamma \to+\infty}\frac{-m\ln\gamma}{\log_2 \gamma}-\left(1+\ln\frac{c_2}{c_2+1}\right)\lim_{\gamma \to+\infty}\frac{\beta}{N\log_2 \gamma}-\left(\frac{m}{\theta NT}-1\right)\lim_{\gamma \to+\infty}\frac{\ln\left(1+\frac{c_1\Omega(1+c_2)}{m}\gamma\right)}{\log_2 \gamma} \\
		&=1-\log_2e\left(1-\ln\left(1+\frac{1}{c_2}\right)\right)\lim_{\gamma\to+\infty}\frac{Q^{-1}(\epsilon(\gamma))}{\sqrt{N}\log_2\gamma}.  
	\end{aligned}
\end{equation}

\end{figure*}

Note that, as $\gamma\to+\infty$, we have $\beta\to+\infty$. Therefore, $1-m-\theta T \beta<0$ holds. Based on this condition, we derive Eq. \eqref{add14} at the bottom of this page, where $c_9$ is an arbitrarily positive constant. Based on this result, we obtain Eq. \eqref{tt004}, which is also provided at the bottom of this page.

For $ 1>\frac{m}{\theta NT\log_2 e}$, by combining Eq. \eqref{limitgamma} with Eq. \eqref{tt004}, we obtain Eq. \eqref{newupp34}, which is at the bottom of this page. Note that, since $c_9$ is an arbitrarily positive constant, $\ln\left(1+\frac{1}{c_9}\right)$ is also an arbitrarily positive constant. 

For $1\leq \frac{m}{\theta NT\log_2 e}$, we will divide the proof into two parts. 

\subsubsection{$1\leq \frac{m}{\theta NT\log_2 e}\leq1+\frac{1}{\theta NT\log_2 e}$ and $\frac{m}{\theta NT\log_2 e}\geq1+\frac{2}{\theta NT\log_2 e}$}
\ 
\newline \indent  
The first part is $1\leq \frac{m}{\theta NT\log_2 e}\leq1+\frac{1}{\theta NT\log_2 e}$ and $\frac{m}{\theta NT\log_2 e}\geq1+\frac{2}{\theta NT\log_2 e}$, which implies $0\leq m-\theta N T \log_2e\leq 1$ and $m-\theta NT \log_2e \geq 2$. We will utilize the conclusion that the following inequality holds for $x>0$ with $s\leq 1$ and $s\geq 2$:
\begin{equation}
	\Gamma(s,x)\geq e^{-x}(1+x)^{s-1}. \nonumber
\end{equation}
This can be easily demonstrated using Jensen's inequality, as $(t+x)^{s-1}$ is a convex function of $t$ with a $s\leq 1$ and $s\geq 2$. As a result, we have 
\setcounter{equation}{148}
\begin{equation}\label{tt003}
	\begin{aligned}
		\int_x^{\infty} y^{s-1}e^{-y}{\rm d}y&=e^{-x}\int_0^{\infty}(t+x)^{s-1}e^{-t}{\rm d}t\\
		&\geq e^{-x}\left(\int_0^{\infty} (t+x)e^{-t}{\rm d}t\right)^{s-1}\\
		&=e^{-x}(1+x)^{s-1}.
	\end{aligned}
\end{equation}
Now we have proved that $\Gamma(s,x)\geq e^{-x}(1+x)^{s-1}$ holds for $s\leq 1$ and $s\geq 2$. Therefore, by substituting Eq. \eqref{tt003} into Eq. \eqref{tt004}, we obtain Eq. \eqref{newupp35}, which appears at the bottom of this page.

\subsubsection{$ 1+\frac{1}{\theta NT\log_2 e}<\frac{m}{\theta NT\log_2 e}<1+\frac{2}{\theta NT\log_2 e}$}
\ 
\newline \indent  
The second part considers $ 1+\frac{1}{\theta NT\log_2 e}<\frac{m}{\theta NT\log_2 e}<1+\frac{2}{\theta NT\log_2 e}$, which implies $ 1<m-\theta N T \log_2e<2$. For $\Gamma(s,x)$ with $1<s<2$, according to \cite{incomplete}, we have
\setcounter{equation}{150}
\begin{equation}\label{tt005}
\Gamma(s,x)>\frac{(x+v_s)^s-x^s}{v_s s }e^{-x},
\end{equation}
where $v_s$ is defined after Eq. \eqref{addcon1}. By substituting Eq. \eqref{tt005} into Eq. \eqref{tt004}, we can derive Eq. \eqref{tt008} using similar procedures as shown in Eq. \eqref{slb2}. Eq. \eqref{tt008} is presented at the bottom of this page. Now we have completed the proof of Eq. \eqref{newcorou3}.

\begin{figure*}[b]
	\hrulefill
\setcounter{equation}{156}
\begin{equation}\label{add15}
	\left(x+\frac{1}{x}\right)^2+\ln\left(x+\frac{1}{x}\right)^2>x^2+\ln\left(x+\frac{1}{x}\right)^2\geq -2\ln\left(\sqrt{2\pi}\epsilon(\gamma)\right).
\end{equation}
	\setcounter{equation}{159}
	\begin{equation}\label{newgain4}
		\begin{aligned}
			\lim_{\gamma\to+\infty}\frac{Q^{-1}(\epsilon(\gamma))}{\sqrt{N}\log_2\gamma}&\geq \lim_{\gamma\to+\infty}\frac{\sqrt{W\left(\frac{\epsilon^{-2}(\gamma)}{2\pi}\right)}}{\sqrt{N}\log_2\gamma}-\lim_{\gamma\to+\infty}\frac{1}{Q^{-1}(\epsilon(\gamma))\cdot\sqrt{N}\log_2\gamma } \\&=\frac{c_4}{\log_2e}\varpi .
		\end{aligned}
	\end{equation}
	
	\setcounter{equation}{163}
\begin{equation}
	g'(x)=\frac{-(\ln 2)^{-1}\left(\gamma (1+\gamma x)\sqrt{(1+\gamma x)^2-1} -\sqrt{\frac{1}{N}}Q^{-1}(\epsilon)\gamma\right)}{(1+\gamma x)^3\sqrt{\big( 1-(1+\gamma x)^{-2}\big)}}. \label{deri}
\end{equation}
\setcounter{equation}{166}
\begin{equation}
	\int_{0}^{\infty}f(x)e^{-\theta NT  r(x)} dx = \sqrt{\frac{2\pi}{\theta NT |r''(x^*)|}}f(x^*)e^{-\theta NT r(x^*)}\left(1+O\left((\theta NT)^{-1}\right)\right). \label{appF1}
\end{equation}
\end{figure*}

\section{Proof of Lemma 1}
Based on the definition of $\varpi$, we obtain
\setcounter{equation}{152}
\begin{equation}
\lim_{\gamma\to+\infty}\frac{1}{\sqrt{N}}\frac{\partial \sqrt{-\log_2\epsilon(\gamma)}}{\partial \log_2 \gamma}=\varpi\sqrt{\frac{1}{2\log_2e}}.
\end{equation}
Under the assumption that $\varpi$ is finite, we have
\begin{subequations}\label{gain0}
\begin{align}
	&\lim_{\gamma\to+\infty}\frac{1}{\sqrt{N}}\sqrt{\frac{-\log_2\epsilon(\gamma)}{\log_2^2\gamma}}\\
	=&\lim_{\gamma\to+\infty}\frac{\sqrt{-\log_2\epsilon(\gamma)}}{\sqrt{N}\log_2\gamma} \label{le2ex1} \\
	=&\lim_{\gamma\to+\infty} \frac{\frac{\partial \sqrt{-\log_2\epsilon(\gamma)}}{\partial \log_2 \gamma} }{\sqrt{N}+\frac{\partial \sqrt{N}}{\partial \log_2 \gamma} \log_2\gamma } \label{le2ex2}\\
	=&\varpi\sqrt{\frac{1}{2\log_2e}}\lim_{\gamma\to+\infty}\frac{1}{1+\frac{\partial \sqrt{N}}{\partial \log_2 \gamma}\frac{\log_2\gamma}{\sqrt{N}}}. \nonumber 
\end{align}
\end{subequations}
Eq. \eqref{le2ex2} holds according to L'Hospital's rule. As defined in Lemma 1, we let 
\begin{equation}
	c_4=\lim_{\gamma\to+\infty}\frac{1}{1+\frac{\partial \sqrt{N}}{\partial \log_2 \gamma}\frac{\log_2\gamma}{\sqrt{N}}}. \nonumber
\end{equation}
 Since $\lim\limits_{\gamma\to+\infty}\Psi(\gamma)=+\infty$, we have
\begin{equation}
	\frac{\partial \sqrt{N}}{\partial \log_2 \gamma}\frac{\log_2\gamma}{\sqrt{N}}\geq 0. \nonumber 
\end{equation} 
Therefore, we conclude that $c_3\in[0,1]$.

Next, we let $Q^{-1}(\epsilon(\gamma))=x$, which can also be expressed as $\epsilon(\gamma)=Q(x)$. According to Eq. \eqref{gain1}, we have
\begin{equation}
x^2+\ln x^2\leq-2\ln\left(\sqrt{2\pi}\epsilon(\gamma)\right). \label{gain2}
\end{equation}

By substituting $\epsilon(\gamma)$ for $v$ in Eq. \eqref{newgain11}, we obtain
\begin{equation}
-\sqrt{W\left(\frac{\epsilon^{-2}(\gamma)}{2\pi}\right)} \leq x\leq  \sqrt{W\left(\frac{\epsilon^{-2}(\gamma)}{2\pi}\right)}.\label{newgain1}
\end{equation}

Since $x>0$, we have $0<x\leq  \sqrt{W\left(\frac{\epsilon^{-2}(\gamma)}{2\pi}\right)}$. Similarly, based on Eq. \eqref{gain1}, we obtain Eq. \eqref{add15}, which is presented at the bottom of this page.

Based on the definition of $W(\cdot)$ and Eq. \eqref{add15}, we have
\setcounter{equation}{157}
\begin{equation}
	\begin{aligned}
		&x+\frac{1}{x}\geq   \sqrt{W\left(\frac{\epsilon^{-2}(\gamma)}{2\pi}\right)} \text{ or } \\
		 &x+\frac{1}{x}\leq   -\sqrt{W\left(\frac{\epsilon^{-2}(\gamma)}{2\pi}\right)}.
	\end{aligned}
 \label{newgain2}
\end{equation}
Since $x>0$, we will utilize $x+\frac{1}{x}\geq \sqrt{W\left(\frac{\epsilon^{-2}(\gamma)}{2\pi}\right)}$ in the following proof.

Based on Eq. \eqref{newgain2}, we obtain
\setcounter{equation}{158}
\begin{subequations}
\begin{align}
	\lim_{\gamma\to+\infty}\frac{Q^{-1}(\epsilon(\gamma))}{\sqrt{N} \log_2\gamma}&\leq \lim_{\gamma\to+\infty}\frac{\sqrt{W\left(\frac{\epsilon^{-2}(\gamma)}{2\pi}\right)}}{\sqrt{N} \log_2\gamma}\nonumber \\
	&=\lim_{\gamma\to+\infty}\frac{\sqrt{-2\log_2\epsilon(\gamma)}}{\sqrt{N}\log_2\gamma}\cdot\frac{\sqrt{W\left(\frac{\epsilon^{-2}(\gamma)}{2\pi}\right)}}{\sqrt{-2\log_2\epsilon(\gamma)}}\nonumber \\
	&=\frac{1}{\sqrt{\log_2e}}\lim_{\gamma\to+\infty}\sqrt{\frac{-2\log_2\epsilon(\gamma)}{\log_2^2\gamma}}\cdot\frac{1}{\sqrt{N}}\label{newgain3a}\\
	&=\frac{c_4}{\log_2e}\varpi \label{newgain3b}.
\end{align}
\end{subequations}
Eq. \eqref{newgain3a} holds according to Eq. \eqref{tt20}. Besides, Eq. \eqref{newgain3b} holds according to Eq. \eqref{gain0}.

Similarly, based on Eq. \eqref{newgain2}, we derive Eq. \eqref{newgain4}, which is shown at the bottom of this page. Eq. \eqref{newgain4} holds because as $\gamma\to+\infty$, $Q^{-1}(\epsilon(\gamma))\to+\infty$. Therefore, by applying the squeeze theorem, we have completed the proof.

\section{Proof of Corollary 3}

Based on Lemma 1 in \cite{li1}, we find that $\alpha_A(\theta)$ is an increasing function of $\theta$. Therefore, the inverse function of $\alpha_A(\cdot)$ exists, which is denoted by $\alpha^{-1}_A(\cdot)$. According to Eq. \eqref{eqso1}, the $\theta$ for the AWGN channel can be expressed as 
\setcounter{equation}{160}
\begin{equation}\label{nn1}
\theta=\alpha_A^{-1}(R_N^{\epsilon}).
\end{equation}

From Eq. \eqref{9}, we know that $\chi\leq e^{-\theta L}$. Substituting Eq. \eqref{nn1} into $\chi\leq e^{-\theta L}\leq \chi_{\rm th}$, we have
\begin{equation}
\begin{aligned}
	\alpha_A^{-1}(R_N^{\epsilon})\geq -\frac{1}{L}\ln\chi_{\rm th}.
\end{aligned}
\end{equation}

Since $\alpha_A(\theta)$ is an increasing function of $\theta$, it follows $R_N^{\epsilon}\geq \alpha_A\left(-\frac{1}{L}\ln \chi\right)$. Substituting Eq. \eqref{c1} into this inequality, we obtain
\begin{equation}\label{nn001}
	\begin{aligned}
		NC-\sqrt{NV}Q^{-1}(\epsilon)+\frac{\log_2 N}{2}+&G(N,\gamma,\epsilon)\\
		&\geq  \alpha_A\left(-\frac{1}{L}\ln \chi\right).
	\end{aligned}
\end{equation}
From Eq. \eqref{nn001}, we obtain Eq. \eqref{nn002}.

\section{Proof of Lemma 2}

We put our focus on the expectation term in the definition of EC. For simplicity, let $g(x)=-r(x)$. The derivative of $g(x)$ is provided in Eq. \eqref{deri}, which is presented at the bottom of this page. 

\begin{figure*}[b]
	\hrulefill
	
\setcounter{equation}{167}
\begin{equation}
	g'(x)=\frac{-(\ln 2)^{-1}\left(\Xi+\Xi'x\right) \left( (1+\Xi x)\sqrt{(1+\Xi x)^2-1} -\sqrt{\frac{1}{N}}Q^{-1}(\epsilon) \right)}{(1+\Xi x)^3\sqrt{\big( 1-(1+\Xi x)^{-2}\big)}}. \label{deri2}
\end{equation}
\end{figure*}

we find that $(1+\gamma x)\sqrt{(1+\gamma x)^2-1}$ is an increasing function of $x$, which equals 0 when $x=0$. Therefore, based on Eq. \eqref{deri}, there exists a stationary point $x^*$. When $x=x^*$, $g'(x)=0$, $x<x^*$, $g'(x)>0$, and $x>x^*$, $g'(x)<0$. $x^*$ satisfies the following equation:
\setcounter{equation}{164}
\begin{equation}
(1+\gamma x^*)\sqrt{(1+\gamma x^*)^2-1}=\frac{Q^{-1}(\epsilon)}{\sqrt{N}}. \label{opti}
\end{equation}

The stationary point $x^*$ can be solved from Eq. \eqref{opti}. Specifically, we have 
\begin{equation}
x^*=\frac{1}{\gamma}\left(\sqrt{\frac{1+\sqrt{1+4\ell}}{2}}-1\right),
\end{equation}
which is an interior point of $(0,\infty)$. According to Laplace's method \cite{SPA}, the integral in the definition of EC can be expressed as shown in Eq. \eqref{appF1}, which is presented at the bottom of the last page. By substituting Eq. \eqref{appF1} into Eq. \eqref{ebs}, we derive Eq. \eqref{lemma31}.

\section{Proof of Theorem 8}
Similar to the proof of Lemma 2, the derivative of $g(x)$ is given in Eq. \eqref{deri2}, where we omit the argument of $\Xi$ for simplicity. Eq. \eqref{deri2} is presented at the bottom of this page. By comparing Eqs. \eqref{deri2} and \eqref{deri}, we find that if $\Xi+\Xi' x\geq 0$, the remaining analysis is identical with the proof of Lemma 2. Therefore, we omit the details.

\section{Proof of Corollary 4}
According to \cite{eur}, the EC with the simple ARQ mechanism that allows for unlimited retransmissions is given by
\setcounter{equation}{168}
\begin{equation}
\begin{aligned}
\alpha_S(\theta)&=-\frac{1}{\theta T}\ln \mathbb{E}_{x}\left\{\epsilon+(1-\epsilon)e^{-\theta NT r(x)}\right\} \\
&=-\frac{1}{\theta T}\ln\left( \epsilon+(1-\epsilon)\int_{0}^{\infty}e^{-\theta NT r(x)} f(x) {\rm d}x\right). \label{appH1}
\end{aligned}
\end{equation}
Similar to the proofs of Lemma 2 and Theorem 8, if the power allocation scheme satisfies $\Xi(x)+\frac{{\rm d}\Xi(x)}{{\rm d}x}x\geq 0$, the integral $\int_{0}^{\infty}e^{-\theta NT r(x)} f(x) {\rm d}x$ can be approximated as shown in Eq. \eqref{appF1} using Laplace's method. By substituting Eq. \eqref{appF1} into Eq. \eqref{appH1}, we obtain Eq. \eqref{coro31}.

\section*{Acknowledgment}

The authors would like to express their gratitude to the editor and anonymous reviewers for their careful reading and constructive comments. In particular, we appreciate their identification of the double-limit issue and their insightful suggestions on addressing it.

\bibliographystyle{IEEEtran}
\bibliography{mybib}

\begin{IEEEbiographynophoto}{Lintao Li} (Graduate Student Member, IEEE) received the B.S. degree in electronic engineering from Xidian University, Xi’an, China, in 2020. He is currently pursuing the Ph.D. degree with the Department of Electronic Engineering, Tsinghua University, Beijing, China. From October 2023 to January 2024, he was a Visiting Ph.D. Student with the Department of Electronic Systems, Aalborg University, Denmark. He received the National Scholarship in 2019 and 2024. His research interests include low-latency communication, goal-oriented communication, and immersive communication.
\end{IEEEbiographynophoto}

\begin{IEEEbiographynophoto}{Wei Chen}
(Senior Member, IEEE) received the B.S. and Ph.D. degrees (Hons.) from Tsinghua University in 2002 and 2007, respectively. Since 2007, he has joined Tsinghua University, where he is currently a tenured full professor. He is also a standing committee member of All-China Youth Federation and the secretary-general of its education board. His research interests are in the areas of information theory, queueing theory, distributed optimization, and networked control.

Prof. Chen received the IEEE Marconi prize paper award in wireless communications in 2009 and the IEEE communications society Asia Pacific board best young researcher award in 2011. He is a recipient of the National May 1st Labor Medal and the China Youth May 4th Medal. He is a member of the National Program for Special Support of Eminent Professionals, also known as 10,000 talent program. He is now serving as an editor for IEEE transactions on wireless communications. He served as editors for IEEE transactions on communications and IEEE wireless communications letters.
\end{IEEEbiographynophoto}

\begin{IEEEbiographynophoto}{Petar Popovski}
	(Fellow, IEEE) received the Dipl.- Ing. (1997) and M.Sc. (2000) degrees in communication engineering from the University of Sts. Cyril and Methodius (UKIM), Skopje, North Macedonia, and the Ph.D. degree (2005) from Aalborg University, Aalborg, Denmark. He is currently a Professor with Aalborg University, where he heads the section on Connectivity and a Visiting Excellence Chair with the University of Bremen, Bremen, Germany. He authored the book ‘‘Wireless Connectivity: An Intuitive and Fundamental Guide.” His research interests include wireless communication and communication theory. He was the recipient of an ERC Consolidator Grant in 2015, Danish Elite Researcher award in 2016, IEEE Fred W. Ellersick prize in 2016, IEEE Stephen O. Rice prize in 2018, Technical Achievement Award from the IEEE Technical Committee on Smart Grid Communications in 2019, Danish Telecommunication Prize in 2020 and Villum Investigator Grant in 2021. He was a Member at Large at the Board of Governors in IEEE Communication Society from 2019 to 2021, the General Chair for IEEE SmartGridComm 2018 and IEEE Communication Theory Workshop 2019.  Prof. Popovski is currently the Editor-in-Chief of IEEE JOURNAL ON SELECTED AREAS IN COMMUNICATIONS and the Chair of the IEEE Communication Theory Technical Committee.

\end{IEEEbiographynophoto}

\begin{IEEEbiographynophoto}{Khaled B. Letaief}
(Fellow, IEEE) is an internationally recognized leader in wireless communications and networks with research interest in artificial intelligence, integrated sensing and communication, mobile cloud and edge computing, tactile Internet, and 6G systems.  In these areas, he has over 700 papers with over 55,670 citations and an h-index of 100 along with 15 inventions, including 11 US patents.

He is a Member of the United States National Academy of Engineering, Fellow of IEEE, Fellow of the Hong Kong Institution of Engineers, Member of India National Academy of Sciences, and Member of the Hong Kong Academy of Engineering Sciences. He is also recognized by Thomson Reuters as an ISI Highly Cited Researcher and was listed among the 2020 top 30 of AI 2000 Internet of Things Most Influential Scholars.

Dr. Letaief is the recipient of many distinguished awards and honors including the 2024 Distinguished Purdue University Alumni Award; 2022 IEEE Communications Society Edwin Howard Armstrong Achievement Award; 2021 IEEE Communications Society Best Survey Paper Award; 2019 IEEE Communications Society and Information Theory Society Joint Paper Award; 2016 IEEE Marconi Prize Paper Award in Wireless Communications; 2011 IEEE Communications Society Harold Sobol Award; 2007 IEEE Communications Society Joseph LoCicero Publications Exemplary Award; and over 20 IEEE Best Paper Awards.

Since 1993, he has been with the Hong Kong University of Science \& Technology (HKUST) where he has held many administrative positions, including Acting Provost, Head of the Electronic and Computer Engineering department, Director of the Wireless IC Design Center, and Director of the Hong Kong Telecom Institute of Information Technology.  
While at HKUST he has also served as Chair Professor and Dean of Engineering.  From September 2015 to March 2018, he joined HBKU as Provost to help establish a research-intensive university in Qatar in partnership with strategic partners that include Northwestern University, Carnegie Mellon University, Cornell, and Texas A\&M.  
Dr. Letaief is well recognized for his dedicated service to professional societies and IEEE where he has served in many leadership positions.  These include founding Editor-in-Chief of the prestigious IEEE Transactions on Wireless Communications. He also served as President of the IEEE Communications Society (2018-19), the world's leading organization for communications professionals with headquarter in New York City and members in 162 countries.  

Dr. Letaief received the BS degree with distinction in Electrical Engineering from Purdue University at West Lafayette, Indiana, USA, in December 1984. He received the MS and Ph.D. Degrees in Electrical Engineering from Purdue University, in Aug. 1986, and May 1990, respectively.  He has also received a Ph.D. Honoris Causa from the University of Johannesburg, South Africa, in 2022.
\end{IEEEbiographynophoto}

\end{document}